%% file: beyondFlatCBC-draft.tex
\documentclass[12pt,oneside,letterpaper]{article}
\usepackage{titlesec}
\titleformat*{\section}{\large\bfseries}
\titleformat*{\subsection}{\normalsize\bfseries}
\usepackage{amssymb}
\usepackage{amsmath}
\usepackage[dvips]{graphicx}
\usepackage{setspace}
\usepackage{stackrel}
\usepackage{amsfonts}
\usepackage{fancyhdr}
\usepackage[dvipsnames]{xcolor}
\usepackage{caption}
\usepackage{graphicx}
\usepackage{rotating}
\usepackage{comment}
\usepackage{color}
\usepackage{cite}
\usepackage{braket}
\usepackage{subcaption}
\numberwithin{equation}{section}
\definecolor{darkgreen}{rgb}{0,0.5,0}
\definecolor{darkblue}{rgb}{0,0,0.6}
\definecolor{purple}{rgb}{0.4,.2,0.7}
\newcommand{\p}{\partial}

\newcommand{\f}{\frac}

\newcommand{\be}{\begin{equation}}
\newcommand{\ee}{\end{equation}}

\newcommand{\mc}{\mathcal}

\usepackage{tikz}
\usetikzlibrary{matrix}
\usetikzlibrary{decorations.markings,calc,shapes,decorations.pathmorphing,patterns,decorations.pathreplacing}
\usetikzlibrary{positioning}

\definecolor{penroseblue}{RGB}{94,153,220}
\definecolor{penrosered}{RGB}{158,28,74}

\usepackage[colorlinks=true,citecolor=darkgreen,linkcolor=black,urlcolor=purple]{hyperref}

\usepackage{pdfsync}
\makeatletter
\newcommand*{\defeq}{\equiv}
\makeatother

\DeclareMathOperator{\Tr}{Tr}
\def\be{\begin{eqnarray}}
\def\ee{\end{eqnarray}}

\newcommand{\bea}{\begin{eqnarray}}
\newcommand{\eea}{\end{eqnarray}}

\def\ben{\begin{equation}}
\def\een{\end{equation}}

\let\a=\alpha \let\b=\beta \let\g=\gamma \let\d=\delta 
   
\let\l=\lambda \let\m=\mu \let\n=\nu  \let\p=\phi \let\r=v
 \let\t=\tau

\let\w=\omega    \let\L=\Lambda

\def\nn{\nonumber}
\let\f=\frac

\def\be{\begin{equation}}
\def\ee{\end{equation}}
\def\ba{\begin{array}}
\def\ea{\end{array}}

\def\ba#1\ea{\begin{align}#1\end{align}}
\def\bs#1\es{\begin{split}#1\end{split}}

\usepackage{amsmath}

\renewcommand{\p}{\partial}

\usepackage[top=1in, bottom = 1in, left = 1in, right = 1in]{geometry}

\allowdisplaybreaks % allow page breaks in displayed eqs

\begin{document}
\onehalfspacing

\begin{center}

~
\vskip5mm

{\LARGE {
Thermal effective actions from\\ \vspace{2mm} conformal boundary conditions in gravity
}}
% thermodynamics of inner horizon cosmologies
\vskip8mm

Batoul Banihashemi, Edgar Shaghoulian, and Sanjit Shashi

\vskip8mm
{ \it UC Santa Cruz\\
Physics Department\\
1156 High Street\\
Santa Cruz, CA 95064}

%\tt{eshaghoulian@ucsc.edu}

\end{center}

\vspace{4mm}

\begin{abstract}
\noindent
Recent work has studied the thermodynamics of gravity subjected to conformal boundary conditions, where the trace of the extrinsic curvature $K$ and the conformal class of metrics are kept fixed. The case $K < 0$ seems tied to solutions with cosmic horizons. To study this situation further, we analyze Einstein--Maxwell theory, where we find solutions with cosmic horizons corresponding to patches inside the inner horizon of the Reissner--Nordstr\"om black hole. We find that these solutions are \emph{necessary} for consistency with the thermal effective action in the putative boundary dual at $O(1)$ chemical potential. Furthermore, we introduce a (positive or negative) cosmological constant and find more cases where solutions with cosmic horizons dominate the ensemble. Transitions between solutions with cosmic horizons and solutions with black hole horizons are mediated by ``extremal" solutions that correspond to patches of (A)dS$_2 \times \Sigma^{d-1}$, for which we discuss an effective JT gravity description with dimensionally reduced 
conformal boundary conditions. Finally, we introduce rotation in some simple cases and find that the bulk theory reproduces the universal velocity-dependent modification of the free energy predicted by the thermal effective action.
\end{abstract}

%\tt{600 Warren Road \#4-3F\\
%Ithaca, NY 14850}

%\tt{Submitted on March 31, 2020}

\thispagestyle{empty}
\pagebreak
\pagestyle{plain}

\setcounter{tocdepth}{2}
{}
\vfill
\clearpage
\setcounter{page}{1}
\vspace*{-20mm}
\tableofcontents
\vfill\pagebreak

\section{Introduction}\label{sec:intro}
In recent work \cite{Anninos:2023epi, Banihashemi:2024yye}, it was shown that gravity with vanishing cosmological constant subjected to conformal boundary conditions---where the boundary data held fixed is the conformal metric and trace of the extrinsic curvature \cite{York:1972sj, York:1986it, York:1986lje, Anderson:2006lqb, An:2021fcq, An:2025rlw}---leads to thermodynamic quantities consistent with a local boundary dual. In particular, the thermodynamics corroborates the predictions of the thermal effective action for the boundary theory when it is placed on a spatial sphere, hyperbolic space, or a flat torus.

The solution relevant for the boundary theory placed on a spatial sphere is, unsurprisingly, given by a patch of the Schwarzschild black hole \cite{Anninos:2023epi}. The solutions for the hyperbolic and flat spaces are somewhat more surprising \cite{Banihashemi:2024yye}. In the hyperbolic case, the solution can be understood as a patch of the geometry one obtains by taking the flat-space limit of the hyperbolic black hole in AdS. In the flat case, the solution is simply a patch of Rindler space. 

The high-temperature thermodynamics with nonvanishing cosmological constant has also been shown to be extensive \cite{Anninos:2024wpy, Banihashemi:2024yye}, with a detailed study of the thermodynamics with positive cosmological constant performed for spherical spatial slices in \cite{Anninos:2024wpy}. Thermodynamics with Dirichlet boundary conditions and positive cosmological constant has also been studied, most recently in \cite{Gorbenko:2018oov, Lewkowycz:2019xse, Coleman:2021nor, Banihashemi:2022jys, Banihashemi:2022htw, Batra:2024kjl, Silverstein:2024xnr}, and exhibits distinct behavior. Aspects of stability in Lorentzian signature can be found in \cite{Anninos:2023epi, Anninos:2024wpy, Liu:2024ymn, Anninos:2024xhc}.

The (linearized) well-posedness of conformal boundary conditions \cite{An:2025rlw} and their consistency with the boundary thermal effective action provides a potential route to understanding holography in finite patches of spacetime (to be contrasted with the approach using Dirichlet boundary conditions \cite{McGough:2016lol,  Giveon:2017nie, Taylor:2018xcy, Hartman:2018tkw, Gross:2019ach, Gross:2019uxi, Guica:2019nzm, Araujo-Regado:2022gvw, Anninos:2022ujl}). They may even provide a way to project the data of an observer in quantum gravity to boundary data on a worldtube surrounding the observer. This would manifest a connection between observers and worldlines  \cite{Anninos:2011af, Anninos:2011zn, Carrozza:2021gju, Carrozza:2022xut, Goeller:2022rsx, Chandrasekaran:2022cip, Loganayagam:2023pfb, Shaghoulian:2023odo, Blacker:2023oan, Witten:2023qsv, Kolchmeyer:2024fly, Abdalla:2025gzn, Harlow:2025pvj, Akers:2025ahe} or observers and boundaries \cite{Shaghoulian:2021cef, Levine:2022wos,  Banihashemi:2022jys, Svesko:2022txo, Maldacena:2024spf} observed in various guises.

To make progress toward these lofty goals, we will focus on the generality of the connection between conformal boundary conditions and thermal effective actions on the boundary. In this paper, we will generalize the analysis from \cite{Anninos:2023epi, Anninos:2024wpy, Banihashemi:2024yye} by studying the case with an electric potential and a cosmological constant, with radial slices hosting an arbitrary maximally symmetric space. This will require studying Einstein--Maxwell theory with a nonzero cosmological constant. For spherical spatial slices, the relevant solutions to the conformal boundary condition problem will correspond to patches of the Reissner--Nordstr\"om-(A)dS geometry, although for hyperbolic and flat slices they will correspond to (analytic continuations and limits of) charged black holes in AdS.
Throughout this paper we will see that there are two distinct types of solutions with horizons that are important for our analysis. One type of solution will contain what we call black hole horizons, where the horizon is a maximin surface in Lorentzian signature, i.e. maximum in time and minimum in space \cite{Wall:2012uf}. In the usual Schwarzschild-like radial coordinate $r$, these are solutions with $r_c > r > r_h$ where $r_h$ is the radius of the horizon and $r_c$ is the radius of the boundary. The other type of solution will contain what we call cosmic horizons, where the horizon is a minimax surface in Lorentzian signature, i.e. minimum in time and maximum in space \cite{Shaghoulian:2021cef}. These are solutions with $r_c < r < r_h$ and include cases where $r_h$ is the radius of the de Sitter horizon but also cases where $r_h = r_-$ is the radius of the inner horizon of a black hole. As there are a lot of cases and analyses to follow, let us summarize the various upshots of this work:
\begin{itemize}
\item We will calculate the solution space and phase structure with conformal boundary conditions in the following cases: Einstein--Maxwell theory with vanishing cosmological constant and Einstein gravity with arbitrary cosmological constant. We will consider the putative boundary theory on a spatial $S^{d-1}$, $\mathbb{H}^{d-1}$, and flat $\mathbb{T}^{d-1}$. 

\item We will analytically solve for the dominant solution in a high-temperature expansion with arbitrary cosmological constant and electric potential. This expansion will exactly agree with the predictions of the thermal effective action of the boundary theory.

\item We will introduce an angular potential to the boundary theory on $\mathbb{T}^{d-1}$. The relevant bulk solutions can be constructed by boosting and compactifying the solutions with horizon geometry $\mathbb{R}^{d-1}$. These solutions will reproduce the universal velocity-dependent modification of the free energy predicted by the thermal effective action of the boundary theory. 

\item In \cite{Banihashemi:2024yye}, we found that the first subleading Wilson coefficient in the thermal effective action was negative, $c_1 < 0$, violating a QFT conjecture from \cite{allameh}. This was interpreted as being due to the fact that the boundary theory is not a local QFT, but is instead coupled to a metric degree of freedom due to the fact that the Weyl mode is not fixed by conformal boundary conditions. In this paper we find $c_1 < 0$ for arbitrary cosmological constant.

\item In \cite{Banihashemi:2024yye}, we found a situation where cosmic horizons are part of the solution space of the boundary value problem, and noticed that in this situation we could also have solutions with $K < 0$. In this paper we will consider many additional examples with cosmic horizons and find again that they allow $K < 0$. We will find that we only have an extensive high-temperature limit with $K < 0$ in the case of a positive cosmological constant. In cases with vanishing or negative cosmological constant, the cosmic horizons appear as interior patches of black holes with inner horizons; they are almost always unstable if $K < 0$. 

\item Whenever we are in a context with both black hole horizons and cosmic horizons, there will be transitions between the two. Tuning to the critical curve which demarcates the transition lands us on an ``extremal" solution: in Einstein--Maxwell theory with spherical spatial slices this is AdS$_2 \times S^{d-1}$, while in Einstein theory with a cosmological constant $\L$ we have AdS$_2 \times \mathbb{H}^{d-1}$ (for $\L < 0$ and hyperbolic spatial slices) and dS$_2 \times S^{d-1}$ (for $\L > 0$ and spherical spatial slices). These solutions can be dimensionally reduced over the transverse space and are described by Jackiw--Teitelboim (JT) gravity, which we will explore with novel boundary conditions (the dimensional reduction of conformal boundary conditions) in Appendix \ref{app:extremal}. 

\item In the case of vanishing cosmological constant and finite electric potential, we find that we \emph{must} include cosmic horizon solutions to reproduce the predictions of the thermal effective action. These solutions correspond to patches of the Reissner--Nordstr\"om black hole inside the inner horizon, and they dominate over black hole solutions above a critical electric potential at high temperature. In the case of negative cosmological constant and hyperbolic spatial slices, we will again see that cosmic horizon solutions dominate over black hole solutions even when taking $K\ell \to d$, which pushes the cutoff to the AdS boundary in the black hole solution. In this case the dominance seems to be due to an uncured divergence that pushes the free energy of the black hole to $+\infty$.

\item In some cases, we will see that we are missing solutions at low temperature. Either there is no candidate solution, or there is a discontinuity in the free energy as the dominant solution disappears. Assuming the geometric description is not breaking down, it is possible that complex solutions or solutions breaking transverse spatial symmetry need to be included to give a continuous free energy.
\end{itemize}

\subsection*{Outline}
In Section \ref{sec:masterEqs} we will set up our boundary value problem for the general case with electric potential, cosmological constant, and boundary theory placed on an arbitrary maximally symmetric space. We will show that patches of the Reissner--Nordstr\"om--(A)dS black hole provide a solution to this boundary value data, given by \eqref{dimensionlessVars}--\eqref{mastereqs3} where we have introduced dimensionless variables to analyze the equations more simply.

In Section \ref{sec:flatCharged} we will analyze the case with $\L = 0$ and nonzero electric potential $\tilde{\Phi} \neq 0$. We will present the solution space and phase diagrams for the boundary theory placed on a sphere ($k=1$), hyperboloid ($k = -1$), and flat torus ($k =0$). The case $k = 1$, whose Penrose diagram is shown in Figure \ref{figs:penroseFlatSphereCharged}, has the richest structure due to the simultaneous existence of black hole and cosmic horizons. The appearance of cosmic horizons in this case is due to the inner horizon that appears for the Reissner--Nordstr\"om solution in flat space. The $k=0$, $k=-1$ solutions have a single horizon and a singularity that lives beyond the cutoff; see Figure \ref{figs:penroseFlatNonSphereCharged} for a Penrose diagram.

In Section \ref{sec:negLambda} we will analyze the case with $\L < 0$ and $\tilde{\Phi} = 0$. We will present the solution space and phase diagram for the cases $k = 0, \pm 1$; see Figure \ref{figs:penroseAdS} for the Penrose diagrams. The case $k=-1$ has the richest structure due to the simultaneous existence of black hole and cosmic horizons. The appearance of cosmic horizons in this case is due to the inner horizon that appears for the negative-mass hyperbolic black hole in AdS. The $k=0$, $k=+1$ solutions have a single horizon and therefore only accommodate black-hole solutions.

In Section \ref{sec:posLambda} we will analyze the case with $\L > 0$ and $\tilde{\Phi} = 0$. We will present the solution space and phase diagram for the cases $k = 0, \pm 1$; see Figure \ref{figs:penrosedS} for the Penrose diagrams. The case $k=1$ has the richest structure due to the simultaneous existence of black hole and cosmic horizons. The existence of both is due to the fact that the relevant solution is Schwarzschild--dS, which has both a black hole and cosmic horizon. The $k=-1$, $k=0$ solutions have a single horizon and therefore only accommodate cosmic solutions.

In Section \ref{sec:highTemp} we will find the analytic solution at high temperature in the general case with $\tilde{\Phi} \neq 0$, $\L \neq 0$ and arbitrary $k =0, \pm 1$, comparing the results to the predictions of thermal effective field theory. In doing so, we compute the associated Wilson coefficients.

In Section \ref{sec:rotation} we will add an angular potential to the boundary theory in the case $k = 0$. The relevant bulk solutions are boosted and compactified black branes, and we will see that their free energy exactly reproduces the prediction of the thermal effective action. 

The appendices contain various results. Appendix \ref{app:bigQ} works out the charge for the solutions we will use in this paper. Appendix \ref{app:dirichlet} discusses the Dirichlet boundary condition for the Maxwell gauge field, which is somewhat subtle due to the conformal boundary conditions chosen for the metric. Appendix \ref{app:onshellAct} simplifies the on-shell action in the general case, which is used throughout the main text to evaluate phase diagrams. Appendix \ref{app:firstlaw} works out the thermodynamic first law obeyed by the solutions in this paper. Appendix \ref{app:extremal} studies the near-extremal near-horizon limit appearing in some of our solution spaces and includes an analysis of JT gravity with boundary conditions inherited by conformal boundary conditions in higher dimensions. Appendix \ref{app:puzzle} discusses proposals for what may fill in missing parts of some of our phase diagrams. 
 
\section{Master equations with general $\tilde{\Phi}$ and $\Lambda$}\label{sec:masterEqs}

The relevant action we will study in this paper is
\begin{equation} \label{I_CBC}
I = -\frac{1}{16\pi G}\int_{\mathcal{M}}d^{d+1}x \sqrt{g}\,\left(R - 2\Lambda -F^2\right) - \frac{1}{8d\pi G}\int_{\partial\mathcal{M}} d^d x \sqrt{h}\,K,
\end{equation}
where $d$ is the dimension of the boundary $\partial \mc{M}$. We will fix boundary conditions
\begin{equation}\label{bdryconditions}
\delta(h^{-1/d}h_{\mu\nu}) = 0,\qquad \delta K = 0,\qquad \d\left(\int A|_{\partial \mathcal{M}} \right) = 0.
\end{equation}
The boundary condition in the gravitational sector is known as conformal boundary conditions, and the boundary term in \eqref{I_CBC} is suited to this case \cite{York:1972sj,York:1986it,Odak:2021axr}. The conformal metric $\overline{h}_{\m\n} \defeq h^{-1/d} h_{\m\n}$ is invariant under Weyl rescaling of $h_{\m\n}$, and it is a tensor density of weight $-2/d$. The boundary condition on the Maxwell gauge field is a Dirichlet one, where we fix the holonomy of the gauge field at the boundary. (For the correct definition in terms of the local gauge field see Appendix \ref{app:dirichlet}.)

We will consider the following conformal class of boundary metrics in this paper:
\be\label{rotbdrymetric}
ds^2 = \l^2(d\tilde{\t}^2 + d\tilde{\Sigma}^2),\qquad (\tilde{\t}, \tilde{\phi}_i) \sim (\tilde{\t} + \tilde{\b}, \tilde{\phi}_i + i \tilde{\b} \tilde{\Omega}_i),
\ee
where $\tilde{\phi}_i$ are some angular coordinates in $\tilde{\Sigma}$ along which we consider rotation. For $\tilde{\Sigma}$ we will consider the unit-radius maximally symmetric spaces $S^{d-1}$, $\mathbb{H}^{d-1}$, and $\mathbb{R}^{d-1}$, or compactifications thereof. The conformal inverse temperature $\tilde{\b}$ and conformal angular potential $\tilde{\Omega}$ are defined by the above equation. Notice that since we fixed the above metric to have a spatial manifold of unit radius, we can interpret $\tilde{\b}$ as the ratio between the proper size of $\tilde{\t}$ (which is $\l \tilde{\b}$) and the proper radius of $\tilde{\Sigma}$ (which is $\l$). This definition requires working in a conformal representative with spacetime-independent $\l$, which is what we will focus on throughout this paper. This makes the fact that $\tilde{\b}$ is a conformal invariant more manifest. 

The thermal effective action for the putative boundary theory is written in terms of the representative boundary metric $\tilde{h}_{\m\n} dx^\m dx^\n = d\tilde{\t}^2 + d\tilde{\Sigma}^2$ as:
\be
\begin{split}\label{grand-Z}
-\log Z(\tilde{\b}, \tilde{\Omega}, \tilde{\Phi}) &= \int_{\tilde{\Sigma}} d^{d-1} x\,\sqrt{\tilde{h}_{\tilde{\Sigma}}} \left[-\f{c_0}{\tilde{\b}^{d-1}} + \frac{1}{\tilde{\b}^{d-3}}\left(c_1 \tilde{R} + c_2 \tilde{F}^2 + c_3 \tilde{\Phi}_0^2\right) + \cdots \right]\\
&\qquad + I_{\text{gapless}} + I_{\text{np}}.
\end{split}
\ee
$\tilde{R}$ and $\tilde{F}$ refer to the Ricci scalar and Kaluza--Klein field strength in the dimensional reduction of the representative boundary metric $d\tilde{\Sigma}^2$; see \cite{Jensen:2012jh, Banerjee:2012iz, horowitz, Benjamin:2023qsc}. $\tilde{\Phi}_0$ is a dimensionless version of the conformal electric potential $\tilde{\Phi}$ to be discussed below. $I_{\text{np}}$ refers to nonperturbative corrections. We will assume there is no contribution $I_{\text{gapless}}$.
To compute $c_2$ and $c_3$, it suffices to consider a rotating black hole and a charged black hole, respectively. At higher orders in the effective field theory, nonlinearities will kick in, e.g.~we will have terms like $\tilde{F}^2 \tilde{\Phi}_0^2$, for which we would need to consider black holes that are both rotating and charged. This partition function can be written in terms of a Hilbert space trace as 
\be \label{Z-conformal}
Z(\tilde{\b}, \tilde{\Omega}, \tilde{\Phi}) = \Tr \exp\left[-\tilde{\b}(\tilde{H} - \tilde{\Omega} J - \tilde{\Phi} Q)\right],
\ee
where $\tilde{H} \equiv \l H^{\rm CBC}$ is the generator of time translations in $\tilde{\t}$ and $H^{\text{CBC}}$ is the boundary Hamiltonian for the action \eqref{I_CBC} appropriate for conformal boundary conditions. For details on conformal boundary conditions and the thermal effective action in this context, see \cite{Banihashemi:2024yye}. 

For most of this paper, we will set $\tilde{\Omega} = 0$. The conformal electric potential $\tilde{\Phi}$ is related to the Maxwell gauge field at the boundary, but it can also be defined as the conjugate to the conserved charge $Q$ as calculated in the bulk (see Appendix \ref{app:bigQ}). Since we define our Maxwell action with coupling $1/G$, the gauge field $A_\m$ is dimensionless, making the conserved current $J^\m$ have mass dimension $d$ and the conserved charge $Q$ have mass dimension 1. Thus, $\tilde{\Phi}$ has mass dimension $-1$. 

A proper solution to the boundary value problem given by \eqref{I_CBC}--\eqref{bdryconditions} is to fill in the bulk geometry starting from some prescribed values for the fixed boundary data. This can be done by considering a solution to the equations of motion following from the action \eqref{I_CBC} which, upon picking some suitable cutoff location for $\partial \mathcal{M}$, realizes the prescribed boundary data. This is what we will do throughout this work. The general space of solutions to \eqref{I_CBC} that we will need is the family of Reissner--Nordstr\"om--(A)dS solutions:
\begin{align} \nonumber
ds^2 &= \,\,f(r) d\tau^2 + \frac{dr^2}{f(r)} + r^2 d\Sigma_k^2,\ \ \ \ d\Sigma_k^2 = \begin{cases}
d\Omega_{d-1}^2,&\ k = 1,\\
dx_{d-1}^2,&\ k = 0,\\
d\mathbb{H}_{d-1}^2,&\ k = -1,
\end{cases}
\\ 
f(r) &= k - \f{\m}{r^{d-2}} + \f{q^2}{r^{2(d-2)}} + \f{r^2}{\ell^2} = k + \frac{r^2}{\ell^2} + \frac{q^2}{r^{2(d-2)}} - \frac{r_h^{d-2}}{r^{d-2}}\left[k + \frac{r_h^2}{\ell^2} + \frac{q^2}{r_h^{2(d-2)}}\right],\label{bulksoln}\\
\,\,\,\,\,A &= \left(\f{q}{r_h^{d-2}}- \f{q}{r^{d-2}}\right)\f{i}{\eta} d\t,\qquad \eta = \sqrt{\f{2(d-2)}{d-1}}.\nonumber
\end{align}
We have picked the usual normalization $\Lambda = -d(d-1)/(2\ell^2) < 0$, with the case $\Lambda > 0$ obtained by the analytic continuation $\ell \rightarrow i \ell$ and $\L = 0$ obtained from the limit $\ell \rightarrow \infty$. The parameter $k$ describes the horizon geometry, $\m$ is the mass parameter, $q$ is the bulk charge parameter, and $r_h$ is the horizon radius (in situations with two horizons it can be either horizon radius). We have written the emblackening factor in two ways by using 
\be\label{murh}
f(r_h) = 0 \implies \m = r_h^{d-2}\left(k + \f{r_h^2}{\ell^2}\right) + \f{q^2}{r_h^{d-2}}.
\ee
For $k=1$, we will also have a horizonless solution, which we will sometimes call the thermal gas:
\be\label{thermalgas}
ds^2 = \left(1+\f{r^2}{\ell^2}\right)d\t^2 + \f{dr^2}{1+\f{r^2}{\ell^2}} + r^2 d\Omega_{d-1}^2,\qquad A = A_\t \, d\t,
\ee
where $A_\t$ is constant. Again, $\ell \rightarrow i \ell$ gives the $\L > 0$ solution and $\ell \rightarrow \infty$ the $\L = 0$ solution.

This family of geometries provides a solution to the boundary value problem \eqref{bdryconditions} upon picking a cutoff $r_c$ for the boundary. For the class of boundary geometries we are considering $ds^2 = \l^2(d\tilde{\t}^2 + d\tilde{\Sigma}^2)$, fixing the conformal class of metrics is equivalent to fixing $\tilde{\b}$, defined in \eqref{rotbdrymetric}. We will always have that $d\tilde{\Sigma}^2 = d\Sigma_k^2$. The metric boundary data at the cutoff $r_c$ is
\begin{equation}\label{sphereflateqns}
\tilde{\beta} = \pm\frac{4\pi}{f'(r_h)} \frac{\sqrt{f(r_c)}}{r_c},\ \ \ \ K = \pm\frac{d-1}{r_c \sqrt{f(r_c)}}\left[f(r_c) + \frac{r_c f'(r_c)}{2(d-1)}\right],
\end{equation}
where the $+$ branch corresponds to taking a black-hole patch of the geometry while the $-$ branch corresponds to taking a cosmic patch. The sign on $\tilde{\b}$ is chosen to ensure $\tilde{\b} > 0$, whereas the sign on $K$ is fixed by the rule that our normal vector to $\partial \mathcal{M}$ is always chosen to point outward (i.e., away from the system). A Dirichlet boundary condition on the gauge field leads to the following equation for the boundary potential $\tilde{\Phi} = - i r_c A_\t/\sqrt{f(r_c)}$ (see Appendix \ref{app:bigQ} to understand these factors relating the gauge field to the electric potential):
\begin{equation}\label{qeq}
\tilde{\Phi} = \left(\frac{q}{r_h^{d-2}} - \frac{q}{r_c^{d-2}}\right) \frac{r_c}{\eta\sqrt{f(r_c)}},\ \ \ \ \eta = \sqrt{\frac{2(d-2)}{d-1}}.
\end{equation}
There is a $\tilde{\Phi} \rightarrow -\tilde{\Phi}$ symmetry (which can be accommodated by a $q \rightarrow -q$ sign flip in the bulk solution) that we will fix by always considering $\tilde{\Phi} \geq 0$. This means for black-hole patches $r_c > r_h$ we will have $q \geq 0$, while for cosmic patches $r_c < r_h$ we will have $q \leq 0$. All together, the equations for the boundary data $\{\tilde{\beta},K,\tilde{\Phi}\}$ in terms of the dimensionful bulk parameters $\{r_h,r_c,q\}$ are given by \eqref{sphereflateqns}--\eqref{qeq}.

To analyze the solution space, we will find it convenient to switch to dimensionless expressions. Taking as our boundary data $\{\tilde{\beta},K\ell,\tilde{\Phi}K\}$, we write the equations for these three parameters in terms of
\begin{equation}
x \equiv \frac{q}{r_h^{d-2}},\ \ \ \ y \equiv \frac{r_c}{\ell},\ \ \ \ z \equiv \frac{r_h}{r_c}.\label{dimensionlessVars}
\end{equation}
The resulting equations for $\L < 0$ are
\begin{align}
\tilde{\beta}
&= \pm \frac{4\pi z}{(d-2)(k-x^2) + d z^2 y^2} \sqrt{\left(1-z^{d-2}\right)\left(k-z^{d-2}x^2\right) + y^2 \left(1 - z^d\right)},\label{mastereqs1}\\
K\ell
&= \pm \frac{1}{y\sqrt{\left(1-z^{d-2}\right)\left(k-z^{d-2}x^2\right) + y^2 \left(1 - z^d\right)}}\nonumber\\
&\qquad\times \left[k(d-1) + z^{2(d-2)}x^2 + \frac{d}{2}\Big(2y^2 - z^{d-2}\left(k + x^2 + z^2 y^2\right)\Big)\right],\label{mastereqs2}\\
\tilde{\Phi} K
&= \pm \frac{x(1-z^{d-2})}{\eta\left[\left(1-z^{d-2}\right)\left(k-z^{d-2}x^2\right) + y^2 \left(1 - z^d\right)\right]}\nonumber\\
&\qquad\times \left[k(d-1) + z^{2(d-2)}x^2 + \frac{d}{2}\Big(2y^2 - z^{d-2}\left(k + x^2 + z^2 y^2\right)\Big)\right].\label{mastereqs3}
\end{align}
Again the $+$ branch corresponds to taking a black-hole patch of the geometry while the $-$ branch corresponds to taking a cosmic patch. The case $\L > 0$ is obtained by taking $\ell \rightarrow i\ell$, which takes $y \rightarrow -i y$. For the black-hole patch we must have $0 \leq x \leq 1$, $y \geq 0$, and $0 \leq z \leq 1$, while for the cosmic patch we must have $x \leq -1$, $y \geq 0$, and $z \geq 1$. (The bound $x \leq -1$ is like an ``inside-out" extremality bound.)

Either the dimensionful or dimensionless set of equations will serve as our ``master equations," in the sense that the various cases we examine as furnishing cosmic geometries are subcases of these equations. The case $\L = 0$ (the subject of Section \ref{sec:flatCharged}) corresponds to taking $\ell \to \infty$, which sends $y \to 0$. The case $\tilde{\Phi} = 0$ (the subject of Sections \ref{sec:negLambda}-\ref{sec:posLambda}) is accommodated by taking $x \rightarrow 0$. (There is a somewhat interesting case $\tilde{\Phi} = \Lambda = k = 0$ where the relevant geometry is Rindler space \cite{Banihashemi:2024yye}. To reach this from $\tilde{\Phi} \neq 0$ requires scaling $z \rightarrow 1$, since in that case $\tilde{\Phi}$ is independent of $x$ but vanishes as $z\rightarrow 1$; to ensure $\tilde{\b}$ and $K$ do not diverge one needs to also scale $x\rightarrow 0$.)

Solving these equations means determining the bulk parameters $r_h$, $r_c$, and $q$ given boundary data $\tilde{\Phi}$, $K$, and $\tilde{\b}$. Doing this analytically in the general case is impractical, although we will be able to solve the equations in a high-temperature expansion in Section \ref{sec:highTemp}. We will also apply numerical methods to the cases $\L = 0$, $\tilde{\Phi} \neq 0$ and $\L \neq 0$, $\tilde{\Phi} = 0$ to find the solution spaces and phase diagrams. This can also be done for the general $\L \neq 0$, $\tilde{\Phi} \neq 0$ case, although we did not pursue this.

An important point should be mentioned about the $\Lambda = 0$, $\tilde{\Phi} \neq 0$ case. Solving \eqref{mastereqs1}, \eqref{mastereqs3} in terms of $x = q/r_h^{d-2}$ and $z = r_h/r_c$ means we actually get an infinity of solutions per choice of $\tilde{\b}$ and $\tilde{\Phi}K$. This is because $\tilde{\b}$, $\tilde{\Phi}$ and $K$ are all independently tunable, so we can change $K$ and $\tilde{\Phi}$ while keeping their product fixed. Alternatively, from the bulk perspective, $q$, $r_h$, and $r_c$ (which go into the relevant dimensionless parameters $x$ and $z$) are all independently tunable in the bulk solution space. This is unlike the $\L \neq 0$, $\tilde{\Phi} = 0$ cases, since the dimensionless quantities are $\tilde{\b}$ and $K\ell$, with $\ell$ fixed by the theory, and so there are only $O(1)$ solutions corresponding to a choice of these dimensionless parameters. Alternatively, from the bulk perspective, the parameters $r_h$, $r_c$ and $\ell$ (which go into the relevant dimensionless parameters $y$ and $z$) are not all independently tunable.

Another important point should be mentioned about the $k=0$ case, where we consider compactifying one of the spatial directions onto a circle. While it seems that we have a freedom in the space of bulk solutions to vary the compactification size, this can be trivially accounted for by a diffeomorphism 
$r \rightarrow \l r$, since our solutions have a transverse spatial metric of the form $r^2 dx_{d-1}^2$. In other words, while it seems that we can take a bulk solution and increase $\b$ (the size of the thermal circle on the boundary) while simultaneously increasing the compactification scale to keep the ratio the same, this solution is actually diffeomorphic to the starting solution. In the fully compactified case $\mathbb{T}^{d-1}$ that we will consider, this means that we can normalize the length of one of the spatial directions in the bulk to have size $1$. This reflects the same structure in the boundary, where scale invariance means that we can fix a similar normalization. This is precisely what occurs in the case of the BTZ black hole.\footnote{Interestingly, for the uncharged $k=0$ case with $\Lambda = 0$ considered in \cite{Banihashemi:2024yye}, the relevant bulk solution is Rindler space and the transverse factors do not have a factor of $r^2$; thus we do not have the argument above, but in that case we also do not have a freedom in changing the temperature of the solution!}

Separately from this family of solutions \eqref{bulksoln}, we will also have a handful of ``extremal" solutions. For $k = 1$ and $\L = 0$, we have AdS$_2 \times S^{d-1}$ as a solution:
\be
\qquad \,\, ds^2 = \left(\f{r^2}{\ell_{2}^2}-1\right)d\t^2 + \f{dr^2}{\f{r^2}{\ell^2_{2}}-1} + (d-2)^2\ell_{2}^2 d\Omega_{d-1}^2,\ \ \ \ A = \f{r-\ell_2}{\ell_{2}}\f{i}{\eta} d\t.\label{ads2s2}
\ee
For $k = -1$ and $\L < 0$, we have AdS$_2 \times \mathbb{H}^{d-1}$:
\be
\quad ds^2 = \left(\f{r^2}{\ell_2^2}-1\right)d\t^2 + \f{dr^2}{\f{r^2}{\ell_2^2}-1} + (d-2)\ell_2^2 d\mathbb{H}_{d-1}^2,\ \ \ \ \Lambda = -\f{(d-1)}{2\ell_2^2}.\label{ads2h2}
\ee
Finally, for $k = 1$ and $\L > 0$, we have dS$_2 \times S^{d-1}$ as a solution:
\be
 \hspace{-10mm}ds^2 = \left(1 - \f{r^2}{\ell_2^2}\right)d\t^2 + \f{dr^2}{1-\f{r^2}{\ell_2^2}} + (d-2)\ell_2^2d\Omega_{d-1}^2,\ \ \ \ \Lambda = \f{(d-1)}{2\ell_2^2}.\label{ds2s2}
\ee
In all cases, we have $\t \sim \t + 2\pi \ell_2$ to ensure a smooth Euclidean geometry. These solutions have an interesting 2-dimensional description in terms of JT gravity, which we will explore in Appendix \ref{app:extremal}. 

\section{$\tilde{\Phi} \neq 0$, $\Lambda = 0$: charged flat space}\label{sec:flatCharged}

We start with Einstein--Maxwell theory ($\tilde{\Phi} \neq 0$) with a vanishing cosmological constant ($\Lambda = 0$):
\begin{equation} \label{I_EM}
I = -\frac{1}{16\pi G}\int_{\mathcal{M}}d^{d+1}x \sqrt{g}\,\left(R - F^2\right) - \frac{1}{8d\pi G}\int_{\partial\mathcal{M}} d^d x \sqrt{h}\,K.
\end{equation}
The equations of motion following from this action are solved by \eqref{bulksoln} with $\ell\rightarrow \infty$, which simply modifies the emblackening factor to 
\be \label{charged-metric}
 f(r) \equiv k + \frac{q^2}{r^{2(d-2)}} - \frac{r_h^{d-2}}{r^{d-2}}\left[k + \frac{q^2}{r_h^{2(d-2)}}\right].
\ee
$k \in \{-1,0,1\}$ indicates the horizon geometry, $r_h$ is the horizon radius, and $q$ is the bulk charge. The equations of our boundary value problem are given by taking the master equations \eqref{mastereqs1}--\eqref{mastereqs3} and sending $\ell \to \infty$, which by \eqref{mastereqs2} sends $y \rightarrow 0$ and reduces \eqref{mastereqs1} and \eqref{mastereqs3} respectively to
\begin{align}
\tilde{\beta}
&= \pm\frac{4\pi z}{(d-2)(k-x^2)}\sqrt{(1-z^{d-2})(k-z^{d-2}x^2)},\label{eqFlatCharge1}\\
\tilde{\Phi}K
&= \pm\frac{x}{2\eta(k-z^{d-2}x^2)}\left[2k(d-1) + 2z^{2(d-2)}x^2 - dz^{d-2}(k+x^2)\right].\label{eqFlatCharge2}
\end{align}
Squaring the first equation gives a quadratic equation in $x^2$, which we can solve to obtain:
\begin{align}
x_1^2 = k + \frac{4\pi(z^{d-2} - 1)}{(d-2)^2 \tilde{\beta}^2}\left[2\pi z^d + \sqrt{4\pi^2 z^{2d} + k(d-2)^2 \tilde{\beta}^2 z^2}\right],\label{xzChargedFlat}\\
x_2^2 = k + \frac{4\pi(z^{d-2} - 1)}{(d-2)^2 \tilde{\beta}^2}\left[2\pi z^d - \sqrt{4\pi^2 z^{2d} + k(d-2)^2 \tilde{\beta}^2 z^2}\right].\label{xzChargedFlat2}
\end{align}
The second solution is spurious for $k = 0,1$. For $k=0$, the $\tilde{\b}$ equation scales as $1/|x|$, and so $x^2$ can be solved for explicitly and only matches \eqref{xzChargedFlat}. For $k=1$, in the black-hole case we have $0<z<1$ and so $\tilde{\b} \propto (k-x_2^2)^{-1} < 0$, while in the cosmic case we have $z > 1$ and therefore $\tilde{\b} \propto -(k-x_2^2)^{-1}< 0$---clearly contradicting the positivity of $\tilde{\b}$.

We can plug both of these solutions into the equation for $\tilde{\Phi}K$. But before doing so, we recall that the sign of $q$ (and thus that of $x$) matches that of $\tilde{\Phi}$ for black-hole horizons but is opposite to $\tilde{\Phi}$ for cosmic horizons. This sign difference cancels the $\pm$ in \eqref{eqFlatCharge2}. Thus, by plugging \eqref{xzChargedFlat} into \eqref{eqFlatCharge2} we get one equation that encompasses both cases:
\begin{align}
&\left.\tilde{\Phi} K\right|_{x = x_1} = \frac{(d-2)\left[k^2(d-2) + d - 2z^{d-2}\right] \tilde{\b}^2 - 4 \pi k z^{d-2} b}{2\eta(d-2)^2 \tilde{\b}^3 z}\sqrt{(b + 4\pi z^d)\left[b + 4\pi(z^d - z^2)\right]},\nonumber\\
&b = \sqrt{4\pi^2 z^{2d} + k(d-2)^2 \tilde{\b}^2 z^2} - 2\pi z^d,\ \ \ \ \eta = \sqrt{\frac{2(d-2)}{d-1}}.\label{phikfundgentopSec3}
\end{align}
This equation describes black holes when $0 < z < 1$ and cosmic geometries when $z > 1$. Due to our choice $\tilde{\Phi}\geq 0$ discussed around \eqref{qeq}, the region $\tilde{\Phi}K > 0$ corresponds to choosing $K > 0$, whereas $\tilde{\Phi}K < 0$ corresponds to choosing $K < 0$.

For $k = - 1$, we can also plug \eqref{xzChargedFlat2} into \eqref{eqFlatCharge2}. This yields
\begin{equation}
\left.\tilde{\Phi}K\right|_{x = x_2} = \frac{(d-2)\left(d - 1 - z^{d-2}\right) \tilde{\b}^2 + 2 \pi k z^{d-2} (b+4\pi z^d)}{\eta(d-2)^2 \tilde{\b}^3 z}\sqrt{b(b+4\pi z^2)}.\label{phikfundgentop2}
\end{equation}
In this section, we will (numerically) solve these equations for $z$ for each of the three cases of $k$. We will also examine the thermodynamics of the classical solutions by calculating the on-shell Euclidean action $I_{\text{cl}}$, derived in Appendix \ref{app:onshellAct} for the general case ($\tilde{\Phi},\Lambda \neq 0$). By setting $y = 0$ in \eqref{dimensionlessActGen}, the formula reduces to 
\begin{align}
I_{\text{cl}} &= -\frac{\tilde{\beta}\,\text{Vol}[\Sigma_k]}{8\pi G|K|^{d-1}} |h(x,z)|^{d-1}\left[(\tilde{\Phi}K) \frac{\eta z^{d-2}|x|}{h(x,z)} + \frac{h(x,z)}{d}\right],\nonumber\\
h(x,z) &= \pm \frac{2k(d-1) + 2z^{2(d-2)}x^2 - dz^{d-2}(k + x^2)}{2\sqrt{(1-z^{d-2})(k-z^{d-2}x^2)}}.\label{flatChargedAct}
\end{align}

\subsection{Spherical solutions}\label{sphericalcharged}

\input{figs/flatCharged/texCode/penroseDiagrams/penroseFlatSphereCharged}

We start by considering a conformal class of geometries on the boundary with representative $S^1 \times S^{d-1}$. In this case, we have the horizonless solution \eqref{thermalgas} to Einstein--Maxwell theory, which we can write in terms of the boundary parameters as
\be
ds^2 = r_c^2 d\tilde{\t}^2 + dr^2 + r^2 d\Omega_{d-1}^2 ,\quad \tilde{\t} \sim \tilde{\t} + \tilde{\b},\quad A = i \tilde{\Phi}\, d\tilde{\t},\label{sphereFlatVac}
\ee
where $r \in (0,r_c)$ and
\begin{equation}
r_c = \frac{d-1}{K}.\label{sphereFlatVacCutoff}
\end{equation}
Note that this solution only makes sense if $r_c > 0$, i.e. if $K > 0$. As we have fixed to $\tilde{\Phi} \geq 0$, this geometry therefore only solves the boundary value problem for $\tilde{\Phi}K \geq 0$.

We also have the Reissner--N\"ordstrom solution, which is given by taking $k = 1$ and $\ell \rightarrow \infty$ in \eqref{bulksoln}. The Penrose diagram of the Lorentzian geometry for $\m > 0$ is shown in Figure \ref{figs:penroseFlatSphereCharged}. This geometry has both an outer and inner horizon, respectively at $r = r_+$ and $r = r_-$, which are the real solutions to the following equation:
\begin{equation}
\mu = r_\pm^{d-2} + \frac{q^2}{r_\pm^{d-2}} \implies r_{\pm}^{d-2} = \f{\m}{2} \left(1 \pm \sqrt{1-\frac{4q^2}{\m^2}}\right).
\end{equation}
There are two ways to use the Reissner--N\"ordstrom solution to satisfy our conformal boundary conditions. These correspond to taking different patches of the full Lorentzian solution (and then continuing to Euclidean signature), with either the outer or inner horizon treated as part of the geometry. Specifically, what we call black hole geometries are patches in which the boundary is outside of the outer horizon ($r_c > r_+ = r_h$), while what we call cosmic geometries are those in which the boundary is between the inner horizon and the singularity ($0 < r_c < r_- = r_h$). See Figure \ref{figs:penroseFlatSphereCharged} for a picture.

In terms of the dimensionless variable $z = r_h/r_c$, the relevant equation to solve is given by \eqref{phikfundgentopSec3} with $k = 1$. %:
%\begin{align}
%&\tilde{\Phi} K = \frac{(d-2)\left(d - 1 - z^{d-2}\right) \tilde{\b}^2 - 2 \pi z^{d-2} b}{\eta(d-2)^2 \tilde{\b}^3 z}\sqrt{(b + 4\pi z^d)\left[b + 4\pi(z^d - z^2)\right]},\nonumber\\
%&b = \sqrt{4\pi^2 z^{2d} + (d-2)^2 \tilde{\b}^2 z^2} - 2\pi z^d,\ \ \ \ \eta = \sqrt{\frac{2(d-2)}{d-1}}.\label{phikfundgentopSphere}
%\end{align}
While we cannot solve this equation for $z$ analytically, we can access the solution space numerically. This is presented for $d = 3$ in Figure \ref{figs:solutionsFlatSphereCharged}. %Each region is labeled as $(n_{\text{BH}},n_{\text{CH}})$, where $n_{\text{BH}}$ and $n_{\text{CH}}$ are the numbers of black hole and cosmic solutions, respectively.

\input{figs/flatCharged/texCode/solutionSpaces/solutionsFlatSphereCharged}

In addition to this family of solutions, however, there is another solution to our boundary value problem corresponding to a patch of the extremal AdS$_2 \times S^{d-1}$ solution \eqref{ads2s2}:
\begin{equation}
\begin{split}
ds^2 &= \left(\f{r^2}{\ell_2^2}-1\right)d\t^2 + \f{dr^2}{\f{r^2}{\ell_2^2}-1} + (d-2)^2\ell_2^2d\Omega_{d-1}^2,\ \ \ \ A = \,\f{r-\ell_2}{\ell_2}\f{i}{\eta} \, d\t,\\
\t &\sim \t + 2\pi \ell_2,\ \ \ \ r \in (\ell_2, r_c).
\end{split}\label{ads2s2patch}
\end{equation}
The boundary data $\{\tilde{\beta},K, \tilde{\Phi}\}$ is related to the bulk parameters by the equations
\begin{equation}
\tilde{\b} = \f{2\pi\sqrt{r_c^2/\ell_2^2-1}}{d-2} ,\qquad K = \frac{1}{\ell_2^2} \frac{r_c}{\sqrt{r_c^2/\ell_2^2-1}},\qquad \tilde{\Phi} = \f{(d-2)(r_c-\ell_2)}{\eta\sqrt{r_c^2/\ell_2^2-1}} .
\end{equation}
We have one tunable dimensionless parameter in the bulk, $y_* \equiv r_c/\ell_2$, in terms of which we can write the following dimensionless equations
\be
\tilde{\b} = \f{2\pi\sqrt{y_*^2-1}}{d-2},\qquad \tilde{\Phi}K = \f{y_*(d-2)(y_*-1)}{\eta(y_*^2-1)}.
\ee
This specifies a curve in the plane spanned by $\tilde{\Phi}K$ and $\tilde{\b}$ given by 
\be \label{crit-KPhi1}
\tilde{\Phi} K = \frac{d-2}{\eta} \frac{\sqrt{4 \pi ^2+(d-2)^2\tilde{\beta}^2 }}{2 \pi + \sqrt{4 \pi ^2+(d-2)^2\tilde{\beta}^2}} \in \left(\frac{d-2}{2\eta},\frac{d-2}{\eta}\right).
\ee
For $d = 3$, this is precisely the curve in gold in Figure \ref{figs:solutionsFlatSphereCharged} that begins at $(\tilde{\b}, \tilde{\Phi}K) = (0, 1/2)$. This curve is where the black hole and cosmic solutions interchange. Since the black hole becomes a cosmic geometry smoothly with $r_c$ crossing $r_+$, the bulk solution must approach the extremal black hole along this boundary, since otherwise there would be a discontinuity due to the region between $r_-$ and $r_+$. Another way to see this is that taking $r_c \rightarrow r_+$ at finite $\tilde{\Phi}$ implies $q \rightarrow r_+^{d-2}$ and therefore $\m \rightarrow 2q$ (or $r_- \rightarrow r_+$). This is thus a near-horizon near-extremal limit. As such, \eqref{crit-KPhi1} can also be obtained by setting $z=1$, $k=1$ in \eqref{phikfundgentopSec3}.\footnote{It is important to note that $K$ does not diverge in this limit; instead, we have
\begin{equation}
K < \frac{d-2}{2\eta}\left(\frac{1}{\tilde{\Phi}} + \frac{\eta^2 \tilde{\Phi}}{r_+^2}\right),\label{Krangecharged}
\end{equation}
with $K$ monotonically increasing from $0$ at $r_c = \infty$ to this upper bound at $r_c = r_+$. This is another sign that the solution approaches the extremal black hole, since any nonextremal horizon will be approximated by Rindler space as $r_c \rightarrow r_+$ and in that limit will yield $K \rightarrow \infty$.} The reason $\tilde{\b}$ remains finite along this curve when taking the limit from the family of black hole solutions is that we are zooming in toward the horizon simultaneously with the solution approaching the extremal black hole. Zooming in toward the horizon pushes the temperature to increase while approaching extremality pushes the temperature to decrease, and the combined effect stabilizes at some finite value captured by our AdS$_2 \times S^{d-1}$ geometry above. For a direct, detailed analysis of this limit, see Appendix \ref{app:extremal}. The on-shell action of this family of solutions is given in \eqref{actionextremalflat}.

As discussed toward the end of Section \ref{sec:masterEqs}, a specification of the dimensionless parameters $\tilde{\Phi}K$ and $\tilde{\b}$ leads to an infinity of solutions in terms of the bulk parameters $r_h$, $r_c$ and $q$, since $\tilde{\Phi}$ and $K$ can be independently tuned.

\subsubsection{Analytic features of the solution space}
There is a critical curve above which we have no black hole solutions. This maximum value of $\tilde{\Phi}K$ is obtained by taking $z= 0$, $k=1$ in \eqref{phikfundgentopSec3}, which gives 
\be\label{crit-KPhi2}
 \tilde{\Phi} K = \f{d-1}{\eta}.
\ee
Setting $d=3$ gives $\tilde{\Phi}K = 2$, in agreement with Figure \ref{figs:solutionsFlatSphereCharged}.

To see why $\tilde{\Phi}K$ is maximized at $z = 0$, we first recall that solving the $\tilde{\b}$ equation \eqref{eqFlatCharge1} for $x^2$ gives \eqref{xzChargedFlat}, which for $k = 1$ is
\begin{equation} \label{eq:x-spherical}
x^2(z;\tilde{\beta}) = 1 + \frac{4\pi(z^{d-2} - 1)}{(d-2)^2 \tilde{\beta}^2}\left[2\pi z^d + \sqrt{4\pi^2 z^{2d} + (d-2)^2 \tilde{\beta}^2 z^2}\right].
\end{equation}
Then from \eqref{phikfundgentopSec3} (with $k=1$) for the black hole solutions, we have 
\be \label{KPhizeq}
2\eta \tilde{ \Phi} K = x(z;\tilde{\beta}) \frac{k(z) }{g(z)},
\ee
where we have defined
\begin{equation}
k(z) \equiv 2 (d- 1) - d z^{d-2} + \left(2z^{d-2} - d\right ) z^{d-2} x^2(z;\tilde{\beta}),\ \ \ \ g(z) \equiv 1- z^{d-2} x^2(z;\tilde{\beta}).
\end{equation}
The $z$-derivative of $k(z)/g(z)$ in \eqref{KPhizeq} is manifestly negative, so it takes its maximum value of $2(d-1)$ at $z=0$. We also know that for the BH solutions, $0\leq x \leq 1$ because $q \leq r_+^{d-2}$. As is evident from \eqref{eq:x-spherical}, the upper bound $x = 1$ is obtained at $z = 1$ and $z=0$. Thus, the entire expression \eqref{KPhizeq} is maximized at $z = 0$, yielding the bound \eqref{crit-KPhi2}.

Interestingly, there is also a critical electric potential $\Phi = 1/\eta$ that separates two qualitatively different branches in AdS/CFT \cite{Chamblin:1999tk}. There, the regime $\Phi > 1/\eta$ has only a large black hole solution while $\Phi < 1/\eta$ has both large and small black hole solutions. For any electric potential there is also the thermal AdS solution with a constant gauge field. However, for $\Phi > 1/\eta$, the large black hole dominates for all temperatures, so there is no confining phase transition. This is most similar to our extremal curve \eqref{crit-KPhi1} (shown in gold in Figure \ref{figs:solutionsFlatSphereCharged}), which separates two black hole solutions from one black hole solution (at high temperature), but when we study the phase structure later in this section we will not find an electric potential above which a solution with a horizon always dominates.

Another critical curve is given by $\tilde{\Phi}K = 0$. This regime requires further specification, because it can either mean $\tilde{\Phi} = 0$ or $K = 0$. The case of $\tilde{\Phi} = 0$ has been analyzed in \cite{Anninos:2023epi, Banihashemi:2024yye}. The key result that we need is that there is a maximum inverse temperature,
\begin{equation}
\tilde{\beta}_{\text{max}} = \frac{4\pi}{d-2}\left(\frac{2}{d}\right)^{1/(d-2)}\sqrt{1-\frac{2}{d}},
\end{equation}
above which there are no black hole solutions, while for $\tilde{\b} < \tilde{\b}_{\text{max}}$ there are two black hole solutions. This gives the boundary between the $(2,1)$ and $(0,1)$ regions in Figure \ref{figs:solutionsFlatSphereCharged} near $\tilde{\Phi} K = 0$, which for $d = 3$ is at $\tilde{\beta}_{\text{max}} = 8\pi/\sqrt{27} \approx 4.837$. Notice that there seems to be an extraneous cosmic solution as we take the uncharged limit $\tilde{\Phi} \rightarrow 0$ at $K > 0$. Tracking this solution shows that $r_h$, $r_c$ crash into the singularity and therefore decouple from the physical solution space. (Similarly, taking the limit from the side $K < 0$ also shows the one and only solution decouples, indicating no solutions for $K < 0$ as required.)

Now, let us consider $K = 0$. Taking $r_c \rightarrow \infty$ with other parameters fixed gives a $K = 0$ black hole solution. This limit, however, also sends $\tilde{\b} \rightarrow 0$. That this is a high-temperature limit may seem a little strange, but it is simply because we compare the size of the thermal circle to the transverse sphere, and in asymptotically flat spaces the transverse space has an $r^2$ warp factor whereas time does not (in sharp contrast with asymptotically AdS spacetimes). So, the asymptotic solution only accommodates $\tilde{\b} = 0$, and we are free to ignore it.

We also find a cosmic solution with $K = 0$. By plotting $\tilde{\Phi}K(z,\tilde{\b})$ at fixed $\tilde{\b}$ using \eqref{phikfundgentopSec3} (with $k=1$), we see that this solution connects $K < 0$ and $K > 0$ branches of the cosmic solution, in that $\tilde{\Phi} K $ smoothly flips sign at some $z > 1$. This zero of $\tilde{\Phi}K$ is at some $z$ solving
\be
\tilde{\b}_0 = \f{2 \pi z^{d-1} }{d-1-z^{d-2}}\sqrt{\f{2 z^{d-2} - d}{d-2}}.
\ee
Observe that $\tilde{\b}_0 \in (0,\infty)$ if and only if $z^{d-2} \in (d/2, d-1)$, so the range of $z$ in the solution space for $K = 0$ is quite limited. This result, coupled with the fact that the $K \neq 0$ branches for the cosmic solutions are connected through $K = 0$ and the monotonicity of $K$ with respect to $z$, gives for our cosmic solutions
\begin{equation}\label{zrange}
\begin{split}
K>0 &\implies z^{d-2} \in \left(1, d-1\right),\\
K = 0 &\implies z^{d-2} \in \left(\frac{d}{2}, d-1\right),\\
K<0 &\implies z^{d-2} \in \left(\frac{d}{2}, \infty\right).
\end{split}
\end{equation}

We can also analyze a couple of simple features of the $(3,0)$ region, in particular where it intersects our extremal curve of AdS$_2 \times S^2$ solutions. Here, we will work out the values of $\tilde{\b}$ at the left and right endpoints of this interval, visible in the zoomed-in region of Figure \ref{figs:solutionsFlatSphereCharged}.
The left endpoint $\tilde{\b}_L$ can be obtained through $d(\tilde{\Phi}K)/dz|_{z=1} = 0$, which implies
\begin{equation}
8 \pi ^3 d+2 \pi \tilde{\beta}_{\text{L}} ^2 (d-2)^2+\left[\tilde{\beta}_{\text{L}} ^2 (d-2)^2-8 \pi ^2 (d-1)\right] \sqrt{\tilde{\beta}_{\text{L}} ^2 (d-2)^2+4 \pi ^2}=0.\label{constraintLeftCrit}
\end{equation}
This can be written as a cubic equation in $\tilde{\b}_{\text{L}}^2$ whose solutions do not cleanly simplify. There is only one positive solution to the above equation, given by $\tilde{\b}_L \approx 6.784$ in $d=3$. 

The right endpoint $\tilde{\b}_R$ can be found by simultaneously solving $d(\tilde{\Phi}K)/dz|_{z=z_0} = 0$ and $\tilde{\Phi}K|_{z=z_0} = \tilde{\Phi}K|_{z=1}$ for some $z_0 < 1$ and $\tilde{\b}_R$. For $d = 3$, we find $z_0 \approx 0.970$ and $\tilde{\beta}_{\text{R}} \approx 6.893$.

\subsubsection{Phase structure}

Let us now discuss the dominant phases of the ensemble, 
beginning with $\tilde{\Phi}K \geq 0$. The solutions we consider are various patches of the Reissner--Nordstr\"om metric discussed above and the patch $r \in (0, r_c)$ of hot flat space \eqref{sphereFlatVac}. The on-shell action for the solutions with horizon is given by \eqref{flatChargedAct}. The on-shell action for hot flat space (which we will refer to as the ``thermal gas" for uniformity with other sections) is straightforward to calculate. The bulk does not contribute, and we have $K = (d-1)/r_c$, so the full answer is given by
\be\label{thermalgasAct}
I_{\text{gas}} = -\f{\tilde{\b}\,\text{Vol}[S^{d-1}]}{8\pi G}\frac{r_c^d K}{d}= -\f{\tilde{\beta}\,\text{Vol}[S^{d-1}]}{8\pi G K^{d-1}}\f{(d-1)^d}{d} .
\ee
The thermal gas action is also a simple limit of the general action \eqref{flatChargedAct}. This answer is independent of the electric potential. The part multiplying $\tilde{\b}$ can be interpreted as a Casimir energy, since, as we will see, this solution dominates at large $\tilde{\b}$.

The on-shell action for the patch of an extremal AdS$_2 \times S^{d-1}$ solution given in \eqref{ads2s2patch} can also be calculated analytically. To do so, we use $R - F^2 = \f{2}{(d-2)\ell_2^2}$ to get
\begin{align}
I_{\text{bulk}} &= -(d-2)^{d-2}\f{\ell_2^{d-1}\text{Vol}[S^{d-1}]}{4 G } \left(\f{r_c }{\ell_2}-1\right),\\
I_{\text{bdry}} &= -(d-2)^{d-2}\f{\ell_2^{d-1} \text{Vol}[S^{d-1}]}{4 G} \f{(d-2)r_c}{d \ell_2}.
\end{align}
To write this in terms of boundary parameters, we use 
\be\label{l2bdry}
\f{r_c}{\ell_2} = \f{K \ell_2}{\sqrt{-1+ K^2 \ell_2^2}},\ \ \ \ 
\ell_2 = K^{-1} \sqrt{1+ \f{4\pi^2}{\tilde{\b}^2(d-2)^2}},
\ee
to get
\begin{align}
I_{\text{total}} = -(d-2)^{d-2}\f{\text{Vol}[S^{d-1}]}{4 G K^{d-1}}&\left[1+ \f{4\pi^2}{\tilde{\b}^2(d-2)^2}\right]^{(d-1)/2}\left[\f{(d-1)\sqrt{\tilde{\b}^2(d-2)^2+4\pi^2}}{d\pi}-1\right].\label{actionextremalflat}
\end{align}
To compare these solutions and plot them in a plane spanned by $\tilde{\Phi} K $ and $\tilde{\b}$, we should recall that, as discussed toward the end of Section \ref{sec:masterEqs}, a specification of the dimensionless parameters $\tilde{\Phi}K$ and $\tilde{\b}$ leads to an infinity of solutions in terms of the bulk parameters $r_h$, $r_c$ and $q$. Breaking this degeneracy is not important, as the actions of all the solutions have been written to depend on $\tilde{\b}$, $\tilde{\Phi} K$, and an overall $1/K^{d-1}$. Thus, we can accurately represent a phase diagram in the $ \{\tilde{\Phi} K, \tilde{\b}\} $ plane even though each point includes an infinity of physical solutions. This is of course guaranteed by dimensional analysis. We also recall that the extremal AdS$_2 \times S^{d-1}$ solutions only exist on the curve specified by \eqref{crit-KPhi1}.

\input{figs/flatCharged/texCode/phasePlots/phasesFlatSphereCharged}

By minimizing $I$ over the solutions, we find a phase structure, plotted for $d = 3$ in Figure \ref{figs:phasesFlatSphereCharged}. We also probe the stability of the dominant phase by computing the specific heat,
\begin{equation}
C_K = -\tilde{\beta}^2 \frac{\partial^2 I_{\text{cl}}}{\partial\tilde{\beta}^2},\label{specificHeat}
\end{equation}
and checking if $C_K > 0$. Whenever there are two cosmic solutions or two black hole solutions, it is always the larger of the two, i.e.~the one with larger $r_h$, that is stable and dominant between the two. However, in the $(3,0)$ region the largest black hole is unstable.

We find a Hawking--Page transition to the thermal gas occurs at fixed $\tilde{\Phi}K\geq 0$ as $\tilde{\b}$ is increased. Denoting the transition point as $\tilde{\beta}_{\text{HP}}$, note that the numerics are consistent with the analytic value $\tilde{\beta}_{\text{HP}}(\tilde{\Phi}K = 0) = 32\pi/27$ computed with the Schwarzschild geometry \cite{Anninos:2023epi}.

The kink in the phase diagram is where $\tilde{\beta}_{\text{HP}}(\tilde{\Phi}K)$ equals the critical inverse temperature beyond which the dominant geometry ceases to exist. Below this kink, we have a first-order Hawking--Page transition between the dominant high-temperature solution and the thermal gas. However, for $\tilde{\Phi}K$ above the kink, we have a phase transition because the dominant high-temperature geometry fails to exist for $\tilde{\beta} > \tilde{\beta}_{\text{HP}}(\tilde{\Phi}K)$. This is a rather strange ``zeroth-order" phase transition, which may signal a breakdown of the geometric description. We will see this occur in other cases as well, so we will comment on it in Appendix \ref{app:puzzle}. Importantly, the zeroth-order transition only occurs at low temperature in (part of) the cosmic branch; the black-hole branch and high temperatures on any branch are well-behaved. 

As for $\tilde{\Phi}K < 0$, there is only one cosmic solution and no vacuum, and so the former dominates the ensemble by default. Interestingly, there is a curve separating a region where this solution is stable ($C_K > 0$) from one where it is unstable ($C_K < 0$). At high temperature, this crossover is calculable by the subextensive solution in Section \ref{sec:highTemp} and is found to be
\begin{equation}
\tilde{\Phi}K = -\sqrt{\frac{(d-1)(d-2)}{4}}.
\end{equation}

\subsection{Hyperbolic solutions}\label{subsec:FlatChargedHyp}

In this section we consider a conformal class of boundary geometries with representative $S^1 \times \mathbb{H}^{d-1}$. In the uncharged case, the bulk solution used in \cite{Banihashemi:2024yye} was a flat-space limit of the negative-mass hyperbolic black hole in AdS. Picking a negative mass ensured that we obtained a geometry with a horizon in the flat-space limit. This is because the negative-mass hyperbolic black hole in AdS has two horizons, and the inner horizon survives the flat-space limit $\ell \rightarrow \infty$. The resulting geometry and patch used in \cite{Banihashemi:2024yye} is a cosmic geometry, shown in Figure \ref{figs:penroseFlatNonSphereCharged}, and it reproduces the predictions of the thermal effective action on $\mathbb{H}^{d-1}$.

\input{figs/flatCharged/texCode/penroseDiagrams/penroseFlatNonSphereCharged}

\input{figs/flatCharged/texCode/solutionSpaces/solutionsFlatHyperCharged}

Upon introducing charge, the relevant solution is given by \eqref{bulksoln} with $k = -1$ and $\ell \rightarrow \infty$. This is just the flat-space limit of the charged AdS hyperbolic black hole. The Penrose diagram is again given by Figure \ref{figs:penroseFlatNonSphereCharged}. The relevant equation we need to solve to satisfy our boundary data is again \eqref{phikfundgentopSec3}, now with $k = -1$. The solution space is shown on the left in Figure \ref{figs:phasesFlatHyperCharged}. The $\tilde{\beta}$ beyond which no solutions exist can be found as $\tilde{\Phi} K \rightarrow 0^+$, for which \eqref{phikfundgentopSec3} simplifies. Either by this equation or consistency with the uncharged case $\tilde{\Phi} \rightarrow 0$ \cite{Banihashemi:2024yye}, we get 
\be
\tilde{\Phi}K = 0: \ \ \ \ \tilde{\b}_c = \f{4\pi}{\sqrt{d(d-2)}} \left(2 - \f 2 d\right)^{1/(d-2)}.\label{critbetahyp}
\ee
For $d = 3$, this value is $16\pi/\sqrt{27} \approx 9.674$, which agrees with our numerical plot. 

As in the case $k=1$, some solutions decouple in the uncharged limit $\tilde{\Phi} \rightarrow 0$. One of the solutions in the $(0,2)$ region has $r_h$ and $r_c$ run into the singularity and the solution becomes unphysical as we approach from $K > 0$. Starting from $K < 0$, this again happens if $\tilde{\b} < \tilde{\b}_c$, otherwise the solution remains physical. This agrees precisely with the result for $\tilde{\Phi} = 0$ \cite{Banihashemi:2024yye}.

The phase diagram follows immediately from the solution space and is represented on the right in Figure \ref{figs:phasesFlatHyperCharged}. The lack of a putative saddle at low temperatures is problematic; we will comment on this in Appendix \ref{app:puzzle}. 

\subsection{Flat solutions}\label{sec:flatflat}

Here, we consider a conformal class of boundary geometries with representative $S^1 \times \mathbb{T}^{d-1}$:
\begin{equation}
ds^2 = \lambda^2(d\tilde{\tau}^2 + dx_i^2),\qquad \tilde{\tau} \sim \tilde{\tau} + \tilde{\beta},\ \ \ \ x_i \sim x_i + \tilde{L}_i,
\end{equation}
where $i$ is summed from $1$ to $d-1$ and we fix $\min \tilde{L}_i = 1$. Normalizing one of the lengths to unity means that the other lengths are dimensionless and measured in units of min $\tilde{L}_i$. Thus, we can think of this set of parameters as the dimensionless ratios needed to characterize the conformal class of metrics.

In the uncharged case, the relevant bulk solution was given by a compactification of Rindler space \cite{Banihashemi:2024yye}. However, introducing an electric potential while preserving such a metric ansatz is not possible. Instead, the relevant solution is the flat-space limit of the charged AdS black brane. This is given by \eqref{bulksoln} with $k = 0$ and $\ell \rightarrow \infty$, but with the transverse directions compactified. The charged AdS black brane has two horizons, and in the flat-space limit the outer horizon goes off to infinity and we preserve the inner horizon. This is similar to what happened for $k = -1$, and in fact the full bulk geometry is described by the same Penrose diagram as the hyperbolic horizon geometry (Figure \ref{figs:penroseFlatNonSphereCharged}). We therefore again have no black hole solutions in this case and can map out the solution space by solving \eqref{phikfundgentopSec3} numerically with $k = 0$. This space is represented in Figure \ref{figs:solutionsFlatPlanarCharged}.

\input{figs/flatCharged/texCode/solutionSpaces/solutionsFlatPlanarCharged}

In the uncharged limit $\tilde{\Phi} \rightarrow 0$ we again find that some solutions decouple: the solution from the $(0,1)$ region and one solution from the $(0,2)$ region have $r_h$ and $r_c$ crash into the singularity and become unphysical. This reproduces the $\tilde{\Phi} = 0$ case \cite{Banihashemi:2024yye}, which has zero solutions for $K < 0$ and one Rindler solution and one thermal gas solution for $K > 0$.\footnote{As the uncharged solution is Rindler space, it is meaningless to refer to it as a black hole or cosmic solution; it sits somewhere in between. In fact, taking the flat-space limit starting from $\L < 0$, $k = 0$ in Section \ref{adsflat} would land on the Rindler geometry from a black-hole solution, whereas taking the flat-space limit starting from $\L > 0$, $k=0$ in Section \ref{dsflat} would land on the Rindler geometry from a cosmic solution.} Interestingly, the electric potential $\tilde{\Phi}$ is independent of $q$, so the $\tilde{\Phi} \rightarrow 0$ limit has to be taken as $z \rightarrow 1$. However, to keep the temperature and extrinsic curvature from diverging in this limit, we also need to take $x\rightarrow 0$.

In addition to the cosmic solutions, we have horizonless geometries due to the toroidal compactification. These are the same geometries used in the uncharged case \cite{Banihashemi:2024yye} and correspond to having one of the spatial cycles cap off in the bulk. In addition, there is a constant electric potential. For example, if the $x_1$ direction caps off, we have
\be
ds^2 = d\t^2 + dr^2 + r^2 dx_1^2 + dx_a^2,\ \ \ \ a = 2,\dots,d-1, \ \ \ \ A = A_\t d\t,\qquad r \in (0, r_c)\,\label{flatsoliton}.
\ee
There are $d-1$ of these solutions, one for each spatial cycle, and they exist for arbitrary $\tilde{\b}$ (where $\t \sim \t + \tilde{\b}$) and $\tilde{\Phi} K \geq 0$.

\input{figs/flatCharged/texCode/phasePlots/phasesFlatPlanarCharged}

Just as in the hyperbolic ($k=-1$) case, we have two critical curves in Figure \ref{figs:solutionsFlatPlanarCharged}. There is one at $\tilde{\Phi}K = 0$ separating the $(0,1)$ region from the rest of the diagram. The other critical curve again separates the $(0,2)$ and $(0,0)$ regions. However, this time the latter curve does not terminate at $\tilde{\Phi}K = 0$ at a finite temperature. Instead, it approaches $\tilde{\Phi}K = 0$ as $\tilde{\beta} \to \infty$. We can actually find the analytic form of this curve, because the equation \eqref{phikfundgentopSec3} simplifies dramatically for $k = 0$ as this implies $b = 0$:
\begin{equation}
\tilde{\Phi}K = \frac{2\pi z(d-2z^{d-2})}{\eta(d-2)\tilde{\beta}}\sqrt{(z^{d-2} - 1)z^{d-2}}.\label{phikbetaplanar}
\end{equation}
The right-hand side has zeroes at $z = 1$ and $z = (d/2)^{1/(d-2)}$. Furthermore, $\tilde{\Phi}K$ is positive precisely between these two values of $z$. As such, there must be a $z = z_{\text{max}}$ in that range at which the right-hand side is maximized. This is
\begin{equation}
z_{\text{max}} = \left[\frac{d^2 + 2d - 4 + (d-2)\sqrt{d^2 + 4}}{4(2d-3)}\right]^{1/(d-2)}.
\end{equation}
Plugging this back into \eqref{phikbetaplanar} yields the critical curve, which is simply $\tilde{\Phi}K \propto 1/\tilde{\beta}$.

The phase structure is obtained by comparing the on-shell action of the stable cosmic solution to the horizonless geometries. The on-shell action of the horizonless geometry with cycle $x_\a$ capping off is the same as the case with vanishing electric potential, since the constant gauge field does not contribute to $F^2$ and the geometry is unchanged:
\be
I_\a= -\frac{(2\pi)^{d-1}}{4dG K^{d-1}} \frac{\tilde{\b}{\rm Vol}[\mathbb{T}^{d-1}]}{\tilde{L}_{\a}^{d}},\qquad \text{Vol}[\mathbb{T}^{d-1}] \equiv \tilde{L}_1 \cdots \tilde{L}_{d-1}.
\ee
Thus the horizonless geometry of least action is the one where $\a$ indexes the smallest spatial cycle. Comparing this against the cosmic horizon solution gives the phase diagram in Figure \ref{figs:phasesFlatPlanarCharged}. As in the spherical ($k = 1$) case, as we increase $\tilde{\b}$ we have a first-order Hawking--Page transition for small $\tilde{\Phi}K$ below a kink, and above the kink we get a ``zeroth-order" transition.

\section{$\tilde{\Phi} = 0$, $\L < 0$: anti-de Sitter space}\label{sec:negLambda}
\input{figs/AdSUncharged/texCode/penroseDiagrams/penroseAdS}

In this section we examine pure gravity with a negative cosmological constant $\L < 0$:
\begin{equation}
I = -\frac{1}{16\pi G}\int_{\mathcal{M}}d^{d+1}x \sqrt{g}\,\left(R - 2\L\right) - \frac{1}{8d\pi G}\int_{\partial\mathcal{M}} d^d x \sqrt{h}\,K.
\end{equation}
The equations of motion following from this action are solved by \eqref{bulksoln} with $q= 0$, which simply modifies the emblackening factor to 
\be \label{ads-metric}
 f(r) \equiv k + \f{r^2}{\ell^2} - \frac{r_h^{d-2}}{r^{d-2}}\left(k + \f{r_h^2}{\ell^2}\right),\ \ \ \ \Lambda = -\frac{d(d-1)}{2\ell^2}.
\ee
$k \in \{-1,0,1\}$ indicates the horizon geometry, $r_h$ is the horizon radius, and $\ell$ is the AdS radius. The equations of our boundary value problem are given by taking the master equations \eqref{mastereqs1}--\eqref{mastereqs3} and taking $q\rightarrow 0$, which sets $x = \tilde{\Phi}K = 0$ and leaves two equations:
\begin{align}
\tilde{\beta} &= \pm \frac{4\pi z}{k(d-2) + d z^2 y^2}\sqrt{k(1-z^{d-2}) + y^2(1-z^d)},\label{cceq1}\\
K\ell &= \pm \frac{1}{y\sqrt{k(1-z^{d-2}) + y^2(1-z^d)}}\left[k(d-1) + \frac{d}{2}\left(2y^2 - z^{d-2}(k + z^2 y^2)\right)\right].\label{cceq2}
\end{align}
Again, the $+$ and $-$ branches respectively describe black holes and cosmic geometries.

Just as in the $\Lambda = 0$, $\tilde{\Phi} \neq 0$ case, we can solve \eqref{cceq1} for one of the bulk parameters (in this case, $y$). We have the following pair of solutions:
\begin{align}
y^2 = y_1^2 &\equiv \frac{1}{d^2 z^2 \tilde{\beta}^2} \bigg[8\pi^2 (1-z^d) - kd(d-2)\tilde{\beta}^2 \bigg.\nonumber\\
&\hspace{2.25cm}\Bigg.+ \sqrt{\Big(8\pi^2(1-z^d)\Big)^2 + 16\pi^2 k d \tilde{\beta}^2 \Big(2(1-z^d) - d(1-z^2)\Big)}\Bigg],\label{y1sol}\\
y^2 = y_2^2 &\equiv \frac{1}{d^2 z^2 \tilde{\beta}^2} \bigg[8\pi^2 (1-z^d) - kd(d-2)\tilde{\beta}^2 \bigg.\nonumber\\
&\hspace{2.25cm}\Bigg.- \sqrt{\Big(8\pi^2(1-z^d)\Big)^2 + 16\pi^2 k d \tilde{\beta}^2 \Big(2(1-z^d) - d(1-z^2)\Big)}\Bigg].\label{y2sol}
\end{align}
The second solution is spurious for $k = 0$ since $y_2^2 = 0$, although we will be able to solve this case exactly in Section \ref{adsflat}. These solutions are plugged into \eqref{cceq2} to reduce the system to a single equation. The hyperbolic case $k = -1$ will allow for both black hole and cosmic solutions, as is clear from the Penrose diagram Figure \ref{figs:penroseAdS}(b). 

We will also examine the thermodynamics, for which we use the on-shell Euclidean action. By setting $x = 0$ in the general answer \eqref{dimensionlessActGen}, we get
\begin{align}
I_{\text{cl}} &= -\frac{\tilde{\beta}\,\text{Vol}[\Sigma_k]}{8\pi G |K|^{d-1}}|h(y,z)|^{d-1}\left[-\frac{y^2 |1-z^d|}{\sqrt{k(1-z^{d-2}) + y^2 (1-z^d)}} + \frac{h(y,z)}{d}\right],\nonumber\\
h(y,z) &= \pm \frac{2k(d-1) + 2dy^2 - dz^{d-2}(k + z^2 y^2)}{2\sqrt{k(1-z^{d-2}) + y^2(1-z^d)}}.\label{adsAct}
\end{align}
For certain ranges of parameters the solutions \eqref{y1sol}--\eqref{y2sol} can become unphysical, but we will always ensure $y > 0$ before plugging into the on-shell action when scanning over the solution space. The $+$ and $-$ branches respectively describe black holes and cosmic geometries.

\subsection{Spherical solutions}

The bulk geometries in the case $k = 1$ are patches of the spherical AdS black hole. The Penrose diagram is given in Figure \ref{figs:penroseAdS}(a). 

By solving the equations numerically, we obtain the solution space shown in Figure \ref{figs:solutionsAdSSphere}. There are precisely two regions---a $(2,0)$ one (with $K\ell > d$) in which we have a large and small black hole and a $(0,0)$ one with no solutions with horizons. This is exactly like in AdS with a Dirichlet boundary at infinity \cite{Hawking:1982dh} and flat space at finite cutoff (with either Dirichlet \cite{York:1986it} or conformal \cite{Anninos:2023epi,Banihashemi:2024yye} boundary conditions). Indeed, the critical values of $\tilde{\beta}$ as $K\ell \to d$ and $K\ell \to \infty$ respectively approach that of AdS with an asymptotic Dirichlet boundary and flat Schwarzschild with conformal boundary conditions:\input{figs/AdSUncharged/texCode/solutionSpaces/solutionsAdSSphere}
\begin{equation}
\tilde{\beta}_{\text{max}}(K\ell \to d) = \frac{2\pi}{\sqrt{d(d-2)}},\ \ \ \ \tilde{\beta}_{\text{max}}(K\ell \to \infty) = \frac{4\pi}{d-2}\left(\frac{2}{d}\right)^{1/(d-2)}\sqrt{1-\frac{2}{d}}.
\end{equation}
We can also show that a Hawking--Page transition occurs in this context for all $K\ell \geq d$, marking a transition between the large black hole solution and a thermal gas phase. To show this, we need to compare the free energies of the two phases. The relevant geometry for the thermal gas phase is the AdS vacuum,
\input{figs/AdSUncharged/texCode/phasePlots/phasesAdSSphere}
\begin{equation}
ds^2 = \left ( 1 + \frac{r^2}{\ell^2} \right ) d\t^2 + \frac{dr^2}{1 + \frac{r^2}{\ell^2}} + r^2 d\Omega_{d-1}^2,\qquad r \in (0,r_c).
\end{equation}
This solution only exists for $K\ell > d$. By inverting the $+$ branch of the $K$ equation \eqref{cceq2} (setting $z = 0$ to get the vacuum AdS geometry), we find that the cutoff radius is given by the following:
\begin{equation}
\frac{r_c}{\ell} = y_* = \sqrt{\frac{2d(d-1) - K\ell\left(K\ell - \sqrt{(K\ell)^2 - 4(d-1)}\right)}{2\left((K\ell)^2 - d^2\right)}}.
\end{equation}
The on-shell action is
\begin{equation}
I_{\text{gas}} = \frac{\tilde{\beta}\,\text{Vol}[S^{d-1}]}{8\pi G}\left[\frac{r_c^{d+1}}{\ell^2 \sqrt{1 + r_c^2/\ell^2}} - \frac{r_c^d K}{d}\right] = -\frac{\tilde{\beta}\,\text{Vol}[S^{d-1}]}{8\pi G} \frac{(d-1)r_c^{d-1}}{d\sqrt{1 + r_c^2/\ell^2}}.
\label{thermalgasads}\end{equation}
Pulling out an overall $\tilde{\b}/(G K^{d-1})$ and writing the rest of the expression in terms of the dimensionless quantity $K\ell$ gives us
\begin{equation}
I_{\text{gas}} = -\frac{\tilde{\beta}\,\text{Vol}[S^{d-1}]}{8\pi dGK^{d-1}} \frac{(d-1)(K\ell)^{d-1} y_*^{d-1}}{\sqrt{1+y_*^2}}.
\end{equation}
Minimizing between this and \eqref{adsAct} for $k=1$ produces the phase diagram, which we present for $d=3$ in Figure \ref{figs:phasesAdSSphere}. The Hawking--Page inverse temperature increases with $K\ell$, but it should approach the AdS value with a Dirichlet boundary at infinity as $K\ell \to d$ and that of Schwarzschild with conformal boundary conditions as $K\ell \to \infty$:
\begin{equation}
\tilde{\beta}_{\text{HP}}(K\ell \to d) = \frac{2\pi}{d-1},\ \ \ \ \tilde{\beta}_{\text{HP}}(K\ell \to \infty) = \frac{4\pi}{d}\left[\frac{4(d-1)}{d^2}\right]^{1/(d-2)}.
\end{equation}
Indeed, for $d = 3$ these values are respectively $\pi\approx 3.142$ and $32\pi/27 \approx 3.723$, which match the transition curve in the phase diagram.

\subsection{Hyperbolic solutions}

The bulk geometries in the case $k = -1$ are patches of the hyperbolic AdS black hole. We will see that in this case we can have cosmic horizon solutions, due to the appearance of an inner horizon in the hyperbolic black hole with negative mass; see Figure \ref{figs:penroseAdS}(b).

The metric is given by
\be \label{hyper-AdS}
ds^2 = f(r) d\t^2 + \f{dr^2}{f(r)} + r^2 d\mathbb{H}_{d-1}^2,\ \ \ \ f(r) = -1-\frac{\mu}{r^{d-2}}+\f{r^2}{\ell^2},
\ee
where $d\mathbb{H}_{d-1}^2$ is the line element of $\mathbb{H}^{d-1}$ and $\mu$ is the mass parameter. This geometry has two horizons for $-\f{2}{d-2}\left(\f{d-2}{d}\right)^{d/2}\ell^{d-2} < \m < 0$, while for $\m \geq 0$ it has just one. The extremal solution corresponds to $\m$ saturating the lower bound. In all cases we have
\input{figs/AdSUncharged/texCode/solutionSpaces/solutionsAdSHyper}
\be
\mu = r_h^{d-2}\left(\frac{r_h^2}{\ell^2} - 1\right).
\ee

The boundary value equations to solve are given by \eqref{cceq1}--\eqref{cceq2} with $k = -1$:
\be \label{dimless-hyper-data}
\tilde{\b}=\pm \frac{4\pi z \sqrt{z^{d-2}-1+y^2(1-z^d)}}{d y^2 z^2-d+2},\ \ \ \ K \ell= \pm \frac{2(1-d)+d z^{d-2}+dy^2(2-z^d)}{2y \sqrt{z^{d-2}-1+y^2(1-z^d)}}.
\ee
In this case, we have the bound $y>\sqrt{(d-2)/d}$ for a black-hole solution and $y<\sqrt{(d-2)/d}$ for a cosmic solution. These bounds come from the fact that $r_h = \ell$ demarcates the situation with two horizons from the situation with one horizon, and the fact that $r_c > r_h$ for a black-hole solution and $r_c < r_h$ for a cosmic solution. 

This pair of equations is reduced to a single equation by solving the $\tilde{\b}$ equation for $y$ and plugging into the $K\ell$ equation, which gives \eqref{cceq2}--\eqref{y2sol} with $k = -1$.

There is another solution to our boundary value problem corresponding to a patch of the extremal AdS$_2 \times \mathbb{H}^{d-1}$ solution \eqref{ads2h2}:
\be
ds^2 = \left( \f{r^2}{\ell_2^2}-1\right)d\t^2 + \f{dr^2}{\f{r^2}{\ell_2^2}-1} + (d-2)\ell_2^2d\mathbb{H}_{d-1}^2,\qquad \t \sim \t + 2\pi \ell_2,\ \ \ \ r \in ( \ell_2, r_c). \label{ads2h2patch}
\ee
The curvature scale $\ell_2$ is related to the cosmological constant via $\Lambda = -(d-1)/(2\ell_2^2)$. 

The boundary data $\{\tilde{\beta},K\}$ is related to the bulk parameters by the equations
\begin{equation}
\tilde{\b} = \f{2\pi\sqrt{r_c^2/\ell_2^2-1}}{\sqrt{d-2}} ,\qquad K = \frac{1}{\ell_2^2} \frac{r_c}{\sqrt{r_c^2/\ell_2^2-1}}.
\end{equation}
Since $\ell_2$ is fixed by the theory (namely, the value of $\Lambda$), the only tunable parameter is $r_c$, and we see that these solutions are only relevant for a curve specified by
\be\label{extremalcurveads2h2}
K\ell = \sqrt{\frac{4 \pi ^2 d}{(d-2)\tilde{\b}^2} +d},
\ee
where we used the relation between the AdS$_2$ curvature scale and the AdS$_{d+1}$ curvature scale $\ell^2 = d \ell_2^2$. This is precisely the extremal gold curve in Figure \ref{fig:solutionsAdSHyper}. The equation for this curve can also be obtained by setting $z=1$ in \eqref{cceq2}, which as usual is a near-horizon near-extremal limit of the general family of black hole solutions. The fact that we have $\tilde{\b} \rightarrow 0$ as $K\ell \rightarrow +\infty$ is consistent with the flat space limit, where we only have a cosmic solution. 

Unlike the $\L = 0$ Reissner--Nordstr\"om black hole, the black hole along this critical curve is always the same. That is because there is only one extremal black hole with horizon radius $r_+ = \ell \sqrt{(d-2)/d}$, in contrast to the family of extremal black holes that appear along the critical line given in \eqref{criteqnsflatcharged}.

\subsubsection{Analytic features of the solution space}
There is a critical line that can be obtained by plugging $z=0$ (and $k = -1$) into \eqref{cceq2}--\eqref{y2sol},
\be
K\ell = d,
\ee
marking the disappearance of a black hole solution as we increase $K\ell$. This is the only value of $K\ell$ at which we have a black hole solution at all $\tilde{\b}$, and as $K\ell = d$ corresponds to $r_c = \infty$, this is consistent with the AdS/CFT result of having such black holes at all temperatures. 

There is a critical point located at the interface of the (2,0), (1,1), and (0,2) regions. It can be obtained by solving
\be
\left.\f{d}{dz}(K\ell)\right|_{z=1} = 0,
\ee
yielding a cubic equation for $\tilde{\b}$ with one real solution. In $d=3$ it is given by $\tilde{\b} \approx 5.914$. The critical lines and critical point were obtained exactly in analogy with the $\L = 0$, $\tilde{\Phi} \neq 0$ case. 

There is also a vertical line at $\tilde{\b}_c \approx 9.674$, which in general dimensions is at
\be
\tilde{\b}_c = \frac{4\pi}{\sqrt{d(d-2)}} \left(2 - \frac{2}{d}\right)^{1/(d-2)}.
\ee
The $\ell \rightarrow \infty$ limit of the solution space leads to one cosmic solution for $K > 0$ with $\tilde{\b}<\tilde{\b}_{c}$ and one cosmic solution for $K < 0$ with $\tilde{\b} > \tilde{\b}_{c}$, agreeing with the $\L = 0$ analysis in \cite{Banihashemi:2024yye}.

\subsubsection{Phase structure}

\input{figs/AdSUncharged/texCode/phasePlots/phasesAdSHyper}

The dominant phases of the ensemble are determined by minimizing the on-shell action \eqref{adsAct} for $k=-1$ over the handful of solutions. In this case there is no competing thermal gas geometry. However, we do have an extremal AdS$_2 \times \mathbb{H}^{d-1}$ solution given in \eqref{ads2h2patch}. To evaluate its on-shell action, we use $R - 2\L = - 2/\ell_2^2 $ to get
\begin{align}
I_{\text{bulk}} &= \left(\f{d-2}{d}\right)^{\f{d-1}{2}}\f{\ell^{d-1}\text{Vol}[\mathbb{H}^{d-1}]}{4 G } \left(\f{r_c }{\ell_2}-1\right),\\
I_{\text{bdry}} &= -\left(\f{d-2}{d}\right)^{\f{d-1}{2}}\f{\ell^{d-1} \text{Vol}[\mathbb{H}^{d-1}]}{4 G} \f{r_c}{d \ell_2}.
\end{align}
We use $r_c/\ell_2 = K \ell_2/\sqrt{-1+ K^2 \ell_2^2}$ to write this in terms of boundary parameters as
\be
I_{\text{total}} = \left(\f{d-2}{d}\right)^{\f{d-1}{2}}\f{\ell^{d-1} \text{Vol}[\mathbb{H}^{d-1}]}{4 G}\left(-1 + \f{(d-1)K \ell}{d^{3/2}\sqrt{K^2\ell^2/d-1}}\right).
\ee
We recall that this solution only exists on the curve specified by \eqref{extremalcurveads2h2}.

In the $(2,0)$ and $(0,2)$ regions the black hole/cosmic solution with the larger horizon radius dominates. In the $(1,1)$ region, the cosmic solution dominates over the black hole solution. In fact, when crossing the boundary between $(1,1)$ and $(0,2)$ at fixed $\tilde{\beta}$, the black hole solution becomes the cosmic solution with the smaller radius, and by crossing the boundary of $(1,1)$ and $(2,0)$ at fixed $K$, the cosmic solution becomes the black hole with the larger radius. Thus, the dominance takes over continuously as these boundaries are crossed.

\subsection{Flat solutions}\label{adsflat}

We now consider a conformal class of boundary geometries with representative $S^1 \times \mathbb{T}^{d-1}$, where $\mathbb{T}^{d-1}$ is a rectangular torus, so we have no spatial twists or time-space twists corresponding to an angular velocity. The boundary data is specified as explained at the beginning of Section \ref{sec:flatflat}, which can be encapsulated by $\tilde{\b}$ and $\tilde{L}_i$, where $i$ goes from $1$ to $d-1$. We fix $\min \tilde{L}_i = 1$. 

The bulk geometries with horizons are patches of the toroidal compactification of the planar AdS black hole. The Penrose diagram is the same as the $k = 1$ case and is again given in Figure \ref{figs:penroseAdS}(a). The equations \eqref{cceq1}--\eqref{cceq2} with $k=0$ admit an analytic solution, given in terms of dimensionful parameters by 
\begin{equation}
r_h = \frac{4\pi \ell}{\tilde{\b}d^2}\left[K\ell - \sqrt{(K\ell)^2 - d^2}\right],\ \ \ \ r_c = r_h \left(\f{2\pi }{4\pi - r_h\tilde{\b}K}\right)^{1/d}.
\end{equation}
The planar solution captures the leading term in the analytic high-temperature solution for $k \neq 0$ from Section \ref{sec:highTemp}. We have $r_c > r_h$ as necessary for black-hole solutions. The on-shell action is given by 
\be
I_{\text{cl}} = -c_0 \frac{\tilde{\b}{\rm Vol}[\mathbb{T}^{d-1}]}{\tilde{\b}^{d}},\ \ \ \ \text{Vol}[\mathbb{T}^{d-1}] \equiv \tilde{L}_1 \cdots \tilde{L}_{d-1}.
\ee
It will soon become clear why we write the action in this funny way. 

\input{figs/AdSUncharged/texCode/solutionSpaces/solutionsAdSPlanar}

In addition to these black hole solutions, we have horizonless geometries due to the toroidal compactification. These are simply patches of the AdS soliton geometry \cite{Horowitz:1998ha}, and there are $d-1$ of them depending on which spatial cycle caps off in the interior. For example, if $x_1$ caps off in the interior, the solution is given by
\be
ds^2 = r^2 d\t^2 + dr^2 + f(r)dx_1^2 + r^2 dx_a^2,\ \ \ \ f(r) = \f{r^2}{\ell^2} - \f{r_0^d}{r^{d-2}\ell^2}\,\qquad r\in (r_0,r_c).
\ee
where $a = 2,\dots,d-1$ and $\t \sim \t + \tilde{\b}$. Like the black brane, these solutions exist for arbitrary $\tilde{\b}$ and $K \ell \geq d$. The phase structure is obtained by comparing the on-shell action of the toroidally compactified black brane to the horizonless geometries. The on-shell action of the horizonless geometry with cycle $x_\a$ capping off is the same as the on-shell action of the black brane under the identification $\tilde{L}_\a \leftrightarrow \tilde{\b}$, since the Euclidean geometries are the same:
\be
I_\a= -c_0 \frac{\tilde{\b}{\rm Vol}[\mathbb{T}^{d-1}]}{\tilde{L}_{\a}^{d}},\qquad \text{Vol}[\mathbb{T}^{d-1}] \equiv \tilde{L}_1 \cdots \tilde{L}_{d-1}.
\ee
Therefore the horizonless geometry of minimal action is the one where $\a$ indexes the smallest spatial cycle. Minimizing this against the solution with a horizon gives the phase diagram on the right in Figure \ref{figs:solutionsAdSPlanar}. This case has a modular-invariant structure similar to that studied in the usual AdS/CFT case \cite{Shaghoulian:2015lcn} and as studied in $\L = 0$ with conformal boundary conditions \cite{Banihashemi:2024yye}.

\section{$\tilde{\Phi} = 0$, $\L > 0$: de Sitter space}\label{sec:posLambda}

In this section, we examine pure gravity with a positive cosmological constant $\L > 0$:
\begin{equation}
I = -\frac{1}{16\pi G}\int_{\mathcal{M}}d^{d+1}x \sqrt{g}\,\left(R - 2\L\right) - \frac{1}{8d\pi G}\int_{\partial\mathcal{M}} d^d x \sqrt{h}\,K.
\end{equation}
The equations of motion following from this action are solved by \eqref{bulksoln} with $q= 0$ and $\ell \rightarrow i \ell$, which simply modifies the emblackening factor to 
\be \label{dS-metric}
 f(r) \equiv k - \f{r^2}{\ell^2} - \frac{r_h^{d-2}}{r^{d-2}}\left(k - \f{r_h^2}{\ell^2}\right),\ \ \ \ \Lambda = \frac{d(d-1)}{2\ell^2}.
\ee
The equations for the boundary data can be found by applying the analytic continuation $\ell \to i\ell$, which also implies $y \to -iy$, to \eqref{cceq1} and \eqref{cceq2}.
\begin{align}
\tilde{\beta} &= \pm \frac{4\pi z}{k(d-2) - d z^2 y^2}\sqrt{k(1-z^{d-2}) - y^2(1-z^d)},\label{cceq1ds}\\
K\ell &= \pm \frac{1}{y\sqrt{k(1-z^{d-2}) - y^2(1-z^d)}}\left[k(d-1) - \frac{d}{2}\left(2y^2 + z^{d-2}(k - z^2 y^2)\right)\right].\label{cceq2ds}
\end{align}
We find the solutions to the $\tilde{\b}$ equation by analytically continuing the solutions \eqref{y1sol}--\eqref{y2sol}:
\begin{align}
y^2 = -y_1^2 &\equiv -\frac{1}{d^2 z^2 \tilde{\beta}^2} \bigg[8\pi^2 (1-z^d) - kd(d-2)\tilde{\beta}^2 \bigg.\nonumber\\
&\hspace{2.25cm}\Bigg.+ \sqrt{\Big(8\pi^2(1-z^d)\Big)^2 + 16\pi^2 k d \tilde{\beta}^2 \Big(2(1-z^d) - d(1-z^2)\Big)}\Bigg],\label{y1solds}\\
y^2 = -y_2^2 &\equiv -\frac{1}{d^2 z^2 \tilde{\beta}^2} \bigg[8\pi^2 (1-z^d) - kd(d-2)\tilde{\beta}^2 \bigg.\nonumber\\
&\hspace{2.25cm}\Bigg.- \sqrt{\Big(8\pi^2(1-z^d)\Big)^2 + 16\pi^2 k d \tilde{\beta}^2 \Big(2(1-z^d) - d(1-z^2)\Big)}\Bigg].\label{y2solds}
\end{align}
The first solution is spurious for $k = -1$ (it gives $y^2 < 0$) while the second solution is spurious for $k=0$ (it gives $y^2 = 0$). We will be able to solve the $k=0$ case exactly in Section \ref{dsflat}.

The on-shell action for this set of solutions is given by setting $x = 0$ and continuing $y^2 \rightarrow -y^2$ in \eqref{dimensionlessActGen}, which leads to
\begin{align}
I_{\text{cl}} &= -\frac{\tilde{\beta}\,\text{Vol}[\Sigma_k]}{8\pi G |K|^{d-1}}|h(y,z)|^{d-1}\left[\frac{y^2 |1-z^d|}{\sqrt{k(1-z^{d-2}) - y^2 (1-z^d)}} + \frac{h(y,z)}{d}\right],\nonumber\\
h(y,z) &= \pm \frac{2k(d-1) - 2dy^2 - dz^{d-2}(k - z^2 y^2)}{2\sqrt{k(1-z^{d-2}) - y^2(1-z^d)}}.\label{dsAct}
\end{align}
As in the $\L < 0$ case, for certain ranges of parameters the solutions \eqref{y1solds}--\eqref{y2solds} can become unphysical, but we will always ensure $y > 0$ when scanning over the solution space.
\input{figs/dSUncharged/texCode/penroseDiagram/penrosedS}

\subsection{Spherical solutions}

Our primary set of solutions is the family of de Sitter--Schwarzschild solutions with spherical spatial slices. These are given by the emblackening factor \eqref{dS-metric} with $k = 1$, and their on-shell action is given by \eqref{dsAct} with $k=1$. These solutions can be of black-hole type or cosmic type as long as $\m > 0$, as shown in Figure \ref{figs:penrosedS}(a). Relevant patches on both sides of the boundary have been emphasized in the context of dS$_3$ with Dirichlet boundary conditions in \cite{Coleman:2021nor}. For $\m < 0$, we only have cosmic-type solutions, as shown in Figure \ref{figs:penrosedS}(b). The phase diagram and solution space have previously been calculated in $d=2, 3$ in \cite{Anninos:2024wpy}.

\input{figs/dSUncharged/texCode/solutionSpaces/solutionsdSSphere}

The boundary-value equations to be solved are given by \eqref{cceq1ds}--\eqref{cceq2ds} with $k=1$:

\be \label{dimless-hyper-data-ds}
\tilde{\b}=\pm \frac{4\pi z \sqrt{1-z^{d-2}-y^2(1-z^d)}}{d-2-d y^2 z^2},\ \ \ \ K \ell= \pm \frac{2(d-1)-d z^{d-2}-dy^2(2-z^d)}{2y \sqrt{1-z^{d-2}-y^2(1-z^d)}}.
\ee
This pair of equations is reduced to a single equation by solving the $\tilde{\b}$ equation for $y$ and plugging into the $K\ell$ equation, which gives \eqref{cceq2ds}--\eqref{y2solds} with $k = 1$.

For $\m = 0$, this family of solutions reduces to empty de Sitter space. A patch of this geometry will serve as the ``thermal gas,"
\be
ds^2 = \left(1-\frac{r^2}{\ell^2}\right)d\t^2 + \f{dr^2}{1-\frac{r^2}{\ell^2}} + r^2d\Omega_{d-1}^2,\qquad \t \sim \t + \b,\ \ \ \ r \in (0, r_c),\label{dsThermalGas}
\ee
where $r_c < \ell$ and the Euclidean time coordinate can be compactified with any periodicity $\b$. Therefore, our boundary parameters are
\be
\tilde{\b} = \f{\sqrt{1-r_c^2/\ell^2}}{r_c}\b, \ \ \ \ K = \f{d-1-d r_c^2/\ell^2}{r_c \sqrt{1-r_c^2/\ell^2}}.
\ee
Solving the $K$ equation above gives
\begin{equation}
\frac{r_c}{\ell} = y_* = \sqrt{\frac{2d(d-1) + K\ell\left[K\ell - \sqrt{(K\ell)^2 + 4(d-1)}\right]}{2\left[(K\ell)^2 + d^2\right]}}.
\end{equation}
The on-shell action is given by 
\be
I_{\text{gas}} = -\f{\tilde{\b}(d-1) r_c^{d-1}\text{Vol}[S^{d-1}]}{8\pi d G \sqrt{1-r_c^2/\ell^2}},
\ee
which can be understood as an analytic continuation $\ell \rightarrow i\ell$ of the on-shell action for the thermal gas in AdS \eqref{thermalgasads}. Pulling out an overall $\tilde{\b}/(G K^{d-1})$ and writing the rest of the expression in terms of the dimensionless quantities $y_*$ and $K\ell$ yields
\begin{equation}\label{dsthermalgasact}
I_{\text{gas}} = -\frac{\tilde{\beta}\,\text{Vol}[S^{d-1}]}{8\pi GdK^{d-1}} \frac{(d-1)(K\ell)^{d-1} y_*^{d-1}}{\sqrt{1-y_*^2}}.
\end{equation}
The thermal gas accommodates all $K \ell \in \mathbb{R}$. 

There is another solution to our boundary value problem corresponding to a patch of the extremal $\text{dS}_2 \times S^{d-1}$ solution \eqref{ds2s2}:
\be
ds^2 = \left(1 - \f{r^2}{\ell_2^2}\right)d\t^2 + \f{dr^2}{1-\f{r^2}{\ell_2^2}} + (d-2)\ell_2^2d\Omega_{d-1}^2,\qquad \t \sim \t + 2\pi \ell_2,\ \ \ \ r \in ( r_c,\ell_2). \label{ds2s2patch}
\ee
The curvature scale $\ell_2$ is related to the cosmological constant via $\Lambda = (d-1)/(2\ell_2^2)$. Another acceptable patch is $r \in (-\ell_2, r_c)$, but due to the $r \rightarrow -r$ symmetry we restrict to the one written above. The boundary data $\{\tilde{\beta},K\}$ is related to the bulk parameters by the equations
\begin{equation}
\tilde{\b} = \f{2\pi\sqrt{1-r_c^2/\ell_2^2}}{\sqrt{d-2}} ,\qquad K = \frac{1}{\ell_2^2} \frac{r_c}{\sqrt{1 - r_c^2/\ell_2^2}}.
\end{equation}
Since $\ell_2$ is fixed by the theory (i.e.~the value of $\Lambda$), the only tunable parameter is $r_c$, and we see that these solutions are only relevant for a curve specified by
\be\label{extremalcurveds2s2}
K\ell = \sqrt{\frac{4 \pi ^2 d}{(d-2)\tilde{\b}^2} - d},
\ee
where we have used the relation between the dS$_2$ curvature scale and the $dS_{d+1}$ curvature scale $\ell^2 = d \ell_2^2$. For $d=3$, this is precisely the extremal dS$_2 \times S^2$ gold curve observed in Figure \ref{figs:solutionsdSSphere}. This equation can also be obtained by setting $z=1$ in \eqref{cceq2ds}, which as usual is a near-horizon near-extremal limit of the general family of black hole solutions. 

\subsubsection{Analytic features of the solution space}

In this case, we are not able to extract any analytic formulas for curves apparent in the numerics besides the extremal curve, whose formula is given in \eqref{extremalcurveds2s2}. In particular, we could not calculate the curve separating the $(0,0)$ region from the rest of the solution space, although we know that as $K \ell \rightarrow \infty$ it should asymptote to the $\L = 0$, $\tilde{\Phi} = 0$ answer $\tilde{\b} = \frac{4\pi}{d-2}\left(\frac{2}{d}\right)^{1/(d-2)}\sqrt{1-\frac{2}{d}}$, which for $d=3$ gives $8\pi/\sqrt{27} \approx 4.837$.

The limit $\ell \rightarrow \infty$ with $K > 0$ correctly reproduces the $\L = 0$ solution space, with two black holes existing above a critical temperature. The $\ell \rightarrow \infty$ limit with $K < 0$ seems to indicate two cosmic solutions and a thermal gas solution. However, all of these solutions decouple. The thermal gas solution decouples because fixing $K < 0$ means $r_c$ has to scale with $\ell$ (recall we pick up the interior patch since it is a thermal gas solution), so $r_c \rightarrow \infty$ as we take the flat-space limit. The two cosmic solutions decouple as well, with $r_c$, $r_h\,\rightarrow \infty$.

\subsubsection{Phase structure}

\input{figs/dSUncharged/texCode/phasePlots/phasesdSSphere}

The dominant phases of the ensemble are determined by minimizing the on-shell actions for the solutions with horizons \eqref{dsAct} (with $k=1$) and the thermal gas \eqref{dsthermalgasact} over the handful of solutions. We also have an extremal dS$_2 \times S^{d-1}$ solution given in \eqref{ds2s2patch}. To evaluate its on-shell action, we use $R - 2\L = 2/\ell_2^2 = 2d/\ell^2$ to write
\begin{align}
I_{\text{bulk}} &= -\left(\f{d-2}{d}\right)^{\f{d-1}{2}}\f{\ell^{d-1}\text{Vol}[S^{d-1}]}{4 G } \,\left(1-\f{r_c }{\ell_2}\right),\\
I_{\text{bdry}} &= -\left(\f{d-2}{d}\right)^{\f{d-1}{2}}\f{\ell^{d-1} \text{Vol}[S^{d-1}]}{4 G}\, \f{r_c}{d \ell_2}.
\end{align}
We use $r_c/\ell_2 = K \ell_2/\sqrt{1+ K^2 \ell_2^2}$ to write this in terms of the boundary parameters as 
\be
I_{\text{total}} = -\left(\f{d-2}{d}\right)^{\f{d-1}{2}}\f{\ell^{d-1} \text{Vol}[S^{d-1}]}{4 G}\left[1 - \f{(d-1)K \ell}{d^{3/2}\sqrt{1+K^2\ell^2/d}}\right].
\ee
The dominant phases are represented in Figure \ref{figs:phasesdSSphere}. The larger solution (in the sense of larger $r_h$) is always the dominant solution. There is again a Hawking--Page transition, as in the other $k = 1$ cases, but this time for all $K\ell$. As $K\ell \to \infty$, we again recover the flat-space answer $\tilde{\beta}_{\text{HP}} = \frac{4\pi}{d}\left[\frac{4(d-1)}{d^2}\right]^{1/(d-2)}$, which is $32\pi/27 \approx 3.723$ in $d = 3$.

\subsection{Hyperbolic solutions}

\input{figs/dSUncharged/texCode/solutionSpaces/solutionsdSHyper}

The relevant bulk geometries in this case have an emblackening factor given by \eqref{dS-metric} with $k=-1$. The Penrose diagram is given in Figure \ref{figs:penrosedS}(b) and indicates that we can only have cosmic solutions. Furthermore, there is no thermal gas solution.

The solution space is quite simple and given on the left in Figure \ref{fig:solutionsdSHyper}, with the phase diagram shown on the right. The flat-space limit $\ell \to \infty$ recovers the correct solution space and phase diagram. For $K > 0$, we get one cosmic solution below a critical $\tilde{\b}$ (given in \eqref{critbetahyp}) and no solutions above. For $K < 0$, an analysis of the solutions shows that the cosmic solution in the $(0,1)$ region and one of the solutions in the $(0,2)$ region decouple as $r_c$, $r_h \to \infty$, leaving one solution above the critical $\tilde{\b}$ in \eqref{critbetahyp} and no solutions below.

\subsection{Flat solutions}\label{dsflat}

\input{figs/dSUncharged/texCode/solutionSpaces/solutionsdSPlanar}

The relevant bulk geometries in this case have an emblackening factor given by \eqref{dS-metric} with $k=0$. Notice that this emblackening function is precisely the negative of the emblackening function for $\L < 0$, $k=0$. The Penrose diagram is given in Figure \ref{figs:penrosedS}(b) and indicates that we can only have cosmic solutions. We consider a rectangular toroidal compactification of this flat plane. This section proceeds precisely as Section \ref{adsflat}, except now the horizon and thermal gas solutions are available for all $K \ell \in \mathbb{R}$, so we will be brief.

The equations admit an analytic solution, given in terms of dimensionful parameters by 
\begin{equation}
r_h = \frac{4\pi \ell}{\tilde{\b}d^2}\left[-K\ell + \sqrt{(K\ell)^2 + d^2}\right],\ \ \ \ r_c = r_h \left(\f{2\pi }{4\pi - r_h\tilde{\b}K}\right)^{1/d}.
\end{equation}
We have $r_h > r_c$, as necessary for cosmic solutions.

The planar solution captures the leading term in the analytic high-temperature solution for $k \neq 0$ from Section \ref{sec:highTemp}. The on-shell action is given by 
\be
I_{\text{cl}} = -c_0 \frac{\tilde{\b}{\rm Vol}[\mathbb{T}^{d-1}]}{\tilde{\b}^{d}},\ \ \ \ \text{Vol}[\mathbb{T}^{d-1}] \equiv \tilde{L}_1 \cdots \tilde{L}_{d-1}.
\ee
We also have $d-1$ horizonless geometries due to the toroidal compactification, one for each of the $d-1$ cycles that can cap off in the interior. For example, if $x_1$ caps off in the interior, the solution is given by
\be
ds^2 = r^2 d\t^2 + dr^2 + f(r)dx_1^2 + r^2 dx_a^2,\ \ \ \ f(r) = -\f{r^2}{\ell^2} + \f{r_0^d}{r^{d-2}\ell^2},\qquad r\in (r_0,r_c).
\ee
where $a = 2,\dots,d-1$ and $\t \sim \t + \tilde{\b}$.

The phase structure is obtained by comparing the on-shell action of the toroidally compactified black brane to the horizonless geometries. The on-shell action of the horizonless geometry with cycle $x_\a$ capping off is the same as the on-shell action of the black brane under the identification $\tilde{L}_\a \leftrightarrow \tilde{\b}$, since the Euclidean geometries are the same:
\be
I_\a= -c_0 \frac{\tilde{\b}{\rm Vol}[\mathbb{T}^{d-1}]}{\tilde{L}_{\a}^{d}},\qquad \text{Vol}[\mathbb{T}^{d-1}] \equiv \tilde{L}_1 \cdots \tilde{L}_{d-1}.
\ee
Therefore the horizonless geometry of minimal action is the one where $\a$ indexes the smallest spatial cycle. Minimizing this against the solution with a horizon gives the phase diagram in Figure \ref{fig:phasesdSPlanar}. Like the AdS $k=0$ case, this has a modular-invariant structure.

As usual, the flat-space limit $\ell \rightarrow \infty$ reproduces the expected structure, with the $K < 0$ thermal gas and cosmic solutions decoupling.

\section{High-temperature analysis}\label{sec:highTemp}

In this section we will analyze the high-temperature limit of our boundary-value problem. We will revert to the dimensionful equations \eqref{sphereflateqns}--\eqref{qeq}, which we will solve in an expansion in $\tilde{\b}^{-1}$ for the dominant solution. This solution will encode the Wilson coefficients of the thermal effective theory.

To reduce the system to two equations, we can solve the $\tilde{\Phi}$ equation \eqref{qeq} for $q$ and plug into the $\tilde{\b}$ and $K$ equations \eqref{sphereflateqns}. 
For the remaining bulk parameters $r_h$ and $r_c$ we will use the following ansatz:
\be \label{ansatzhcK}
r_h = \frac{1}{\tilde{\beta}}\sum_{i=0}^\infty a_i \tilde{\beta}^i,\ \ \ \ r_c = \frac{1}{\tilde{\beta}}\sum_{i=0}^\infty w_i \tilde{\beta}^i.
\ee
We will plug \eqref{ansatzhcK} into the $\tilde{\beta}$ and $K$ equations \eqref{sphereflateqns} and solve for the coefficients order by order in $\tilde{\beta}$. We will find that this ansatz accommodates both the dominant solution, whose entropy is extensive (i.e. $S \sim 1/\tilde{\b}^{d-1}$ and so $a_0 \neq 0$), and subdominant solutions which have subextensive entropy (and so $a_0 = 0$). Our purpose in finding the subdominant solutions at high temperature is as a consistency check on our numerics. 

We will solve for the extensive solution with nonzero $\tilde{\Phi}$ and $\L$, which is more general than what was analyzed in the previous sections. The subextensive solutions will be restricted to the cases analyzed previously, i.e.~either $\L$ nonzero or $\tilde{\Phi}$ nonzero.

\subsection{The extensive solution}

The leading order constraint from the $\tilde{\b}$ equation \eqref{sphereflateqns} gives
\be
s\, n\, d\, a_0=4\pi \ell \sqrt{s(A-1)},\ \ \ \ A\equiv\left(\frac{a_0}{w_0}\right)^d,
\ee
where $s$ denotes the sign of the cosmological constant (which we will keep nonzero and take a zero-cosmological-constant limit at the end) and $n$ represents the chosen branch of the equations \eqref{sphereflateqns}, with our chosen convention being $n=-1$ for the black hole branch and $n=+1$ for the cosmic branch. (Note that this is opposite to the signs in the $\tilde{\beta}$ and $K$ equations \eqref{sphereflateqns}, but this turns out to be a convenient choice for the current discussion.) Thus an extensive solution for negative $\Lambda$ ($s=-1$) exists only if $n=-1$ and $A<1$, corresponding to a black hole horizon for which $r_h<r_c$. For positive $\Lambda$ ($s=1)$, there is an extensive solution if $n=1$ and $A>1$, corresponding to a cosmic horizons with $r_h > r_c$. Therefore, in all cases we have $sn=1$.

The leading order constraint from the $K$ equation yields
\be
s\, n\, d(2-A)=2K \ell \sqrt{s(A-1)}\xrightarrow{sn=1} d(2-A)=2K \ell \sqrt{s(A-1)}.
\ee
For $\Lambda < 0$, for which we found $A<1$, it is clear from above that $K$ should be positive, which is consistent with the solution being a black hole. For $\Lambda>0$, we may have a solution with either sign of $K$, which is consistent with a cosmic geometry. The results for the leading coefficients are
\be
A=\frac{2}{sd^2}\left[(K\ell)^2+s d^2- K\ell \sqrt{(K\ell)^2 +s d^2}\right], \ \ \ \ a_0=\frac{4\pi \ell}{s d^2} \left[ \sqrt{(K\ell)^2+s d^2} -K\ell \right]. \label{A-a0}
\ee
For $s = -1$, we see that there is no extensive black hole solution for $K\ell < d$.
From the higher order terms we find $a_1=w_1=0$ and 
\begin{equation} \label{extensivea2w2}
\begin{split}
a_2&=\frac{d-2}{4\pi \ell}\frac{1}{\sqrt{s(A-1)}}\left[ k\ell^2 s\left(1-A^{-(d-2)/d}\right)+\frac{A-1}{A^{(d-2)/d}-1}\eta^2\tilde{\Phi}^2 \right ],\\
w_2&=\frac{A^{-2+1/d} \sqrt{s(A-1)} }{4 \pi d \ell \, s} \left[\frac{d (d-3) A^{2(d-1)/d}-(d-1) (d-4)A-2 (d-2)}{s(A-1)}k \ell^2\right.\\
&\hspace{2cm}\left.+ \frac{2 (d-2) \left[-d (d-1) A^{2/d}+ ((d-2) (d-1)+2)A+2 (d-2)\right]}{(d-1) \left(1-A^{-(d-2)/d}\right)^2} \tilde{\Phi}^2\right].
 \end{split}
\end{equation}
Using the expressions for $a_0$ and $a_2$, we will compute the Wilson coefficients of the thermal effective theory in Section \ref{sec:Wilson}. 

Taking the flat limit $\ell \to \infty$ gives $A\to 1+sd^2/(4K^2\ell^2)$, and so we get
\begin{equation}
\begin{split}
a_0 &= w_0 = \frac{2\pi}{K},\\
a_2 &= \frac{1}{8\pi K}\left[k(d-2)^2 + \frac{8(d-2)(\tilde{\Phi}K)^2}{d-1}\right],\\
w_2 &= \frac{1}{8\pi K}\left[k(d-1)(d-2) + \frac{8(d-3)(\tilde{\Phi}K)^2}{d-1}\right].\label{coeffextensiveflat-K}
\end{split}
\end{equation}
In these equations, $K$ is positive, which can be derived directly from the $\L = 0$ equations or understood from the flat limit of the $\Lambda <0$ case. Since the leading coefficients are the same, the type of the solution is determined by the relative size of $a_2$ and $w_2$:
\begin{equation}
\begin{split}
&a_2 < w_2 \iff (\tilde{\Phi}K)^2 < \frac{k(d-1)(d-2)}{8},\\
& a_2 > w_2 \iff (\tilde{\Phi}K)^2 > \frac{k(d-1)(d-2)}{8}.
\end{split}
\end{equation}
We therefore see that the actual solution space for $K > 0$ depends on the horizon geometry. For flat ($k = 0$) and hyperbolic ($k = -1$) horizons, we always have that $a_2 > w_2$, meaning that the extensive solution is cosmic. For spherical ($k = 1$) horizons we get
\begin{equation}
\qquad \tilde{\b} \rightarrow 0 \implies \left\{\begin{split}
&\tilde{\Phi}K \in \left(\frac{\sqrt{(d-1)(d-2)}}{2\sqrt{2}},\infty\right) \textcolor{white}{0} \implies \text{cosmic horizon},\\
&\tilde{\Phi}K \in \left(0,\frac{\sqrt{(d-1)(d-2)}}{2\sqrt{2}}\right) \textcolor{white}{\infty} \implies \text{black hole}.
\end{split}\right.
\end{equation}
Thus, the spherical high-temperature solution in the flat limit has a cosmic horizon at larger values of $\tilde{\Phi}K$ and has a black hole horizon at smaller values of $\tilde{\Phi}K$. The transition is mediated by an extremal AdS$_2 \times S^{d-1}$ solution. This is consistent with the high-temperature limit of our phase diagram in Figure \ref{figs:phasesFlatSphereCharged}.

\subsection{The subextensive solutions}

\subsubsection{$\Lambda \neq 0$ and $\tilde{\Phi}=0$: de Sitter and anti-de Sitter}

We begin with the uncharged case with nonzero cosmological constant. We first note that there are no subextensive solutions for planar horizons ($k = 0$). This can be seen from the dimensionless equations
\begin{equation} \label{dimless-k=0}
\tilde{\beta}
= ns\, \frac{4\pi}{d z y} \sqrt{s(z^d-1)},\ \ \ \ K\ell
= ns \, \frac{d}{\sqrt{s(z^d-1)}} \left(1 - \frac{z^d}{2}\right),
\end{equation}
where again $s$ denotes the sign of the cosmological constant, and $n=-1$ for the black hole and $n=1$ for the cosmic branch.
Therefore we must have $ns=1$, since $y,z ,\tilde{\beta}> 0$. This means there are black hole (cosmic) solutions when $\Lambda$ is negative (positive).

The second equation in \eqref{dimless-k=0} fully determines $z$ in terms of $K$. The $\tilde{\beta}$ equation then implies that $y \sim 1/\tilde{\beta}$, and so the only solutions for $r_c = y\ell$ and $r_h = yz\ell$ are terms that precisely scale as $1/\tilde{\beta}$, i.e. the extensive terms in \eqref{ansatzhcK}. Thus to find subextensive solutions, we must restrict to curved horizon geometries ($k \neq 0$).

We set $a_0=0$ in \eqref{ansatzhcK} so that the Bekenstein--Hawking entropy is subextensive. By expanding the equations \eqref{sphereflateqns} we find $w_0=a_1=0$ and
\be
a_2 =-n \frac{(d-2)k \ell w_1}{4\pi\sqrt{ k \ell^2 -s w_1^2 }}.
\ee
Since $r_h<r_c$ (since $r_h = O(\tilde{\beta}^1)$ while $r_c = O(\tilde{\beta}^0)$), this is a black hole solution and hence $n=-1$. This solution exists when $K\ell>d$ for $\L<0$ and for all values of $K$ when $\L>0$. Furthermore, positivity of $r_h$ and $r_c$ requires that $k=1$, so this solution must have a spherical horizon. The $K$ equation gives
\be
K\ell w_1 \sqrt{\ell^2-s w_1^2}= (d-1) \ell^2 -s d w_1^2,
\ee
which leads to one solution for real $w_1$:
\be
w_1^2=\ell^2 \frac{s K\ell \left[K\ell -\sqrt{(K\ell) ^2 +4 (d-1) s}\right]+2 d(d-1) }{2 (K\ell) ^2+ 2s d^2} .
\ee
Altogether, we see that there are no subextensive solutions in the hyperbolic ($k=-1$) and planar ($k=0$) cases for either sign of cosmological constant. This is consistent with Figures \ref{fig:solutionsAdSHyper} and \ref{figs:solutionsAdSPlanar} for $\Lambda < 0$ and Figures \ref{fig:solutionsdSHyper} and \ref{fig:phasesdSPlanar} for $\Lambda > 0$. For the spherical case ($k=1$), we have a single subextensive solution for either sign of $\Lambda$ (but only for $K\ell > d$ if $\Lambda < 0$), consistent with Figures \ref{figs:solutionsAdSSphere} and \ref{figs:solutionsdSSphere}.

\subsubsection{$\Lambda =0$ and $\tilde{\Phi}\neq0$: charged flat space}

We now find the subextensive solutions in the $\L = 0$, $\tilde{\Phi} \neq 0$ case. We will find it easier to use the dimensionless equation \eqref{phikfundgentopSec3}. Based on the ansatz \eqref{ansatzhcK}, we must have that
\begin{equation}
z = \sum_{i=0}^\infty \zeta_i \tilde{\beta}^i.
\end{equation}
The high-temperature expansion of the dimensionless equation implies the following constraint on the first coefficient:
\begin{equation}
\zeta_0^{d/2} \left(\zeta_0^{d-2} - \frac{d}{2}\right)\sqrt{\zeta_0^{d-2}-1} = 0 \implies \zeta_0 = 0,1,\left(\frac{d}{2}\right)^{1/(d-2)}.
\end{equation}
The solution $\zeta_0 = 1$ corresponds to the extensive solution discussed above; indeed, by expanding \eqref{phikfundgentopSec3} further, we find that
\begin{equation}
\zeta_0 = 1 \implies \zeta_1 = 0 \implies \zeta_2 = -\frac{k(d-2)}{16\pi^2} + \frac{(\tilde{\Phi}K)^2}{2\pi^2(d-1)},
\end{equation}
consistent with \eqref{coeffextensiveflat-K}. Additionally, it turns out that the constraint on $\zeta_2$ only has a solution for $\tilde{\Phi}K \geq 0$. The other two solutions must be the subextensive ones. Let us identify them. First, we consider
\begin{equation}
\zeta_0 = \left(\frac{d}{2}\right)^{1/(d-2)} \implies \zeta_1 = -\frac{\tilde{\Phi}K}{\pi}\sqrt{\frac{2}{d^3(d-1)}}.\label{cosmicsubextflat}
\end{equation}
As $\zeta_0 > 1$, this solution is always cosmic. Furthermore, it exists for all real $\tilde{\Phi}K$, including negative values. The other possible solution is
\begin{equation}
\zeta_0 = 0 \implies \frac{d-1}{\eta}\sqrt{k - \frac{4\pi\sqrt{k}\zeta_1}{d-2}} = \tilde{\Phi}K.\label{bhsmallflat}
\end{equation}
As $\tilde{\beta}$ is small, we must have that $z < 1$, meaning that this solution is a black hole. Furthermore, we only have a nontrivial real solution to the constraint on $\zeta_1$ for spherical horizon geometry ($k = 1$) and for $\tilde{\Phi}K > 0$. In this case, the solution is
\begin{equation}
\zeta_1 = \frac{d-2}{4\pi}\left[1 - \frac{\eta^2}{(d-1)^2}(\tilde{\Phi} K)^2\right],
\end{equation}
and the positivity of $z$ implies that this solution is only physical for $\tilde{\Phi}K < (d-1)/\eta$.

So, to summarize, we have at most three solutions including the extensive one. For planar and hyperbolic geometries, the only subextensive solution is the cosmic one \eqref{cosmicsubextflat}, which exists for all $\tilde{\Phi}K$. In the spherical case, we have not only this cosmic solution, but also the small black hole \eqref{bhsmallflat} for $0 \leq \tilde{\Phi}K < (d-1)/\eta$. These counts match those read off from the solution spaces in Section \ref{sec:flatCharged} (Figures \ref{figs:solutionsFlatSphereCharged}, \ref{figs:phasesFlatHyperCharged}, and \ref{figs:solutionsFlatPlanarCharged}).

\subsection{Wilson coefficients of the thermal effective action}
\label{sec:Wilson}

Let us now compute some of the Wilson coefficients appearing in \eqref{grand-Z}, specifically $c_0$, $c_1$, and $c_3$. One way to do so is to compute the bulk Euclidean action at the dominant saddle and in a high-temperature expansion \cite{Banihashemi:2024yye}. Alternatively, we may compute the entropy $S$ through the Bekenstein--Hawking area law, which for a bulk solution with horizon radius $r_h$ reads as
\begin{equation}
S = \frac{\text{Vol}[\Sigma]}{4G}r_h^{d-1}.
\end{equation}
The entropy $S = (1-\tilde{\beta}\partial_{\tilde{\beta}})\log Z$ computed from the thermal effective action \eqref{grand-Z} is\footnote{Recall from the discussion around \eqref{grand-Z} that $\tilde{\Phi}_0$ is a dimensionless version of the conformal electric potential $\tilde{\Phi}$ used throughout this work. The usual Maxwell coupling $1/e^2$ is related to the coefficient $c_J$ of the boundary two-point function of the dual conserved current via $e^2 \sim 1/c_J$. Thus, up to trading the coupling $1/G$ for $1/e^2$, we can relate our computed bulk quantities to boundary quantities once we have $c_J$. Absent this two-point function calculation, we simply rescale as $\tilde{\Phi}_0 = \tilde{\Phi} K/d$ (i.e. by a factor $\ell = d/K$), since $\tilde{\Phi}$ and $\tilde{\Phi}_0$ agree in the limit of asymptotic Dirichlet boundary in AdS.}
\begin{equation}
S = \int d^{d-1} x\,\sqrt{\tilde{h}_\Sigma} \left[\frac{d c_0}{\tilde{\beta}^{d-1}} - \frac{d-2}{\tilde{\beta}^{d-3}}\left(c_1 \tilde{R} +c_2 \tilde{F}^2+ c_3 \tilde{\Phi}_0^2\right) + \cdots\right].\label{grand-S0}
\end{equation}
The solution that reproduces the scalings of these terms is precisely the one for which $r_h$ diverges as $1/\tilde{\beta}$ and the odd coefficients in \eqref{ansatzhcK} vanish. The Bekenstein--Hawking law gives
\begin{equation}
S = \frac{\text{Vol}[\Sigma]}{4G} \left [ \frac{a_0^{d-1}}{\tilde{\beta}^{d-1}} + (d-1)\frac{a_0^{d-2}a_2}{\tilde{\beta}^{d-3}} + \cdots \right ].
\end{equation}
With both of these equations in hand, we can directly relate the Wilson coefficients to the coefficients of $r_h$. The parameter $a_0$ directly determines $c_0$, while $a_0$ and $a_2$ encode $c_1$, $c_2$, and $c_3$, but the precise relation depends on the details of the background fields $\tilde{R}$, $\tilde{F}$, and $\tilde{\Phi}_0$. For example, suppose that $\tilde{F}$ and $\tilde{\Phi}_0$ are constant on $\Sigma$. Furthermore, assume that $\Sigma$ is maximally symmetric with unit radius and curvature parameter $k$, implying $\tilde{R} = k(d-1)(d-2)$. Then, consistency with the thermal effective theory would imply that 
\begin{equation}
-\frac{(d-1) a_0^{d-2}a_2}{4G} = c_1 k (d-1)(d-2)^2 + c_2 (d-2)\tilde{F}^2 + c_3 (d-2)\tilde{\Phi}_0^2.
\end{equation}
In our case, $\tilde{F} = 0$, so we cannot extract $c_2$. Using \eqref{A-a0} and \eqref{extensivea2w2}, we find the following Wilson coefficients for $\L < 0$, valid for $K \ell >d$:
\begin{align}
c_0 &= \frac{\ell^{d-1}}{4d G} \left[\frac{4\pi}{d^2}\left(K\ell - \sqrt{(K\ell)^2 - d^2}\right)\right]^{d-1},\nonumber\\
\L < 0:\ \ \ \ 
c_1 &= \frac{\ell^{d-1}}{4d(d-2)G}\left[\frac{4\pi}{d^2}\left(K\ell - \sqrt{(K\ell)^2 - d^2}\right)\right]^{d-3}\left[1 - \left(\frac{K\ell + \sqrt{(K\ell)^2 - d^2}}{2\sqrt{(K\ell)^2 - d^2}}\right)^{\frac{d-2}{d}}\right]\nonumber,\\
c_3 &= \frac{d-2}{2dG K^2} \left(\frac{4\pi \ell}{d^2}\right)^{d-3} \left(K\ell - \sqrt{(K\ell)^2 - d^2}\right)^{d-1}\frac{1}{\left(\frac{2\sqrt{(K\ell)^2 - d^2}}{K\ell + \sqrt{(K\ell)^2 - d^2}}\right)^{\frac{d-2}{d}} - 1}.
\end{align}
Notice that setting $K\ell = d$, which pushes us to the AdS boundary, gives values for $c_0$ and $c_3$ which agree with the asymptotic AdS/CFT values calculated in \cite{allameh}. For $c_1$, we have a divergent term in the square brackets, which if we drop also agrees with the asymptotic AdS/CFT value. 

We can readily take the flat-space limit of these expressions by taking $\ell \rightarrow \infty$. Doing so yields the following Wilson coefficients:
\begin{equation}
\L = 0:\ \ \ \ c_0 = \frac{(2\pi)^{d-1}}{4dGK^{d-1}},\ \ \ \ c_1 = -\frac{(2\pi)^{d-3}}{16G K^{d-1}},\ \ \ \ c_3 = -\f{(2\pi)^{d-3}d^2}{2 G K^{d-1}} .
\end{equation}
The coefficient $c_0$ was calculated in \cite{Anninos:2023epi} while $c_1$ was calculated in \cite{Banihashemi:2024yye}.

The Wilson coefficients for the theory with positive cosmological constant are given by taking $\ell \rightarrow i\ell$ in the AdS expressions. The resulting expressions are 
\begin{align}
c_0 &= \frac{\ell^{d-1}}{4dG}\left[\frac{4\pi}{d^2}\left(-K\ell + \sqrt{(K\ell)^2 + d^2}\right)\right]^{d-1},\nonumber\\
\Lambda > 0:\ \ \ \ c_1 &= \frac{-\ell^{d-1}}{4d(d-2)G}\left[\frac{4\pi}{d^2}\left(-K\ell + \sqrt{(K\ell)^2 + d^2}\right)\right]^{d-3}\left[1 - \left(\frac{K\ell + \sqrt{(K\ell)^2 + d^2}}{2\sqrt{(K\ell)^2 + d^2}}\right)^{\frac{d-2}{d}}\right],\nonumber\\
c_3 &= -\frac{d-2}{2d G K^2}\left(\frac{4\pi \ell}{d^2}\right)^{d-3} \left(-K\ell + \sqrt{(K\ell)^2 + d^2}\right)^{d-1}\frac{1}{\left(\frac{2\sqrt{(K\ell)^2 + d^2}}{K\ell + \sqrt{(K\ell)^2 + d^2}}\right)^{\frac{d-2}{d}} - 1}.
\end{align}
It is worth highlighting the fact that AdS gravity is only consistent with the thermal effective action for $K\ell > d$, while dS gravity is consistent for all real $K\ell$. Flat-space gravity, on the other hand, is consistent with the thermal effective action only for $K > 0$. In all cases, we find $c_1 < 0$, which is especially stringent for dS where $K\ell \in \mathbb{R}$.

\section{Adding rotation}\label{sec:rotation}
As discussed in \cite{Banihashemi:2024yye}, adding rotation to the ensemble with spherical spatial slices to compute $c_2$ is difficult (for any $\L$), as the relevant solutions are the ones constructed numerically in \cite{Adam:2011dn}. These novel solutions are needed because we need to keep $K$ constant and have a boundary geometry conformal to $S^1 \times S^{d-1}$, so as to accurately compare to our results in the non-rotating case. This is not possible to do by slicing the Kerr black hole, because constant radial slices in Kerr give oblate spheroids. %Computing $c_2$ would require using the novel solutions of \cite{Adam:2011dn}. 

A simpler thing we can do is to introduce rotation when the spatial slices on the boundary are flat. In this case, say with a flat torus, we do not have the effect of the spatial circles becoming oblate due to the rotation (since circles have no intrinsic curvature). 

For compactifications with several length scales, it is clear that the conformal invariants are $d-1$ ratios between proper cycle sizes. For uniformity with \eqref{rotbdrymetric} and the case of a spatial sphere, we will write the boundary metric in the frame
\be
ds^2 = \l^2(d\tilde{\t}^2 + d\tilde{\Sigma}^2) = \l^2(d\tilde{\t}^2 + d\tilde{\phi}_i^2),\quad i = 1,\dots,d-1\label{rotbdrymetrictorus1}
\ee
with
\be
\text{Vol}[\tilde{\Sigma}] = \text{Vol}[S^{d-1}],\qquad (\tilde{\t}, \tilde{\phi}_i) \sim (\tilde{\t}+ \tilde{\b}, \tilde{\phi}_i + i \tilde{\b}\tilde{\Omega}_i).\label{rotbdrymetrictorus2}
\ee
The potentials $\tilde{\b}$ and $\tilde{\Omega}_i$ are defined in this frame by the above periodicities. There are of course many other choices one could have made, e.g. we could have worked in a frame where one of the cycles has $2\pi$ periodicity and then defined $\tilde{\b}$ as the ratio between the proper size of $\tilde{\t}$ and the proper radius of the $2\pi$-periodic cycle. Such a choice is more suitable to a mix of compact and noncompact directions. We could have also worked entirely in terms of the ratios of the cycle lengths, as was done in the toroidal ($k = 0$) analyses earlier in the paper, without fixing a fiducial value for $\text{Vol}[\tilde{\Sigma}]$. But none of these choices will affect the final answer, which will clearly be a conformal invariant. 

%we will continue to define $\tilde{\b}$ as above with $d\tilde{\Sigma}^2 = d\tilde{\phi}_1^2 + d\tilde{\phi}_i^2$ with $\tilde{\phi}_1$ being $2\pi$-periodic and arbitrary periodicities for the $\tilde{\phi}_i$ coordinates. Then we can again interpret $\tilde{\b}$ as a conformal invariant defined by the ratio between the proper size of $\tilde{\t}$ and the proper size of $\tilde{\phi}_1$. 

Let us begin with $\L < 0$ to make contact with the AdS/CFT literature. The $\L \geq 0$ cases will follow immediately. The relevant solutions are compactifications of boosted black branes, also known as rotating black strings. The simplest example of this is the rotating BTZ black hole, although one can construct these solutions in general dimension (see \cite{Hubeny:2011hd}). In Lorentzian signature, we have 
\begin{align} \label{BoostedBraneMetric}
ds^2 &= -f(r)u_\m u_\n dx^\m dx^\n + \frac{r^2}{\ell^2} (u_\m u_\n + \eta_{\m\n})dx^\m dx^\n + \f{dr^2}{f(r)},\\
u_\m &= \f{1}{\sqrt{1-a^2}} \left(-1, a_{1},\dots, a_{d-1}\right),\ \ \ \ f(r) = -\f{\m}{r^{d-2}} + \f{r^2}{\ell^2},
\end{align}
where we define $a^2 \equiv \sum a_i^2$ for simplicity. The spatial coordinates $x^i$ are compactified into circles of arbitrary lengths. The analytic continuation of this metric $t \rightarrow i \t$ gives a complex metric due to the time-space cross terms. As usual, this metric can be made real by analytically continuing the angular velocities as well. In order to have a regular horizon we need to identify the compact coordinates $(x^0,x^1, \dots)$ with $(x^0+i\beta, x^1+i\beta a_1, \dots)$.

We will begin with the case $a = a_1$, i.e. we have velocity only in the $x^1$ direction. To have a smooth Euclidean geometry we need the identification $(x^0,x^1)\sim(x^0+i\beta, x^1+i\beta a)$.
The induced boundary metric at $r=r_c$ from \eqref{BoostedBraneMetric} can be written in the ADM form 
\begin{align}
ds^2&=-N(r_c)^2 (dx^0)^2+g_{11}(r_c)\left(dx^1-\omega(r_c) dx^0 \right)^2 + \frac{r_c^2}{\ell^2}\, (dx^i)^2,\ \ \ \ i=2, \cdots d-1,\nonumber\\
N(r)&=\frac{r}{\ell}\sqrt{\frac{f(r)}{g_{11}(r)}},\ \ \ \ f(r)=\frac{r^2}{\ell^2}\left [ 1- \left ( \frac{r_h}{r} \right )^d\right ],\ \ \ \ g_{11}(r) =\frac{1}{1-a^2}\frac{r^2}{\ell^2}\left [ a^2 \left (\frac{r_h}{r} \right )^d +1-a^2\right ],\nonumber\\
\omega(r) &\equiv -\frac{g_{01}}{g_{11}}=\frac{u_0 u_1 [f(r)-r^2/\ell^2]}{-f(r) u_1^2+r^2(1+u_1^2)/\ell^2} = \frac{a r_h^d}{\left(1-a^2\right) r^d+a^2 r_h^d}.
\end{align}
We now introduce the coordinate $y \equiv x^1-\omega(r_c) x^0$ (not to be confused with the earlier dimensionless $y = r_c/\ell$) so that the Euclidean boundary metric takes the form 
\be
ds^2= N(r_c)^2 d\tau^2+ g_{11}(r_c) dy^2 + \frac{r_c^2}{\ell^2} (dx^i)^2. 
\ee
Regularity of the geometry and periodicity of the coordinates imply the identifications:
\be
(\tau , y , x^i )\sim(\tau, y+L_{y},x^i)\sim(\tau, y,x^i+L_{x^i}),\ \ \ \ (\tau , y , x^i )\sim\left(\tau+\beta, y+i\beta [a-\omega(r_c)],x^i\right),
\ee
and smoothness of the horizon relates $\beta$ to the other parameters through $\beta=4\pi \ell^2/(dr_h \sqrt{1-a^2})$.

We can put the metric in the form \eqref{rotbdrymetrictorus1}--\eqref{rotbdrymetrictorus2} by writing
\begin{align}
ds^2 &= \l^2 \left[\f{N(r_c)^2}{\l^2} \, d\t^2 + \f{ g_{11}(r_c)}{\l^2} dy^2 + \f{r_c^2}{\ell^2 \l^2 }\, (dx^i)^2\right]\nonumber\\
&= \l^2 \left[d\tilde{\t}^2 + d\tilde{\phi}^2 + (d\tilde{x}^i)^2\right],
\end{align}
in which we have identified $\tilde{\phi} = y\sqrt{g_{11}(r_c)}/\lambda$ and $\tilde{x}^i = x_i r_c/(\ell \lambda)$, and 
where $\l$ is defined by the equation 
\be 
\left(\frac{r_c}{\ell}\right)^{d-2} \f{\sqrt{g_{11}(r_c)} L_y L_{x^2}\cdots L_{x^{d-1}} }{\l^{d-1}} = \text{Vol}[S^{d-1}].\label{lambdadef}
\ee
This ensures that $\text{Vol}[\tilde{\Sigma}] \defeq L_{\tilde{\phi}} L_{\tilde{x}^2} \cdots L_{\tilde{x}^{d-1}} = \text{Vol}[S^{d-1}]$, satisfying \eqref{rotbdrymetrictorus2}. Therefore we have
\be
\tilde{\beta}= \frac{N(r_c)}{ \l}\beta.
\ee
The periodicity
\be
(\tilde{\t} , \tilde{\phi},\tilde{x}^i) \sim \left(\tilde{\t} + \tilde{\b}, \tilde{\phi} + i \b\frac{[a- \w(r_c)]\sqrt{g_{11}(r_c)}}{\l}, \tilde{x}^i\right)
\ee
lets us identify the angular potential through \eqref{rotbdrymetrictorus2}:
\be
\b \f{[a- \w(r_c)]\sqrt{g_{11}(r_c)}}{\l} = \tilde{\b}\tilde{\Omega}\implies \tilde{\Omega} = \f{[a-\omega(r_c)]\sqrt{g_{11}(r_c)}}{N(r_c)}.
\ee
%The fact that $\tilde{\Omega}$ is independent of $\l$ means that the expression is invariant under changing the definition of $\tilde{\b}$ by defining a different conformally invariant ratio. 
The boundary data in terms of the variables $\l$, $z=r_h/r_c$ and $a$ are thus given by 
\be \label{boosted-data-general}
\tilde{\beta} = \frac{4\pi\ell}{d}\frac{\sqrt{1-z^d}}{z\sqrt{1 - a^2 +a^2 z^d}}\f{1}{\l},\ \ \ \ 
K\ell = \frac{d(2-z^d)}{2\sqrt{1-z^d}},\ \ \ \ 
\tilde{\Omega} = a\sqrt{1-z^d},
\ee
in addition to the $d-2$ ratios between the spatial cycle sizes $L_{\tilde{x}^i}/L_{\tilde{x}^j}$, which will not play a role. Notice the appearance of a time dilation factor in $\tilde{\b}$, as $1 - a^2 +a^2 z^d= 1-\tilde{\Omega}^2$. From here, the quickest way to see how an angular velocity modifies the thermal effective action is to calculate the Bekenstein-Hawking entropy:
\begin{align}
S &= \f{\sqrt{g_{11}(r_h)} r_h^{d-2}L_y L_{x^2}\cdots L_{x^{d-1}}}{4G\ell^{d-2}}\nn\\
& = \f{(4\pi\ell^2/d)^{d-1} N(r_c)^{d-1}L_y L_{x^2}\cdots L_{x^{d-1}}}{4G \ell^{d-1}\tilde{\b}^{d-1}\l^{d-1}(1-a^2)^{d/2}}\nn\\
& = \f{(4\pi\ell/d)^{d-1} \text{Vol}[S^{d-1}]}{4G \tilde{\b}^{d-1}(1-\tilde{\Omega}^2)^{d/2}}(1-z^d)^{\f{d-1}{2}} \nn\\
&= \f{(4\pi/d)^{d-1} \ell^{d-1} \text{Vol}[S^{d-1}]}{4G} \left(\f{K\ell-\sqrt{K^2\ell^2-d^2}}{d}\right)^{d-1} \f{1}{\tilde{\b}^{d-1}(1-\tilde{\Omega}^2)^{d/2}}.
\end{align}
Since $S = (1-\tilde{\b} \partial_{\tilde{\b}})\log Z(\tilde{\b}, \tilde{\Omega}, K) = d \log Z(\tilde{\b}, \tilde{\Omega}, K)$ for a flat boundary spatial manifold, this leads to the following relation for the free energies
\be\label{zrot}
\log Z(\tilde{\b}, K, \tilde{\Omega}) = \f{\log Z(\tilde{\b}, K, \tilde{\Omega} = 0)}{(1-\tilde{\Omega}^2)^{d/2}}. 
\ee
This is exactly as predicted by the thermal effective action \cite{allameh}. To see this agreement it is important to recall that the Lagrangian density is unchanged upon introducing rotation; all terms built out of $F = dA$ vanish since the Kaluza--Klein gauge field $A$ is constant for flat spatial slices.

Equation \eqref{zrot} can also be understood in terms of length contraction and time dilation in the boundary theory \cite{Shaghoulian:2015lcn, allameh}; the boost parameter in the boundary theory is $\tilde{\Omega}$, and we get a factor $\sqrt{1-\tilde{\Omega}^2}$ multiplying $\b$ and dividing the volume of the spatial manifold. This leads to $d$ factors of $1/\sqrt{1-\tilde{\Omega}^2}$, giving the result above. The fact that the boost parameter $a$ transmuted into the boundary boost parameter $\tilde{\Omega}$ is nontrivial from the perspective of the bulk at finite cutoff.

This result can also be obtained by directly evaluating the on-shell action. To see it in a quick way, notice that since the rotating metric is locally diffeomorphic to the nonrotating metric, their on-shell actions must be the same up to differences in periodicities of the coordinates. Since $\b = 4\pi\ell^2/(d r_h \sqrt{1-a^2})$, there is a factor of $\sqrt{1-a^2}$ difference in periodicities. Denoting the rotating and nonrotating parameters with r and nr subscripts, respectively, 
\be
\log Z(\tilde{\b}_{\text{r}},K_{\text{r}}, \tilde{\Omega}_{\text{r}}) = \f{\log Z(\tilde{\b}_{\text{nr}}, K_{\text{nr}}, 0)}{\sqrt{1-a^2}} = \f{\log Z(\tilde{\b}_{\rm r}, K_{\rm r}, 0)}{(1-\tilde{\Omega}^2)^{d/2}},
\ee
where in the final expression we have used the fact that $\log Z(\tilde{\b}_{\text{nr}}, K_{\text{nr}}, 0) \sim 1/\tilde{\b}_{\text{nr}}^{d-1}$ and \eqref{boosted-data-general} to get $K_{\text{r}} = K_{\text{nr}}$ and $\tilde{\b}^{d-1}_{\text{r}} = \tilde{\b}_{\text{nr}}^{d-1}\sqrt{1-a^2}/(1-\tilde{\Omega}^2)^{d/2}$.

The case $d=2$ is also captured by the above analysis. In that case, the black brane is the BTZ black string, i.e.~the BTZ black hole with the spatial coordinate $\theta$ decompactified. This geometry is globally diffeomorphic to empty AdS$_3$. After boosting and compactifying, the resulting geometry is the rotating BTZ black hole, which has both an inner and an outer horizon. This is unlike the $d > 2$ cases, where no inner horizon appears. This is because the singularity for the black brane with $d > 2$ is a spacelike curvature singularity; no amount of boosting or compactifying can change its spacelike nature, and the geometry therefore cannot accommodate an inner horizon. The singularity for BTZ is instead a conical singularity.

Generalizing to arbitrary angular velocities is straightforward. The relation between the boost parameters $a_i$ and the boundary angular velocities $\tilde{\Omega}_i$ is $\tilde{\Omega}_i = a_i \sqrt{1-z^d}$. This means we can rotate to a frame where the velocity is in a single direction, and our equations become \eqref{boosted-data-general}, now with $a^2 = \sum_i a_i^2$, and $\tilde{\Omega}^2 = \sum_i \tilde{\Omega}_i^2$. Thus we recover
\be
\log Z(\tilde{\b}, K, \tilde{\Omega}_i) = \f{\log Z(\tilde{\b}, K, \tilde{\Omega}_i = 0)}{\left(1-\sum_i\tilde{\Omega}_i^2\right)^{d/2}}. 
\ee
This formula also works for $\L = 0$ and $\L > 0$, as the derivation in those cases proceeds as above.

\section{Conclusions}\label{conclusions}
We have seen that in every case we considered, the results of semiclassical gravity calculations reproduce the predictions of the thermal effective action in a putative boundary dual theory. For a fuller summary of the various upshots of this paper, see the bulleted list in Section \ref{sec:intro}.

An interesting question in the context of $\L < 0$ is the role of counterterms, where we may want to take a limit where the boundary goes off to infinity and compare to AdS/CFT. As discussed in \cite{Banihashemi:2024yye}, the usual sorts of terms (e.g. scalar contractions of the Riemann tensor integrated with measure $\sqrt{g}$) are not allowed, as they violate conformal boundary conditions. We can take an asymptotic limit by tuning $K\ell \rightarrow d$, which is the value of the extrinsic curvature obtained on the boundary of AdS. In this limit, almost all of the usual divergences appear, but, interestingly, the modified coefficient of the extrinsic curvature boundary term is exactly such that the leading divergence is canceled. This is the divergence that requires the cosmological constant counterterm in ordinary AdS/CFT. In particular, AdS$_3$ gravity with conformal boundary conditions is completely finite without the addition of any counterterms. The fact that AdS$_3$ gravity with one-half of the usual GHY counterterm leads to completely finite expressions was first noted in \cite{Banados:1998ys}.

For $d > 2$, however, divergences remain and lead to different results in the $K \ell \rightarrow d$ limit than in ordinary AdS/CFT. For example, in the case of hyperbolic spatial slices $k=-1$, the free energy of the black hole solution diverges to $+\infty$, meaning that the cosmic solution with $K \ell = d$ dominates the phase diagram. 

The lack of counterterms was observed to lead to energies unbounded below in the case $\L = 0$, $\tilde{\Phi} = 0$ \cite{Banihashemi:2024yye}. It was also observed that the black-hole states with negative energy were precisely the black-hole states with negative specific heat. This correlation between negative energy and negative specific heat survives the inclusion of an electric potential but not the inclusion of a cosmological constant. For recent work on slightly modifying these boundary conditions to lead to finite on-shell quantities see \cite{Parvizi:2025shq, Parvizi:2025wsg}.

Our analysis of vanishing cosmological constant with finite electric potential led to discontinuous phase diagrams, which we will comment on in Appendix \ref{app:puzzle} with some proposed solutions to smoothen the phase diagrams. However, we would like to point out that in the $k=0$ and $k=1$ cases, the discontinuities occur at some $O(1)$ electric potential. At lower electric potentials, the Hawking-Page transition curve bends toward lower temperatures as the electric potential is increased. This is the expected physical behavior and is precisely what is seen in AdS/CFT \cite{Chamblin:1999tk}, where this curve asymptotes at some finite electric potential above which one is always in the deconfined phase. It would be interesting to see if solutions of the type proposed in Appendix \ref{app:puzzle} lead to a phase diagram with an analogous feature.

The following curiosity is worth mentioning. In the $\L < 0$ case, the solutions that dominate at high temperature and reproduce the predictions of the thermal effective action are always black-hole solutions. In the $\L > 0$ case, they are always cosmic solutions. In between, for $\L = 0$, they are sometimes cosmic solutions and sometimes black-hole solutions. For example, at finite electric potential with $k=1$, the identity of the high-temperature solution depends on whether the electric potential is large (cosmic) or small (black hole), and at vanishing electric potential it depends on whether $k = -1$ (cosmic) or $k=1$ (black hole). Thus we see the $\L = 0$ theory retain features from both the $\L > 0$ and $\L < 0$ theories.

\subsection*{The thermal effective action and $K < 0$}
Part of the motivation for this work was the connection observed in \cite{Banihashemi:2024yye} between $K < 0$ and cosmic horizons. This occurred for the hyperbolic horizons studied in \cite{Banihashemi:2024yye} and de Sitter space \cite{Anninos:2024wpy}. It is easy to understand this geometrically; when we have a solution with a cosmic horizon, the normal vector to the boundary flips sign as compared to the normal vector for black hole horizons. This, coupled with the fact that there is usually a limit $r_c \rightarrow 0$ where $K \rightarrow -\infty$, means that all $K < 0$ is accessible. 

Interestingly, none of the cases studied here with $K < 0$ and $\L \leq 0$ admitted an extensive high-temperature limit. In fact, except for a small sliver in the $\L = 0$, $\tilde{\Phi} \neq 0$ case, we found that solutions for $K < 0$ were thermodynamically unstable. This is in stark contrast with the $\L > 0$ case \cite{Anninos:2024wpy}, which has an extensive high-temperature limit for all $K \ell \in \mathbb{R}$. 

In some cases, there is a simple way to see why we cannot have an extensive high-temperature entropy with $K < 0$. For example, any case where the bulk action is $R - 2\L$ with $\L < 0$ will have an on-shell action that evaluates to a positive number. Since $K < 0$, the boundary term is also positive. To get an extensive entropy, we need $I \sim 1/\tilde{\b}^{d-1}$, but the signs are such that the entropy $(\tilde{\b}\p_{\tilde{\b}} - 1)I < 0$ if the total on-shell action is positive. More generally, one would need a bulk on-shell action to overwhelm the positive boundary contribution and yield a negative on-shell action. This is what happens in the case of de Sitter space. 

\section*{Acknowledgments}
The authors would like to thank Dionysios Anninos and Eva Silverstein for fruitful conversations. BB is grateful to Abdolali Banihashemi for guidance on numerical computations. The authors are supported in part by DOE grant DE-SC001010 and the Federico and Elvia Faggin Foundation.

\appendix

\section{Review of charged black holes}\label{app:bigQ}

To understand the relation between the parameters in the metric, the physical charge, and the boundary electric potential, we start with the line element
\be \label{general-ds2}
ds^2=-f(r)dt^2+\frac{dr^2}{f(r)}+r^2d\Sigma^2_k,\ \ \ \ f(r) \equiv k - \frac{\mu}{r^{d-2}} + \frac{q^2}{r^{2(d-2)}}-\frac{2\Lambda r^2}{d(d-1)},
\ee
where as in the main text $k=0,\pm1$ determines the horizon geometry. The vacuum Maxwell equation is $\nabla_\mu F^{\mu\nu}=\partial_{\mu}(\sqrt{-g} F^{\mu\nu})/\sqrt{-g}=0$, and for configurations that respect the horizon symmetry the only nonzero component of the field strength tensor (with zero magnetic charge) is $F^{tr}(r)$. Thus, we have
\be
\frac{1}{r^{d-1}}\partial_r (r^{d-1}F^{tr})=0 \implies F^{tr}= \frac{\kappa}{r^{d-1}},
\ee
with $\kappa$ an integration constant. To find $\kappa$ in terms of $q$, we need the Einstein equation and the energy-momentum tensor of the Maxwell field. To set the conventions straight and not get ourselves confused with normalization of the gauge field we consider the action
\be
I=\frac{1}{16 \pi G}\int_{\mc{M}}d^{d+1}x \sqrt{-g} \left (R-2\Lambda - g^{\alpha \mu} g^{\beta \nu }F_{\mu\nu}F_{\alpha \beta } \right).
\ee
Variation with respect to $g^{\mu \nu}$ (up to boundary terms) gives
\be
\delta I = \frac{1}{16 \pi G} \int_{\mc{M}}d^{d+1}x \sqrt{-g} \left ( G_{\mu \nu}+ \Lambda g_{\mu\nu } +\frac{1}{2} g_{\mu\nu} F^2 - 2 g^{\alpha \beta }F_{\mu\alpha} F_{\nu \beta} \right ) \delta g^{\mu \nu}.
\ee
Thus the Einstein--Maxwell equation becomes $G_{\mu \nu} + \Lambda g_{\mu\nu } = -g_{\mu\nu} F^2/2 + 2 g^{\alpha \beta }F_{\mu\alpha} F_{\nu \beta}$, and its contraction with $g^{\mu\nu}$ yields
\be \label{RFF}
R= \frac{d-3}{d-1}g_{\mu \alpha}g_{\nu \beta}F^{\mu \nu} F^{\alpha \beta}+ \frac{2(d+1)}{d-1} \Lambda.
\ee
The Ricci scalar $R$ for the metric \eqref{general-ds2} is 
\be \label{R-general}
R=-(d-2)(d-3)\frac{q^2}{r^{2d-2}} +\frac{2(d+1)}{d-1} \Lambda.
\ee
Note that the $q$-dependent term vanishes in 4-dimensional spacetimes, in accordance with \eqref{RFF}. On the other hand we have
\be
F^2= g_{\mu \alpha}g_{\nu \beta}F^{\mu \nu} F^{\alpha \beta} = 2g_{tt} g_{rr} (F^{tr})^2 = -\frac{2\kappa^2}{r^{2d-2}}.
\ee
Comparing \eqref{R-general} with \eqref{RFF} gives
\be
F^{tr}=\sqrt{\frac{(d-1)(d-2)}{2}}\frac{q}{r^{d-1}} = -F_{tr}=\partial_r A_t,
\ee
which by integration yields
\be
A_t=-\sqrt{\frac{d-1}{2(d-2)}} \left (\frac{1}{r^{d-2}}-\frac{1}{r_h^{d-2}} \right)q,
\ee
where we have set the constant such that the gauge potential vanishes on the horizon at $r_h$. This result is in agreement with equations (12) and (13) of \cite{Chamblin:1999tk}.

The total charge is defined by 
%\batoul{Poisson's toolkit (5.23) and (3.21)}
\be
Q=\frac{1}{8\pi G}\int_{\Sigma_k^{d-1}}F^{\mu\nu}dS_{\mu\nu} = \frac{1}{8\pi G}\int_{\Sigma_k^{d-1}} d^{d-1}x \sqrt{\sigma_{d-1}}F^{\mu\nu} (u_\nu n_\mu - u_\mu n_\nu),
\ee
where $u=f^{-1/2}\partial_t$ and $n=\pm f^{1/2}\partial_r$ are timelike and spacelike unit normals to the surface $\Sigma_k^{d-1}$ on the boundary, located at $r=r_c$. The plus (minus) corresponds to the case where $r_c$ is greater (smaller) than $r_h$. Therefore the total charge becomes
\be
Q=\frac{{\rm Vol}[\Sigma_k^{d-1}] r_c^{d-1}}{8\pi G}\times 2 [-F^{tr}(r_c)u_t n_r]=\pm \frac{{\rm Vol}[\Sigma_k^{d-1}]}{8\pi G} \sqrt{2(d-1)(d-2)}q.
\ee
The $+$ branch (when the boundary radius is larger than horizon radius) matches equation (11) of \cite{Chamblin:1999tk}. The gauge potential at the boundary in terms of the total charge becomes
\be
\Phi \equiv A_t(r_c)= \pm \frac{4 \pi G }{(d-2) {\rm Vol}[\Sigma_k^{d-1}]} \left (\frac{1}{r_\pm^{d-2}}-\frac{1}{r_c^{d-2}} \right)Q,
\ee
where we denote the outer and inner horizon radius by $r_+$ and $r_-$, respectively.

Let us elaborate on the units. Since the Maxwell action appears with coupling $1/G$, we have that the gauge field $A_\m$ is dimensionless. From the above equation, this means that $[Q] = 1/L$, i.e. $Q$ is an inverse length. Normally $Q$ is dimensionless, since it comes from integrating the dimension-$(d-1)$ current $J^\m$ over a $(d-1)$-dimensional surface, but in this case the coupling $J^\m A_\m$ with dimensionless $A_\m$ makes $J^\m$ a dimension-$d$ operator.

\section{Dirichlet boundary conditions for the gauge field}\label{app:dirichlet}
Before studying conformal boundary conditions, let us clarify the situation with Dirichlet boundary conditions for the metric and gauge field. In the variational problems for Einstein theory or Maxwell theory, the object being varied is either the metric $g_{\m\n}$ or the gauge field $A_\mu$. Dirichlet boundary conditions means fixing this local field. It is important in the variational problem that the boundary geometry has coordinates whose range does not change when varying the relevant field, so that when we vary the action we do not pick up a term that varies the range of integration of the boundary term. This is automatic in quantum field theory, as the geometry is a fixed background. But when we consider a theory of gravity, the geometry becomes dynamical; the boundary coordinates having the same range across the space of solutions requires parameterizing things slightly differently than is conventional. For the space of black hole solutions, we write
\be
ds^2 = f(r)\beta_\t^2 dT^2 + \f{dr^2}{f(r)} + r^2 d\Sigma_k^2, \qquad T \sim T + 1.
\ee
We have scaled out $\beta_\t$ from the usual Schwarzschild time $\t \sim \t + \b_\t$. This is so that $T \sim T + 1$ on the full space of solutions, as we wanted. We can now apply the result of the (Dirichlet) variational problem, for which we fix the induced metric on the boundary. In particular, we fix $\sqrt{f(r_c)}\beta_\t$. In the asymptotic problem where $r_c \rightarrow \infty$, this corresponds to fixing $\b_\t$. This is the correct result! But instead of this procedure, what is often done instead is to write the space of solutions the usual Schwarzschild time $\t \sim \t + \b_\t$, and the ``nonlocal" quantity $\int \sqrt{g_{\t\t}} d\t$ is kept fixed. While mathematically equivalent, this makes less immediate contact with the usual variational problem, where local quantities are held fixed. 

For Einstein--Maxwell theory where we vary both the metric and the gauge field, the usual convention is to work in Schwarzschild coordinates and fix the nonlocal quantity of the metric $\b = \int \sqrt{g_{\t\t}} d\t$ while fixing the local gauge field $A_\t$. We can justify this by returning to our coordinate $T = \t/\b_\t$, with which we can straightforwardly use the variational principle. Once we write the answer according to this time coordinate, we have $A_T = A_\t \b_\t$. Thus, the gauge field that should be fixed is $A_T$. Since $\b_\t$ is fixed as part of the metric's Dirichlet conditions (at asymptotic infinity), this is the same variational problem as fixing $A_\t$. This justifies the usual procedure, but in some sense we got lucky. (In the end, one likes to translate the gauge field to the electric potential, and it is $A_\t$ which is proportional to the electric potential, but this is unrelated to ensuring we have a consistent variational problem.) 

Now, let us come to conformal boundary conditions, where it will become clear why we took the time for the pedantry above. We want to fix conformal boundary conditions for the metric and Dirichlet boundary conditions for the gauge field. This does not, however, correspond to fixing $\tilde{\b}$ and $A_\t$, which would be the natural analogue of what is usually done in the completely Dirichlet problem. Instead, we should use our universal time coordinate $T$ and fix the gauge field $A_T$. As discussed before, translating back to $\t$ means we fix $A_\t \b_\t$. Importantly, however, $\b_\t$ is \emph{not} held fixed in our conformal boundary condition problem, which is why $A_\t$ is not the appropriate quantity to keep fixed. 

The last thing we need to sort out is the identification of the electric potential in terms of the gauge field. The electric potential depends on the choice of time, and we are fixing to the reference metric with time $\tilde{\t}\sim \tilde{\t} + \tilde{\b}$ given in \eqref{rotbdrymetric}. Thus, we identify our electric potential as (proportional to) $A_{\tilde{\t}}$. Note that since $\tilde{\t} = \t \sqrt{f(r_c)}/r_c$, the electric potential is
\be
\tilde{\Phi} =-i A_{\tilde{\t}} = -\frac{i A_T}{\tilde{\b}}.
\ee
Since $\tilde{\b}$ and $A_T$ are held fixed in our boundary value problem, so is the electric potential. As usual, one can check that defining the electric potential in this way ensures that $\tilde{\b}\tilde{\Phi}$ is conjugate to the total conserved charge $Q$, which is its fundamental definition. 

Now that we have clarified what the boundary value problem is, we can work with a bulk time $\tilde{\t} = \t \sqrt{f(r_c)}/r_c$. The induced metric and gauge field on the boundary are given as 
\be
ds^2|_{\partial \mc{M}} = r_c^2 \left(d\tilde{\t}^2 + d\Sigma_k^2\right),\qquad A = \left(\f{q}{r_+^{d-2}} - \f{q}{r_c^{d-2}}\right)\f{i r_c}{\eta\sqrt{f(r_c)}} d\tilde{\t}.
\ee
In these coordinates, we can state the problem in a way that is most similar to what is usually done: we fix the periodicity $\tilde{\b}$ and the gauge field $A_{\tilde{\t}}$, with the latter interpreted as $i$ times the electric potential.

\section{Computing the Euclidean on-shell action}\label{app:onshellAct}

In the various cases analyzed in the main text, part of our analysis involves examining the thermodynamics of the solutions. This requires calculating the on-shell Euclidean action, which we will do here for general $\Lambda,\tilde{\Phi} \neq 0$.

The starting point is the Einstein--Maxwell action with nonzero cosmological constant \eqref{I_CBC}, the metric ansatz, and the gauge field. For convenience, we rewrite them here:
\begin{align}
I &= -\frac{1}{16\pi G}\int_{\mathcal{M}}d^{d+1}x \sqrt{g}\,\left(R - 2\Lambda - F^2\right) - \frac{1}{8d\pi G}\int_{\partial\mathcal{M}} d^d x \sqrt{h}\,K,\\
ds^2 &= f(r)d\tau^2 + \frac{dr^2}{f(r)} + r^2 d\Sigma_k^2,\\
f(r) &= k + \frac{r^2}{\ell^2} + \frac{q^2}{r^{2(d-2)}} - \frac{r_h^{d-2}}{r^{d-2}}\left[k + \frac{r_h^2}{\ell^2} + \frac{q^2}{r_h^{2(d-2)}}\right],\\
A &= \left(\f{q}{r_\pm^{d-2}} - \f{q}{r^{d-2}}\right)\f{i}{\eta}\,d\tau,\qquad \eta = \sqrt{\f{2(d-2)}{d-1}}.
\end{align}
Note that $\tau \sim \tau + \beta$, where $\beta = \tilde{\beta} r_c/\sqrt{f(r_c)}$. Furthermore, for the above bulk fields, the on-shell terms in the bulk Lagrangian in the case $\L < 0$ are:
\begin{equation}
\left.\begin{split}
R &= -\frac{d(d+1)}{\ell^2} - \frac{(d-2)(d-3) q^2}{r^{2(d-1)}},\\
\Lambda &= -\frac{d(d-1)}{2\ell^2},\\
F^2 &= -\frac{(d-1)(d-2)q^2}{r^{2(d-1)}},
\end{split}\right\} \implies R - 2\Lambda - F^2 = -\frac{2d}{\ell^2} + \frac{2(d-2)q^2}{r^{2(d-1)}}.
\end{equation}
Now, let us evaluate the integrals. We start with the bulk term. Recall that the black hole solutions have $r_h < r_c$, whereas for the cosmic solutions $r_h > r_c$. We always orient the contour to go from smaller to larger $r$. Thus, after integrating over the $\tau$-cycle and $\Sigma_k$, we can summarize both cases in a single line by writing
\begin{align}
I_{\text{bulk}}
&= -\frac{1}{16\pi G}\int_{\mathcal{M}}d^{d+1}x \sqrt{g}\,\left(R - 2\Lambda - F^2\right)\nonumber\\
&= \pm\frac{\beta \text{Vol}[\Sigma_k]}{16\pi G}\int_{r_h}^{r_c} dr\,r^{d-1}\left(\frac{2d}{\ell^2} - \frac{2(d-2) q^2}{r^{2(d-1)}}\right)\nonumber\\
&= \pm \frac{\tilde{\beta} \text{Vol}[\Sigma_k]}{8\pi G}\left[\frac{1}{\ell^2}\left(r_c^d - r_h^d\right) + \left(\frac{q^2}{r_c^{d-2}} - \frac{q^2}{r_h^{d-2}}\right)\right]\frac{r_c}{\sqrt{f(r_c)}}.\label{bulkact}
\end{align}
As for the boundary term, first recall that the induced metric on the boundary is
\begin{equation}
ds^2 = r_c^2\left(d\tilde{\tau}^2 + d\Sigma_k^2\right),\ \ \ \ \tilde{\tau} \sim \tilde{\tau} + \tilde{\beta}.
\end{equation}
Thus,
\begin{equation}
I_{\text{bdry}} = -\frac{1}{8d \pi G}\int_{\partial\mathcal{M}} d^d x\sqrt{h}\,K = -\frac{\tilde{\beta}\,\text{Vol}[\Sigma_k]}{8d\pi G}r_c^d K.\label{bdryact}
\end{equation}
Adding these together yields
\begin{align}
I_{\text{cl}}
&= - \frac{\tilde{\beta}\text{Vol}[\Sigma_k]}{8\pi G}\left(-\frac{r_c |r_c^d-r_h^d|}{\ell^2\sqrt{f(r_c)}}\pm \eta q \tilde{\Phi} + \frac{1}{d}r_c^d K\right)\nonumber\\
&= \frac{\tilde{\beta}\text{Vol}[\Sigma_k]}{8\pi G}\left(\frac{r_c |r_c^d-r_h^d|}{\ell^2\sqrt{f(r_c)}}-\eta |q \tilde{\Phi}| -\frac{1}{d}r_c^d K\right),
\end{align}
We can rewrite these expressions in terms of the dimensionless variables $\{x,y,z\}$ \eqref{dimensionlessVars} up to an overall factor of $1/(G|K|^{d-1})$:
\begin{equation}
\begin{split}
I_{\text{bulk}} &= \frac{\tilde{\beta}\,\text{Vol}[\Sigma_k]}{8\pi G|K|^{d-1}} \,\,|h(x,y,z)|^{d-1}\left[\frac{y^2 |1-z^d|}{\sqrt{f(x,y,z)}} - (\tilde{\Phi}K)\frac{\eta z^{d-2}|x|}{h(x,y,z)}\right],\\
I_{\text{bdry}} &= -\frac{\tilde{\beta}\,\text{Vol}[\Sigma_k]}{8\pi G|K|^{d-1}} \frac{|h(x,y,z)|^{d-1} h(x,y,z)}{d},
\end{split}\label{dimensionlessActGen}
\end{equation}
where $h$ and $f$ (and of course $\tilde{\b}$ and $\tilde{\Phi}K$) are dimensionless:
\begin{align}
h(x,y,z) &= \pm\frac{1}{\sqrt{f(x,y,z)}}\left[k(d-1) + z^{2(d-2)}x^2 + \frac{d}{2}\Big(2y^2 - z^{d-2}\left(k + x^2 + z^2 y^2\right)\Big)\right],\nonumber\\
f(x,y,z) &= \left(1-z^{d-2}\right)\left(k-z^{d-2}x^2\right) + y^2 \left(1 - z^d\right).
\end{align}
This will be useful for our numerical analyses of the phase structure, where we scan over the boundary data and compare the on-shell actions of the various solutions given by $\{x,y,z\}.$

\section{Thermodynamic first law}\label{app:firstlaw}

In this section, we will focus on $\L = 0$, $\tilde{\Phi} \neq 0$ with $k=1$, although the results straightforwardly generalize. All of the formulas are the natural generalizations of the first laws derived in \cite{Banihashemi:2024yye}. The charge and energy from Appendix \ref{app:bigQ} here and Appendix A in \cite{Banihashemi:2024yye} are given by %\edgar{expression for energy is a bit funny since $K$ has a $\pm$ hiding in it}
\begin{align}
Q &= \pm\f{\text{Vol}[S^{d-1}]}{8\pi G} \sqrt{2(d-1)(d-2)}q,\label{QEn}\\
\tilde{E} &= -\f{(d-1)\text{Vol}[S^{d-1}]r_c^d}{8\pi G}\left(\pm \f{\sqrt{f(r_c)}}{r_c} - \f{K}{d}\right).
\end{align}
The $+$ sign is the appropriate data for black hole solutions with $r_c > r_+$, while the $-$ sign is the appropriate data for cosmic solutions with $r_c < r_-$. As discussed in Appendix A of \cite{Banihashemi:2024yye}, $\tilde{E}$ is the thermal energy, which is related to the conserved energy as $\tilde{E} = r_c E^{\text{CBC}}$. The charge flips sign for $r_c < r_-$ since the field lines point away from the boundary. The entropy is given by the usual Bekenstein--Hawking formula,
\be\label{BHform}
S = (\tilde{\b}\partial_{\tilde{\b}}-1)I= \f{\text{Vol}[S^{d-1}]r_\pm^{d-1}}{4G}.
\ee
Using \eqref{QEn}--\eqref{BHform} with the function $f(r)$ for a charged black hole \eqref{charged-metric}, the total on-shell action evaluates to 
\be\label{grandcanans}
I = -\log Z = \tilde{\b}(\tilde{E} - \tilde{\Phi} Q) - S.
\ee
The thermodynamic first law is written as:
\be
d\tilde{E} = \tilde{T} dS + \tilde{V} dK + \tilde{\Phi} dQ.
\ee
For a first law where $K$ is held fixed we switch to the internal energy:
\be\label{firstlawdbc1}
d\tilde{U} = \tilde{T} dS - K d\tilde{V} + \tilde{\Phi} dQ.
\ee
We therefore have the thermodynamic relations
\be
\left(\f{\p \tilde{U}}{\p S}\right)_{\tilde{V}, Q} = \tilde{\b}^{-1},\quad \left(\f{\p \tilde{U}}{\p \tilde{V}}\right)_{S, Q} = -K,\quad \left(\f{\p \tilde{U}}{\p Q}\right)_{S, \tilde{V}} = \tilde{\Phi}.
\ee
This can be verified using \eqref{sphereflateqns}--\eqref{qeq} and \eqref{BHform} with the function $f(r)$ for a charged black hole \eqref{charged-metric}. The volume is proportional to $r_c$ times the spatial volume of the boundary:
\be
\tilde{V} = \f{(d-1)\text{Vol}[S^{d-1}]}{8\pi G d}\, r_c^d.
\ee

\section{Near-extremal near-horizon limit and JT gravity}\label{app:extremal}
In cases where we have both cosmic horizons and black hole horizons, we have transitions between the two in the space of solutions as we vary the fixed boundary parameters. What always happens in these situation is that $r_c$ and the two horizons all approach each other in a triple coincidence limit. For example, in the $\tilde{\Phi} \neq 0$, $\L = 0$ case of Section \ref{sec:flatCharged}, we have a transition from $r_c > r_+$ to $r_c < r_-$ for a black hole to cosmic transition. That $r_+$ and $r_-$ are coming together suggests that we are approaching an extremal black hole. Furthermore, that $r_c$ is approaching the horizons suggests an AdS$_2$ description. As we have discussed throughout the paper, this limiting curve in our plots is actually dominated by the appropriate extremal solution: AdS$_2 \times S^{d-1}$ for $\tilde{\Phi} \neq 0$, $\L = 0$, $k=1$; AdS$_2 \times \mathbb{H}^{d-1}$ for $\tilde{\Phi} = 0$, $\L < 0$, $k=-1$; and dS$_2 \times S^{d-1}$ for $\tilde{\Phi} = 0$, $\L > 0$, $k=1$. %However, due to the precise way the parameters scale this is not what occurs, as can be guessed by the fact that $\tilde{\b}$ remains finite in these situations, whereas we would expect it to diverge for a zero-temperature solution. 

In this appendix, we analyze the approach to this crossover from the general black hole solutions. We will focus on the case $\tilde{\Phi} \neq 0$, $\L = 0$ with $k=1$, covered in Section \ref{sphericalcharged}.

Crossing the extremal curve in Figure \ref{figs:solutionsFlatSphereCharged} (i.e.~the golden curve) by increasing $\tilde{\Phi} K$ corresponds to a transition where a black hole solution becomes a cosmic solution. As argued earlier, this occurs because $r_c \rightarrow r_+$ simultaneously with the inner horizon approaching the outer horizon ($r_- \rightarrow r_+$). The critical line corresponds to what seems like a somewhat degenerate solution, $r_+ = r_- = r_c$. Let us regulate this to see what occurs as we cross this line. We take equations \eqref{sphereflateqns}--\eqref{qeq} for the black hole case, solve the $\tilde{\Phi}$ one for $q$, and plug into the $\tilde{\b}$ and $K$ equations. Expanding these as $r_c \rightarrow r_+$, we have 
\begin{align}
\tilde{\b} &\approx \frac{8 \pi \eta^{-1} r_+ \tilde{\Phi}}{(d-1)r_+^2-2(d-2)\tilde{\Phi}^2} + O(r_c-r_+), \\
K &\approx \f{(d-1)r_+^2 + 2(d-2)\tilde{\Phi}^2}{4 \eta^{-1}r_+^2\tilde{\Phi}}+O(r_c-r_+), \\
r_- &\approx r_+ - \left(\f{r_+^2}{\eta^2\tilde{\Phi}^2}-1\right)(r_c-r_+),\ \ \ \ r_c = r_+ + (r_c-r_+).
\end{align}
We will only deal with the black hole equations, and $r_-$ in this case means the location of the inner horizon for the black hole solution. We have $r_+ \geq \tilde{\Phi}$ by non-negativity of $\tilde{\b}$, which ensures $r_- \leq r_+$. In fact, defining $\tilde{b}^{-1} \equiv r_+^2/(\eta^2\tilde{\Phi}^2)-1$ lets us write
\be
\tilde{\b} \approx \f{4\pi}{d-2}\, \tilde{b} \sqrt{1+\tilde{b}^{-1}} ,\ \ \ \ r_- \approx r_+ - \tilde{b}^{-1}(r_c-r_+),\ \ \ \ \tilde{\Phi}K \approx \f{\sqrt{(d-2)(d-1)}}{2\sqrt{2}} \f{1+2\tilde{b}}{1+\tilde{b}} .\label{tripcoincidence}
\ee
This parameterization has $\tilde{b}$ scaling monotonically with $\tilde{\b}$ (at leading order in $r_c - r_+$) and shows that as we increase $\tilde{b}$ (or $\tilde{\b}$), the inner horizon approaches the outer horizon more quickly. This is what we would want in order to land on a colder black hole. 

The emblackening factor can be written as 
\begin{align}
f(r) &= \f{(r^{d-2}-r_+^{d-2})(r^{d-2}-r_-^{d-2})}{r^{2(d-2)}}\nonumber\\
&\approx \f{(r^{d-2}-r_+^{d-2})^2}{r^{2(d-2)}} + \f{(d-2)r_+^{d-3}(r^{d-2}-r_+^{d-2})(r_c-r_+)}{\tilde{b} r^{2(d-2)}}.
\end{align}
where we substitute $r_-$ using \eqref{tripcoincidence} and discard all but the $O(r_c - r_+)$ terms. Notice that, regardless of how large $\tilde{b}$ is, we have a single zero at leading order as $r \rightarrow r_+$ as long as $\tilde{b}$ is finite. This zero is responsible for the finite temperature we get in this limit. It is only for $\tilde{b} = \infty = \tilde{\b}$ that we recover the double zero.

We can evaluate the entropy exactly along the critical curve by keeping the leading-order expressions for $\tilde{\b}$ and $K$ (the $O[(r_c - r_+)^0]$ piece) and solving to compute $r_+$ and $\tilde{\Phi}$. We find
\be\label{criteqnsflatcharged}
r_+ \approx \f{2\pi}{\tilde{\b} K} \sqrt{1+\f{(d-2)^2\tilde{\b}^2}{4\pi^2}},\qquad \tilde{\Phi} K %= \f{4\pi^2}{\eta(d-2)\tilde{\b}^2} \left(1+\f{(d-2)^2\tilde{\b}^2}{4\pi^2}- \sqrt{1+\f{(d-2)^2\tilde{\b}^2}{4\pi^2}}\right) 
= \frac{d-2}{\eta} \frac{\sqrt{4 \pi ^2+(d-2)^2\tilde{\beta}^2 }}{2 \pi + \sqrt{4 \pi ^2+(d-2)^2\tilde{\beta}^2 }}
%= \frac{d-2}{\eta} \frac{ \sqrt{1+\f{(d-2)^2\tilde{\b}^2}{4\pi^2}}}{1 + \sqrt{1+\f{(d-2)^2\tilde{\b}^2}{4\pi^2}}}.
\ee
The expression for $\tilde{\Phi}K$ reproduces \eqref{crit-KPhi1}, as it should. Notice that as $\tilde{\b}\rightarrow 0$, we recover our analytic high-temperature solution for $r_+$. The corrections in $\tilde{\b}$ are also consistent upon plugging in the value of $\tilde{\Phi}K$ above into the expansion \eqref{ansatzhcK}. So, our expression for $r_+$ here resums the series. 

To compute the action, we solve \eqref{qeq} for $q$, plug into the general form of the on-shell action \eqref{bulkact}, \eqref{bdryact} (with $\ell\rightarrow \infty$), expand $r_c$ around $r_+$, and plug in \eqref{criteqnsflatcharged}. This exactly reproduces the action \eqref{actionextremalflat} as computed via the extremal AdS$_2 \times S^{d-1}$ solution in that section. Of course all of this indicates that the geometry is approaching that of AdS$_2 \times S^{d-1}$ given in \eqref{ads2s2patch}.

\subsection*{JT gravity in AdS$_2$ with AdS$_3$ conformal boundary conditions}
As is now well studied, near-extremal geometries can be captured by dilaton-gravity theories in two dimensions, see e.g.~\cite{Mertens:2022irh} and \cite{Nayak:2018qej}. In particular, the leading correction to the thermodynamics is captured by Jackiw--Teitelboim (JT) gravity. It is interesting to study this theory with boundary conditions inherited by dimensionally reducing a theory with conformal boundary conditions. Such a 2d theory would be the relevant framework for describing the extremal curves discussed in the previous subsection.

To begin with a simpler problem, we will study JT gravity with boundary conditions inherited by dimensionally reducing the s-wave sector of 3d gravity with conformal boundary conditions. (For a prescient work on conformal boundary conditions in 3d gravity, see \cite{Coleman:2020jte}.) This gives JT gravity without the topological Einstein--Hilbert term in 2d. We will comment on the dimensional reduction of near-extremal black holes with conformal boundary conditions placed in the near-horizon limit in the next subsection.

We will focus our analysis on a dimensional reduction to AdS$_2$, so that we can make some remarks about a boundary description. We write our 3d action as 
\be
I = -\f{1}{16 \pi G} \int d^3 x \sqrt{g} \,(R_{3d}-2) - \f{\a}{8\pi G}\int d^2x \sqrt{h}\,K_{2d}.
\ee
For Dirichlet boundary conditions, we have $\a = 1$, which gives the standard Gibbons--Hawking--York term, while for conformal boundary conditions, we have $\a = 1/2$. (For general $d$ the boundary term for conformal boundary conditions becomes $\a = 1/d$.) As discussed in Section \ref{conclusions}, the action above with $\a = 1/2$ leads to completely finite answers even when the boundary is at infinity, as the modified coefficient of the boundary term captures the effect of the cosmological constant counterterm in AdS/CFT. This finiteness with one-half the usual Gibbons--Hawking--York counterterm was first noted in \cite{Banados:1998ys} and further discussed in \cite{Miskovic:2006tm, Detournay:2014fva, Krishnan:2016mcj}, although these references were not considering conformal boundary conditions as considered here. (The choice of Neumann boundary conditions, where one fixes the conjugate momentum $\pi_{ij} = \f{1}{16\pi G} \sqrt{h} (K_{ij} - K h_{ij})$, leads to the same one-half GHY counterterm.)

We pick a spherically symmetric (and reflection $\phi \rightarrow -\phi$ symmetric) metric ansatz
\be \label{3d-ansatz}
ds^2 = g_{\a\b}^{(2)}(x^\a) dx^\a dx^\b + \Phi^2(x^\a)d\phi^2 = g_{rr} dr^2 + g_{\t\t} d\t^2 + \Phi^2(r,\t) d\phi^2,\qquad \phi \sim \phi + 1.
\ee
We dimensionally reduce this action using 
\be
R_{3d} = R_{2d} - 2 \Phi^{-1} \Box \Phi,\ \ \ \ K_{2d} = K_{1d} + n^\m \nabla_\m \log \Phi,\label{3d2d}
\ee
with $n^\m$ a normal vector to the 1d boundary. The covariant derivatives are with respect to the 2d metric. Observe that, near the asymptotic boundary, we have $K_{1d} = 1$ and $n^\m \nabla_\m \log \Phi = r \p_r(\log r) = 1$, which together make up the condition $K_{2d} = 2$ at the boundary. We write $\sqrt{g} = \sqrt{g^{(2)}}\,\Phi$ and evaluate the $\phi$ integral, obtaining 
\be
I = -\f{1}{16 \pi G} \int d^2 x\sqrt{g^{(2)}}(\Phi R_{2d}-2 \Phi - 2 \Box \Phi) - \f{\a}{8\pi G}\int d\t \sqrt{g_{\t\t}} \left(\Phi K_{1d} +n^\m \nabla_\m \Phi \right).
\ee
 Writing $\Box\Phi$ as a boundary term $n^\m \nabla_\m \Phi = \p_n \Phi$ gives
\be \label{s-wave-action}
I = -\f{1}{16 \pi G} \int d^2 x\sqrt{g^{(2)}}\,\Phi (R_{2d}-2) - \f{\a}{8\pi G}\int d\t \sqrt{g_{\t\t}}\Phi \left(K_{1d}+\f{\a-1}{\a} \p_n \log \Phi\right).
\ee
Notice in the Dirichlet case ($\a = 1$), the derivative-of-$\Phi$ terms cancel out, leaving the usual JT action. (If we include a cosmological-constant counterterm, then that would descend to the cosmological-constant counterterm in JT gravity.) For conformal boundary conditions ($\a = 1/2$) the derivative-of-$\Phi$ term remains. Using the second equation of \eqref{3d2d} to trade out this term gives
\be\label{2dfinform}
I = -\f{1}{16 \pi G} \int d^2 x\sqrt{g^{(2)}}\,\Phi (R-2) - \f{1}{8\pi G}\int d\t \sqrt{g_{\t\t}}\Phi \left(K_{1d}-\f 1 2 K_{2d}\right)
\ee
If we fix $K_{2d} = 2$, which corresponds to the asymptotic AdS value, then this term has the same magnitude of the cosmological constant counterterm needed for finiteness of the action with Dirichlet boundary conditions. This is just the finiteness of conformal boundary conditions at asymptotic infinity alluded to earlier, now in the context of JT gravity. And in fact, in this case our full action becomes precisely that of JT gravity in AdS$_2$ with asymptotic Dirichlet boundary conditions, which can be written as 
\be
I = -\f{1}{16 \pi G} \int d^2 x\sqrt{g^{(2)}}\,\Phi (R-2) - \f{1}{8\pi G}\int \f{du}{\epsilon} \f{\phi_r}{\epsilon} \left(K_{1d}-1\right).\label{doubleaction}
\ee
We have used the standard notation for the Dirichlet problem, where one fixes \cite{Maldacena:2016upp}
\be
ds^2 = \f{du^2}{\epsilon^2},\ \ \ \ \Phi = \f{\phi_r}{\epsilon},\qquad u \sim u + \b.
\ee
However, with conformal boundary conditions, we are fixing $K_{2d}=2$ and the conformal class of 2d boundary metrics (from the 3d perspective). The latter condition fixes the ratio of the size of the thermal circle to the size of the spatial circle, i.e.~we fix the ``inverse conformal temperature" $\tilde{\b}_{2d} = \b/ \phi_r$, which is in any case the relevant dimensionless combination. (We can work with units natural to the 3d picture, with $G$ being the 3d Newton's constant that has dimensions of length. $\Phi$ has dimensions of length and $\epsilon$ is dimensionless.)

Changing $K_{2d}$ from the asymptotic AdS value $K_{2d} = 2$ leads to the cutoff living somewhere in the AdS$_3$ bulk (and therefore somewhere in the AdS$_2$ bulk in the dimensionally reduced picture). In this general case, we can also write our boundary action as
\be
I_{\text{bdry}} = -\f{1}{8\pi G} \int du \f{\Phi^2}{\phi_r} \left(\f 1 2 K_{2d} - \p_n \log \Phi\right)=-\f{1}{8\pi G} \int \f{du}{\phi_r} \left(\f 1 2 \Phi^2 K_{2d} - \Phi \p_n \Phi\right).
\ee
This action is written in terms of fixed data $K_{2d}$ and $\b/\phi_r$ and unfixed dilaton profile $\Phi$. 

Shifting perspectives, another interesting aspect of the one-half GHY term in the context of 3d gravity in the first-order formalism is the following \cite{Banados:1998ys}. In the standard Chern--Simons description of 3d gravity we have two $SL(2,\mathbb{R})$ gauge fields $A$ and $\bar{A}$ defined in terms of the dreibein $e^a$ and spin connection $\w^a = \f 1 2 \epsilon^{abc} \omega_{bc}$ as \cite{Achucarro:1986uwr, Witten:1988hc}
\be
A^a = \w^a + e^a,\qquad \bar{A}^a = \w^a - e^a.
\ee
The Einstein--Hilbert action with negative cosmological constant is given by 
\be
I_{\text{EH}} = -\f{1}{16\pi G} \int_{\mathcal{M}} d^3 x \sqrt{g}(R+2) = I_{\text{CS}}[A] - I_{\text{CS}}[\bar{A}] +\f{1}{16\pi G} \int_{\p\mathcal{M}} \Tr(A \wedge \bar{A}).
\ee
The standard Gibbons--Hawking--York term is given by
\be
I_{\text{GHY}} =- \f{1}{8\pi G} \int_{\p \mathcal{M}} d^2 x \sqrt{h} K = -\f{1}{4\pi G} \int_{\p \mathcal{M}} \Tr(e \wedge \omega) = -\f{1}{8\pi G} \int_{\p \mathcal{M}} \Tr(A \wedge \bar{A}).
\ee
Notice then that if we add the GHY term with one-half the usual coefficient, as is needed for conformal boundary conditions, we get a cancellation between the boundary term arising from the bulk Einstein--Hilbert action in first-order variables and the standard boundary term added for consistency with conformal boundary conditions. Therefore,
\be
I_{\text{CBC}} =- \f{1}{16\pi G} \int_{\mathcal{M}} d^3 x \sqrt{g}(R+2) - \f{1}{16\pi G} \int_{\p \mathcal{M}} d^2x \sqrt{h} K = I_{\text{CS}}[A] - I_{\text{CS}}[\bar{A}].
\ee
In other words, in the Chern--Simons description, there is no boundary term needed \cite{Banados:1998ys}; Chern--Simons theories without boundary term accommodate boundary conditions $A^1 = c A^3$ for arbitrary constant of proportionality $c$. 

We can dimensionally reduce Chern--Simons theory and write it in terms of BF variables. To do so, we assume translation invariance along $x^3\sim x^3 + 1$ and define $B = A^3$. There are terms in the action that look like $\epsilon^{ij} A_i \partial B/\p x^j$, which upon integration by parts leads to a boundary term:
\begin{align}
I_{\text{CS}} &= -\f{1}{16\pi G} \int_{\mathcal{M}_3} \Tr\left(A \wedge dA + \f 2 3 A \wedge A \wedge A\right)\\
&= -\f{1}{16\pi G} \int_{\mathcal{M}_3} d^3 x\, \epsilon^{\m\n\rho} \Tr \left(A_\m \p_\n A_\rho + \f 2 3 A_\m A_\n A_\rho\right)\\
&=-\f{1}{8\pi G}\int_{\mathcal{M}_2} \Tr \left(B F\right) +\f{1}{16\pi G}\int_{\p \mathcal{M}_2} \Tr (A B),
\end{align}
The boundary condition $A^1 = c A^3$ from the Chern--Simons description descends to $A^1 = c B$; see \cite{Saad:2019lba} for a nice discussion. What is usually done at this point is that the coefficient $c$ is fixed by demanding asymptotic (Dirichlet) AdS boundary conditions \cite{Coussaert:1995zp}. However, with conformal boundary conditions we do not fix the metric at infinity; we instead fix the conformal class, which leaves $A^1/B$ unfixed, and the trace of the extrinsic curvature. AdS$_3$ with conformal boundary conditions and its dimensional reduction to AdS$_2$ will be studied in upcoming work \cite{allameh2}.

\subsection*{JT gravity in AdS$_2$ with higher-dimensional conformal boundary conditions}

Now we turn to dimensionally reducing near-extremal black holes in higher dimensions, which is the case relevant to the rest of this paper. The conclusion will be similar to the 3d s-wave reduction: fixing conformal boundary conditions in higher dimensions leads to a boundary condition in JT gravity where one fixes the ratio of the boundary time circle to the dilaton (from fixing the conformal class of metrics) and a combination of the 1d extrinsic curvature and the dilaton (from fixing the trace of the extrinsic curvature). 

We will focus on the charged spherical black hole in flat space. The Euclidean metric for such a black hole in $d+1$ spacetime dimensions $(d\geq 3)$ is
\be \label{eq:f(r)}
ds^2 = f(r) d\tau^2 + \frac{1}{f(r)} dr^2 + r^2 d\Omega_{d-1}^2,\ \ \ \ f(r) = 1 - \frac{2M}{r^{d-2}} + \frac{q^2}{r^{2(d-2)}}.
\ee
The radii of the inner and outer horizons are
\be
r_\pm=\left ( M\pm \sqrt{M^2-q^2} \right)^{\frac{1}{d-2}},
\ee
and the extremal limit corresponds to $M = q$; we take $q>0$.

To obtain the near-horizon geometry of a near-extremal black hole, we first parameterize
\be
q=M\left(1 - \frac{\epsilon^2}{2}\right),
\ee
with $\epsilon \ll 1$, so that the radius of the outer horizon is $r_+^{d-2}\approx M(1+\epsilon)$. In the strictly extremal case, we have $r_+=M^{1/(d-2)}\equiv r_0$. We parameterize the radial direction by $r=r_0(1+\frac{u}{d-2})$, where $\epsilon \leq u \ll 1$. Then, to second order in $\epsilon$ and $u$, the function $f$ in \eqref{eq:f(r)} is $f\approx u^2-\epsilon^2$. Lastly, we define $\rho \equiv (d-2)u/r_0$, in terms of which the line element becomes
\be
ds^2 \approx \left ( \frac{r_0}{d-2} \right )^2 \left [ (\rho^2 - \rho_h^2) d\tau^2+ \frac{d\rho^2}{\rho^2 - \rho_h^2} \right ] +r_0^2 d\Omega_{d-1}^2 , 
\ee
where $\rho_h \equiv (d-2)\epsilon/r_0$. This is the AdS$_2 \times S^{d-1}$ geometry.

We now study perturbations around the near-horizon geometry by considering the metric parametrization
\be \label{eq:D-metric}
ds^2=g_{\mu \nu}dx^\mu dx^\nu = \tilde{g}_{ab} dx^a dx^b + \Phi^{\frac{2}{d-1}}(x^a)d\Omega_{d-1}^2,
\ee
where $x^a$ covers the temporal and radial directions. The Euclidean action for the higher-dimensional gravitational theory (omitting the Maxwell term for now) is
\be \label{eq:action}
I_{\text{grav}} = -\frac{1}{16 \pi G}\int d^{d+1}x \sqrt{g}\, R - \frac{\alpha}{8\pi G} \int d^{d}x \sqrt{h}\, K,
\ee
where $h$ is the determinant of the induced metric on the boundary, $K$ is the trace of the boundary extrinsic curvature and $\alpha=1$ for Dirichlet and $\alpha=1/d$ for conformal boundary conditions. We would like to dimensionally reduce this action using \eqref{eq:D-metric}.

The Ricci scalar $R$ of the $(d+1)$-dimensional spacetime can be written in terms of the dilaton field $\Phi$ and $\tilde{R}$, the Ricci scalar of $ \tilde{g}_{ab}$, as
\be
R=\tilde{R}+(d-1)(d-2)\Phi^{-\frac{2}{d-1}}-2 \tilde{g}^{ab}\tilde{\nabla}_a \tilde{\nabla}_b \log \Phi -\frac{d}{d-1} \tilde{g}^{ab} \tilde{\nabla}_a \log \Phi \tilde{\nabla}_b \log \Phi.
\ee
Using $\tilde{\nabla}_a \tilde{\nabla}_b \log \Phi = (1/\Phi) \tilde{\nabla}_a \tilde{\nabla}_b \Phi -(1/\Phi^2) \tilde{\nabla}_a \Phi \tilde{\nabla}_b \Phi$, we have
\be \label{eq:bulk}
\begin{aligned}[b]
\int d^{d+1}x \sqrt{g}\, R = \text{Vol}[S^{d-1}] \int d^2x \sqrt{ \tilde{g}}\Big[ & \Phi \tilde{R} + (d-1)(d-2) \Phi^{\frac{d-3}{d-1}}\\
&+\frac{d-2}{d-1} \Phi^{-1} \tilde{g}^{ab}\tilde{\nabla}_a\Phi \tilde{\nabla}_b\Phi -2 \tilde{g}^{ab}\tilde{\nabla}_a \tilde{\nabla}_b \Phi \Big],
\end{aligned}
\ee
where $\text{Vol}[S^{d-1}]$ is the volume of S$^{d-1}$. The final term is a total divergence and becomes a boundary integral which we will combine with the boundary term in the action \eqref{eq:action}.

The kinetic term for the dilaton in \eqref{eq:bulk} disappears via a Weyl rescaling 
\be
 \tilde{g}_{ab}=\Phi^{-\frac{d-2}{d-1}}\, g^{(2)}_{ab} ,
\ee
so we have
\be
\sqrt{ \tilde{g}}= \Phi^{-\frac{d-2}{d-1}} \sqrt{g^{(2)}},\ \ \ \ \tilde{R}= \Phi^{\frac{d-2}{d-1}} \left ( R_{2d} + \frac{d-2}{d-1} g_{(2)}^{ab}\nabla_a \nabla_b \log \Phi\right ),
\ee
where $\nabla_a$ above is the covariant derivative with respect to $g^{(2)}_{ab}$. The action \eqref{eq:action} may thus be written in terms of $\Phi$ and $g^{(2)}_{ab}$. 
Noting that\footnote{This is true since the metric $ \tilde{g}_{ab}$ is 2-dimensional. For higher dimensions, there would be extra terms involving the derivatives of the Weyl factor.}
\be
\sqrt{ \tilde{g}} \tilde{g}^{ab}\tilde{\nabla}_a \tilde{\nabla}_b \Phi=\sqrt{g^{(2)}}g_{(2)}^{ab} \nabla_a \nabla_b \Phi,
\ee
we have that the bulk part of the action \eqref{eq:action} evaluates to
\be
-\frac{\text{Vol}[S^{d-1}]}{16\pi G} \int d^2x \sqrt{g^{(2)}} \left[ \Phi R_{2d} +(d-1)(d-2) \Phi^{-\frac{1}{d-1}} -\frac{d}{d-1} \nabla^a \nabla_a \Phi\right].
\ee
To deal with the boundary term in the action \eqref{eq:action}, we take into account the Weyl scaling and get
\be
K=\Phi^{\frac{d-2}{2(d-1)}} \left [K_{1d}+ \frac{d}{2(d-1)} \, n^a \nabla_a \log \Phi \right],\label{Khigherd}
\ee
where $n^a$ is the unit vector orthogonal to the boundary, normalized with respect to $g^{(2)}_{ab}$, and $K_{1d}=\nabla_a n^a$. Combining everything, we obtain the reduced gravitational action
\be \label{eq:2d-action}
\begin{aligned}[b]
I_{\text{grav}} = &-\frac{\text{Vol}[S^{d-1}]}{16\pi G} \int d^2x \sqrt{g^{(2)}} \left [ \Phi R_{2d} +(d-1)(d-2) \Phi^{-\frac{1}{d-1}} \right ]\\
&\ \ -\alpha \frac{\text{Vol}[S^{d-1}]}{8\pi G} \int d\tau \sqrt{g^{(2)}_{\tau \tau }} \left ( \Phi K_{1d}+ \frac{\alpha -1}{2\alpha}\frac{d}{d-1} \partial_n \Phi \right ). 
\end{aligned}
\ee
We observe that the boundary term matches 
the one derived from the 3-dimensional reduction \eqref{s-wave-action} when $d = 2$,\footnote{The apparent mismatch by a factor of $\text{Vol}[S^1]=2\pi$ is due to the normalization of the angle in \eqref{3d-ansatz}, where it had unit periodicity.}
and the last term vanishes for Dirichlet boundary condition. 

The Euclidean Maxwell action,
\be
I_{\text{EM}}=\frac{1}{16\pi G}\int d^{d+1}x \sqrt{g} F_{\mu\nu}F^{\mu\nu},
\ee
needs to be added to \eqref{eq:2d-action} to get the full expression. The only nonzero component of the Maxwell field strength tensor is $F^{tr}$, and near the horizon we have that
\be
F^2 = -\frac{(d-1)(d-2)q^2}{r^{2(d-1)}}\approx -(d-1)(d-2)\frac{M^2}{\Phi^2}.
\ee
After dimensional reduction and Weyl rescaling, we obtain
\be \label{I_F}
I_{\text{EM}}=- (d-1)(d-2) \frac{\text{Vol}[S^{d-1}]}{16\pi G} \int d^2x \sqrt{g^{(2)}} \, M^2 \Phi^{-\frac{2d-3}{d-1}}.
\ee
To get the JT action, we expand the dilaton around its classical value, which relates to the horizon radius of the higher-dimensional black hole---$\Phi= \Phi_0 +\phi$, where $\Phi_0=r_0^{d-1}$ and $\phi \ll \Phi_0$. Also, recall that in the extremal limit the black hole mass parameter $M$ is related to the horizon radius by $M=r_0^{d-2}$. Thus, the second term in \eqref{eq:2d-action} together with the Maxwell term \eqref{I_F} produces an effective cosmological constant. The total action becomes
\be\label{eq:tot-act}
\begin{aligned}[b]
 \frac{8\pi G }{\text{Vol}[S^{d-1}] } I = &- \frac{1}{2} \int d^2x \sqrt{g^{(2)}} \phi \, (R_{2d} - 2\Lambda) - \alpha \int d\tau \sqrt{g^{(2)}} \left ( \phi \, K_{1d} + \frac{\alpha -1}{2\alpha }\frac{d}{d-1} \partial_n \phi \right )\\
&\ \ -\Phi_0 \left [ \frac{1}{2} \int d^2x \sqrt{g^{(2)}} R_{2d}+\alpha \int d\tau \sqrt{g^{(2)}_{\tau \tau}} K_{1d} \right ]\\
&\ \ -(d-1)(d-2) \Phi_0^{-\frac{1}{d-1}}\int \! d^2x \sqrt{g^{(2)}},
\end{aligned}
\ee
where the effective cosmological constant is
\be
\Lambda=(d-1)(d-2)\Phi_0^{-\frac{d}{d-1}}.
\ee
The boundary data held fixed for conformal boundary conditions is given by the trace of the extrinsic curvature \eqref{Khigherd} and the conformal class of metrics, which can be characterized by the ratio $\b/\Phi^{\f{1}{d-1}}$, where $\b$ is the proper length of the boundary time circle.

% The on-sell action is
% \be
% \begin{split}
% I&=\frac{p_x p_y}{18 G}\frac{1}{\sqrt{1-a^2}} \frac{y^2}{z}
% (3(1-z^3)-K\ell \sqrt{1-z^3}) \\
% &=\frac{2\pi^2 \ell^2}{9 G} \frac{\sqrt{1-\tilde{\Omega}^2}}{1-a^2} \frac{y^2}{z}
% (3(1-z^3)-K\ell \sqrt{1-z^3}),
% \end{split}
% \ee
% which after substituting $
% y=\frac{4\pi }{3}\frac{\sqrt{1-z^3}}{\tilde{\beta} z}\frac{\sqrt{1-a^2}}{a^2z^3+1-a^2}=\frac{4\pi }{3}\frac{\sqrt{1-z^3}}{\tilde{\beta} z}\frac{\sqrt{1-a^2}}{1-\tilde{\Omega}^2}$ can be written as 
% \be
% I(\tilde{\beta},K,\tilde{\Omega})= \frac{I(\tilde{\beta},K,0)}{(1-\tilde{\Omega}^2)^{3/2}}.
% \ee

% Generalization to $d$-dimensional boundary is straightforward. We may first perform a rotation before compactification to make the velocity along one of the directions, say the $x$-direction. Then following the same steps as above we will have for the boundary data
% \be \label{boosted-data-general}
% \begin{split}
% &\tilde{\beta}=\frac{4\pi}{3}\sqrt{1-a^2}\frac{\sqrt{1-z^d}}{yz(a^2z^d+1-a^2)},\\
% &K\ell=\frac{d(2-z^d)}{2\sqrt{1-z^d}},\\
% &\tilde{\Omega}= a\sqrt{1-z^d},
% \end{split}
% \ee
% and the periodicities of the other compact directions are
% \be \label{periods-general}
% p_i = \frac{2\pi \ell}{\sqrt{1-a^2}}\sqrt{a^2z^d+1-a^2}, \quad i=2,\cdots, d-1.
% \ee

\section{Smoothening some phase diagrams}\label{app:puzzle}

\input{figs/penroseAnalyticCont/penroseAnalyticCont}

While all of our phase diagrams are precisely consistent with the thermal effective action at high temperature, some cases seem incomplete at low temperatures. In particular, in several instances, solutions that dominate the phase diagram abruptly disappear as the temperature is lowered (although interestingly, this only ever happens on the cosmic branch). Depending on the case, either a thermal gas phase begins dominating or we have no candidate solution. Both are problematic, since in the latter case we do not have a prediction for the value of the free energy, whereas in the former we have a discontinuous jump in the free energy (the so-called ``zeroth-order" transitions).

These sorts of transitions are not possible for an object written as (the logarithm of) a uniformly convergent sum of $C^\infty$ Boltzmann factors $\log(\sum e^{-\beta E})$. The uniform convergence ensures that we can exchange the derivative and sum. In the large-$N$ limit, the free energy can famously develop kinks, due to breakdowns in uniform convergence around phase transitions, but the convexity of the free energy and the existence of a standard large-$N$ limit protect its continuity. Our interpretation of these transitions is therefore that we are missing relevant solutions.

Let us recap the cases where we have issues at low temperature. We will restrict our discussion to $K > 0$ for $\L = 0$, $K\ell > d$ for $\L < 0$, and $K\ell \in \mathbb{R}$ for $\L > 0$. In the charged case with zero cosmological constant ($\tilde{\Phi} \neq 0$, $\L = 0$) discussed in Section \ref{sec:flatCharged}, all cases have a region with two cosmic solutions (with one of them dominating the phase diagram) where the two solutions abruptly disappear as the temperature is lowered. For $k=1$ and $k=0$, this only happens above some $O(1)$ electric potential $\tilde{\Phi}K$ and leads to the dominance of a thermal gas solution (see Figures \ref{figs:phasesFlatSphereCharged} and \ref{figs:phasesFlatPlanarCharged}), whereas for $k=-1$ this occurs at any electric potential and there is no candidate solution at lower temperature (see Figure \ref{figs:phasesFlatHyperCharged}). In the uncharged case with negative cosmological constant ($\tilde{\Phi} = 0$, $\L < 0$), the case $k=-1$ has a cosmic solution in a $(0,1)$ region that dominates at intermediate temperature and then abruptly disappears as the temperature is lowered (see Figure \ref{figs:phasesAdSHyper}), with no candidate solution at low temperature. Finally, in the uncharged case with positive cosmological constant ($\tilde{\Phi} = 0$, $\L > 0$), the case $k=-1$ has a cosmic solution in either a $(0,1)$ or $(0,2)$ region that dominates at intermediate temperature and then abruptly disappears as the temperature is lowered (see Figure \ref{fig:solutionsdSHyper}), with no candidate solution at low temperature.

We now discuss various possibilities that may smooth out our phase diagrams, or at least provide solutions where none currently exist. One possibility is that at low temperatures and large electric potentials, we spontaneously break the transverse spatial symmetry. Notably, our boundary manifold only has spatial symmetry as a conformal isometry, since we only fix the conformal class of metrics, so searching among bulk solutions with this symmetry is perhaps a stronger assumption than in the context of AdS/CFT. Spontaneous breaking of spatial isometries has been observed in the context of AdS/CFT in Einstein--Maxwell-scalar theory (see e.g. \cite{Donos:2011bh}), though there the scalar field is crucial.

There are two other types of solutions we can consider and that we can actually characterize to some extent. The first type of solution is relevant in situations where a dominant cosmic solution in a $(0,1)$ region disappears, which occurs for the uncharged $k=-1$ cases with $\L = 0$ (treated in \cite{Banihashemi:2024yye}), $\L < 0$, and $\L > 0$ (with $K\ell > 0$). In these situations, the solution disappears because $r_c \rightarrow 0$. Clearly the geometric description is breaking down and we cannot trust it here, but let us follow the solution beyond this point. The solution still exists mathematically but has $r_c < 0$ and $r_h > 0$. We can now imagine a contour as drawn in Figure \ref{figs:penroseDouble}. This contour connects $r_c$ and $r_h$ but avoids the singularity by going into the complex plane. Of course, there is a \emph{curvature} singularity at $r = 0$, so to keep curvatures small we would need a large excursion into the complex plane. As explained above, including this solution is motivated by tracking the way in which the relevant solution disappears in the first place.

There is a simple case where we can provide some analytic formulas for these $r_c < 0$ solutions. Consider the uncharged flat case $\tilde{\Phi} = \L = 0$ with $d=4$ and $k=-1$, treated in Appendix B of \cite{Banihashemi:2024yye}. The metric is given by
\be
ds^2 = f(r) d\t^2 + \f{dr^2}{f(r)} + r^2 d\mathbb{H}_3^2,\qquad f(r) = -1 + \f{r_h^2}{r^2}.
\ee
The relevant boundary-value equations that we need to solve are given by
\be
\tilde{\b} = 2\pi r_h \sqrt{\f{f(r_c)}{r_c^2}},\qquad K = \f{1}{r_c \sqrt{f(r_c)}}\left[3 - 2 \left(\f{r_h}{r_c}\right)^2\right].\label{hyptosolve}
\ee
The solutions considered in \cite{Banihashemi:2024yye} correspond to
\be
\tilde{\b} < \sqrt{3}\pi: \ \ \ \ r_h = \f{\pi\left(\pi + \sqrt{\pi^2 +\tilde{\b}^2}\right)-\tilde{\b}^2}{\pi \tilde{\b} K },\ \ \ \ r_c = \sqrt{\f{2}{\pi}} \f{\left(2\pi - \sqrt{\pi^2 + \tilde{\b}^2}\right)\sqrt{\pi + \sqrt{\pi^2 + \tilde{\b}^2}}}{\tilde{\b} K },
\ee
which breaks down as the temperature is lowered $\tilde{\b}\rightarrow \sqrt{3}\pi$, since $r_h$ and $r_c$ both approach the singularity $r=0$. But let's naively continue through this point, flipping the sign of the solution for $r_h$. This is given by
\be
\tilde{\b} > \sqrt{3}\pi: \ \ \ \ r_h = \f{\tilde{\beta}^2 -\pi\left(\pi + \sqrt{\pi^2 +\tilde{\b}^2}\right)}{\pi \tilde{\b} K },\ \ \ \ r_c = \sqrt{\f{2}{\pi}} \f{\left(2\pi - \sqrt{\pi^2 + \tilde{\b}^2}\right)\sqrt{\pi + \sqrt{\pi^2 + \tilde{\b}^2}}}{\tilde{\b} K },
\ee
and solves the equations \eqref{hyptosolve}. It corresponds to having $r_c < 0$ and $r_h > 0$, as illustrated in Figure \ref{figs:penroseDouble}. Including such solutions would smoothen the phase diagrams for all cases where a dominant cosmic solution in a $(0,1)$ region disappears (since it always occurs due to $r_c \rightarrow 0$).

A final possibility, again motivated by tracking how the dominant solution disappears, is to include complex-conjugate pairs of solutions in the path integral. This is relevant for the other problematic situations (see Figures \ref{figs:phasesFlatSphereCharged} and \ref{figs:phasesFlatPlanarCharged}), where a dominant solution is part of a $(0,2)$ or $(1,2)$ region and coincides with the subdominant solution before moving into the complex plane. A pair of real solutions coinciding and moving into the complex plane is a common phenomenon, and it occurs for black holes in AdS/CFT. In that context, we are thankfully dominated by the thermal gas phase when the black hole solutions move into the complex plane. If the minimum temperature of black holes had somehow been higher than the Hawking-Page transition temperature, then we would have found ourselves with similar zeroth-order transitions.

Let us discuss these complex solutions in some more detail. In the course of finding bulk parameters $\{r_h,r_c,q\}$ consistent with a choice of real boundary data $\{\tilde{\beta},K,\tilde{\Phi}\}$, allowing for complex solutions leads us to complex-conjugate pairs of solutions, and their on-shell actions will also be complex conjugate. Given such a pair, their collective contribution to the path integral will be of the form
\begin{equation}
\exp\left[-\frac{1}{G|K|^{d-1}}\left(\rho + i\gamma\right)\right] + \exp\left[-\frac{1}{G|K|^{d-1}}\left(\rho - i\gamma\right)\right] \subset Z.
\end{equation}
If such solutions dominate the path integral, then we have
\be
Z \approx 2 \cos \left(\f{\g}{G|K|^{d-1}}\right) \exp\left[ -\frac{\rho}{G|K|^{d-1}}\right].
\ee
The highly oscillatory prefactor can flip the sign of the partition function. While subleading, this sign is unlikely to be cancelled by any standard one-loop contributions.

\small
\bibliographystyle{jhep}
\bibliography{references.bib}

\end{document}

%% file: figs/flatCharged/texCode/penroseDiagrams/penroseFlatSphereCharged.tex
\begin{figure}
\centering
\begin{tikzpicture}[scale=1.3]
\draw[draw=none,pattern = crosshatch,pattern color=penroseblue] (1,3) .. controls (0.45,2.25) and (0.45,1.75) .. (1,1) to (0,2) -- cycle;
\draw[draw=none,pattern = crosshatch,pattern color=penrosered] (1,1) .. controls (0.45,0.25) and (0.45,-0.25) .. (1,-1) to (0,0) -- cycle;

\draw[-,penrosered!80!black,dashed,very thick] (-1,-1) to (1,1);
\draw[-,penrosered!80!black,dashed,very thick] (-1,1) to (1,-1);

\draw[-,penroseblue!80!black,dashed,very thick] (-1,3) to (1,1);
\draw[-,penroseblue!80!black,dashed,very thick] (-1,1) to (1,3);

\draw[-,penroseblue!80!black,dashed,very thick] (-1,-3) to (1,-1);
\draw[-,penroseblue!80!black,dashed,very thick] (-1,-1) to (1,-3);

\draw[-] (-1,-1) to (-2,0) to (-1,1);
\draw[-] (1,-1) to (2,0) to (1,1);

\draw[-,decorate,decoration={zigzag,amplitude=0.5mm,segment length=2.5mm}] (1,1) to (1,3);
\draw[-,decorate,decoration={zigzag,amplitude=0.5mm,segment length=2.5mm}] (1,-3) to (1,-1);

\draw[-,decorate,decoration={zigzag,amplitude=0.5mm,segment length=2.5mm}] (-1,3) to (-1,1);
\draw[-,decorate,decoration={zigzag,amplitude=0.5mm,segment length=2.5mm}] (-1,-1) to (-1,-3);

\node[penrosered!80!black] at (0.35,0.7) {\large$r_+$};
\node[penrosered!80!black] at (0.35,-0.7) {\large$r_+$};

\node[penroseblue!80!black] at (0.35,2+0.7) {\large$r_-$};
\node[penroseblue!80!black] at (0.35,2-0.7) {\large$r_-$};

\node[rotate=90] at (0,3.25) {\LARGE$\cdots$};
\node[rotate=90] at (0,-3.25) {\LARGE$\cdots$};

\draw[-] (1,3) .. controls (0.45,2.25) and (0.45,1.75) .. (1,1);
\draw[-] (1,1) .. controls (0.45,0.25) and (0.45,-0.25) .. (1,-1);

\node at (0.775,2) {\large$r_c$};
\node at (0.775,0) {\large$r_c$};
\end{tikzpicture}
\caption{The Penrose diagram for the Reissner--Nordstr\"om black hole. A black hole patch and a cosmic patch are depicted as red and blue patches, respectively.}
\label{figs:penroseFlatSphereCharged}
\end{figure}
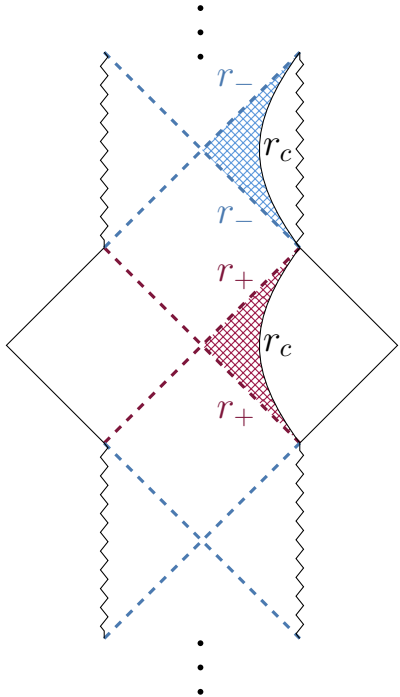

%% file: figs/flatCharged/texCode/solutionSpaces/solutionsFlatSphereCharged.tex
\begin{figure}
\centering
\begin{tikzpicture}
%\node at (0,0) {\includegraphics[scale=0.75]{figs/numPhasesRN/numSolutionsPosK.png}};
\node at (0,0) {\includegraphics[scale=0.75]{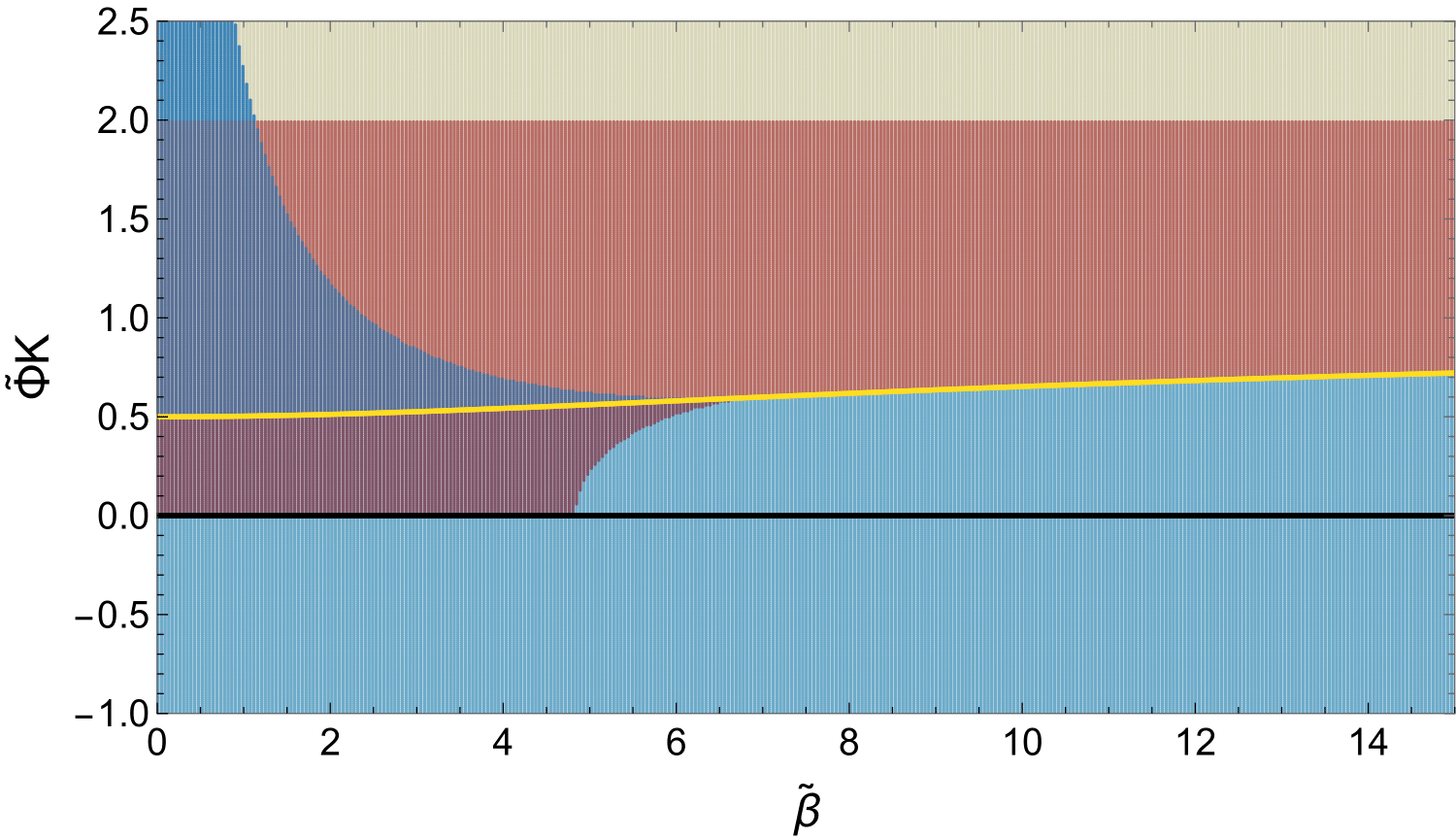}};

\node at (8.75,0) {};

\draw[-,thick,black] (3.95-4+0.15,-0.9+0.925) -- (3.95-4+0.15,-0.55+0.925) to (3.25-4+0.15,-0.55+0.925) to (3.25-4+0.15,-0.9+0.925) -- cycle;

\node at (7.5-4.26,-5-1.35-0.63) {\includegraphics[scale=0.375]{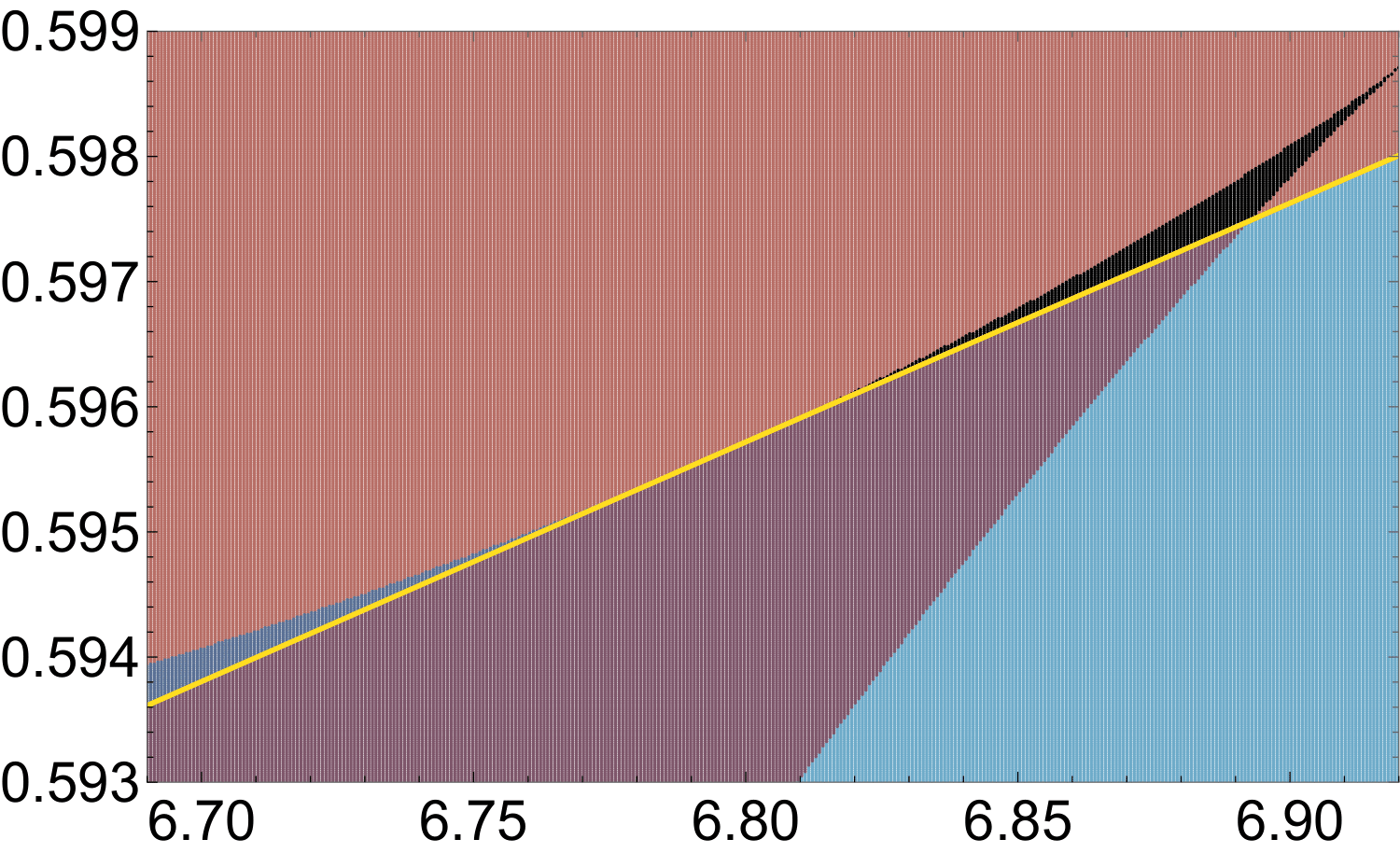}};

\draw[-,thick,black] (3.25-4+0.15,-0.9+0.925) to (4.9-4.25-0.225,-6.625-1.35-0.265-0.55);
\draw[-,thick,black] (3.25-4+0.15,-0.55+0.925) to (4.9-4.25-0.225,-2.525-1.35-0.34-0.75);
\draw[-,thick,black] (3.95-4+0.15,-0.55+0.925) to (11.5-4.25-0.45,-2.525-1.35-0.335-0.75);

\node[white] at (-0.85-3.25+0.775,-1.35+0.875) {\footnotesize\textbf{(2,1)}};
\node[white] at (3.5-0.6+1,-1.35+0.875) {\footnotesize\textbf{(0,1)}};

\node[white] at (-2.35-3.25+0.775,0.45+1.025) {\footnotesize\textbf{(1,2)}};
\node[white] at (2-0.6,0.45+0.1+1.025) {\footnotesize\textbf{(1,0)}};

\node[white] at (-2.64-3.32+0.775,2.285+0.1+1.025) {\footnotesize\textbf{(0,2)}};
\node[black] at (1.71-0.67,2.285+0.1+1.025) {\footnotesize\textbf{(0,0)}};

\node[white] (A) at (8.75-4.25,-3.5-1.35-0.5) {\footnotesize\textbf{(3,0)}};
\draw[->,very thick,white] (10.35-4.25-0.175,-3.5-1.35-0.25-0.75) to[bend right] (9.2-4.25,-3.25-1.35-0.75);

\draw[<->,very thick] (7.3,-0.95) to (7.3,-0.95-1.9375);
\draw[<->,very thick] (7.3,-0.95) to (7.3,-0.95+1.9375*2.5);

\node[rotate=-90] at (7.575,-1.91875) {$n_{\text{TG}} = 0$};
\node[rotate=-90] at (7.575,1.471875) {$n_{\text{TG}} = 1$};

\node[rotate=1.21] at (-4.25,0.325) {\textcolor{yellow}{$\boldsymbol{\mathrm{AdS}_2 \times S^2}$}};

\end{tikzpicture}
\caption{$\tilde{\Phi} \neq 0$, $\L = 0$, $k=1$. The number of each solution with spherical horizon across various values of $(\tilde{\beta},\tilde{\Phi}K)$ for $d=3$. Each color corresponds to a particular pair $(n_{\text{BH}},n_{\text{CH}})$, where $n_{\text{BH}}$ is the number of black hole solutions and $n_{\text{CH}}$ is the number of cosmic solutions. %Note that the thermal gas exists for all temperatures and arbitrary $\tilde{\Phi} K \geq 0$, making it the only solution in the $(0,0)$ region. The thermal gas does not exist for $\tilde{\Phi} K < 0$. 
The gold curve has an equation given by \eqref{crit-KPhi1} and represents an extremal solution across which a cosmic and black hole solution interchange. We also find a numerically small region in which there are three black hole solutions, which we zoom in on above. We have fixed $\tilde{\Phi} \geq 0$, so $\tilde{\Phi}K < 0$ corresponds to $K < 0$. }
\label{figs:solutionsFlatSphereCharged}
\end{figure}
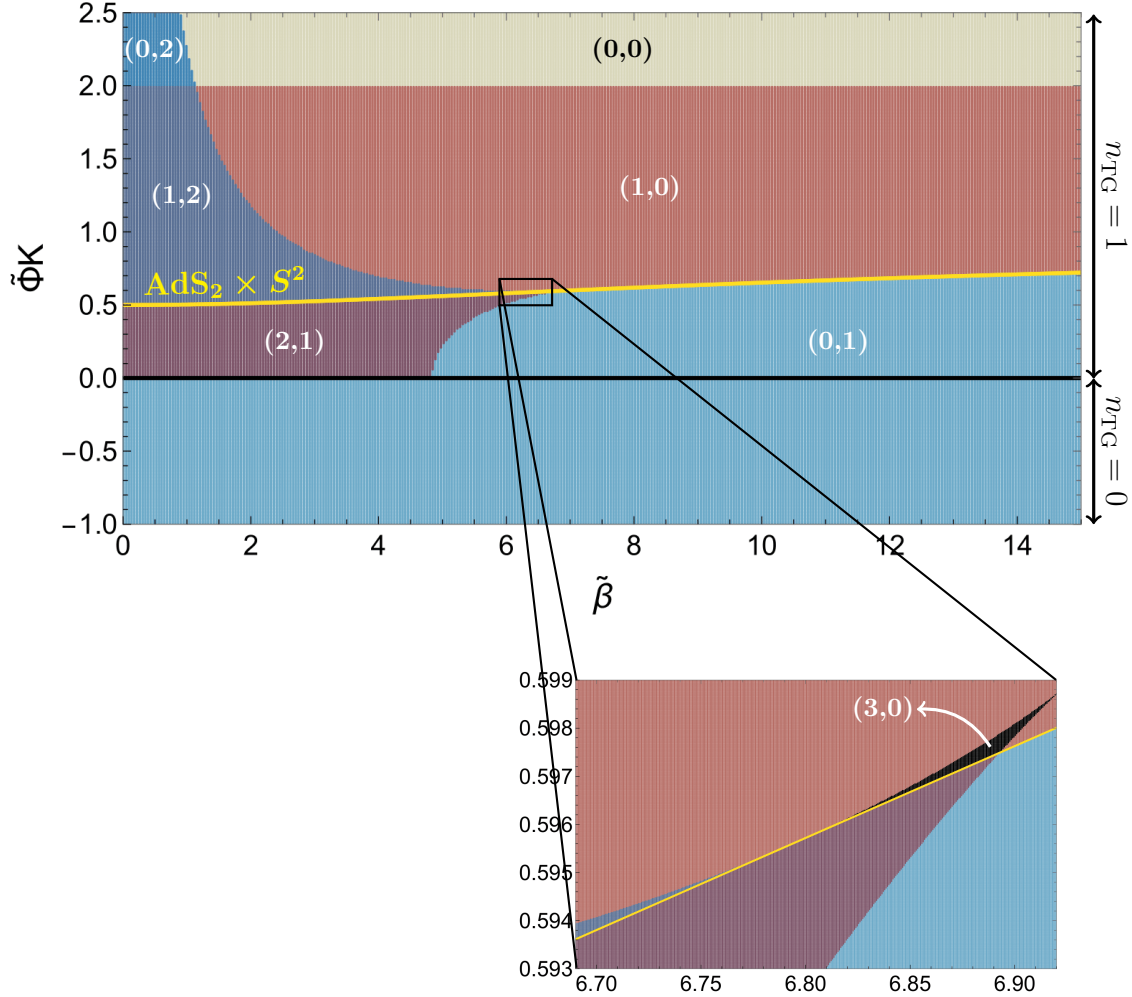

%% file: figs/flatCharged/texCode/phasePlots/phasesFlatSphereCharged.tex
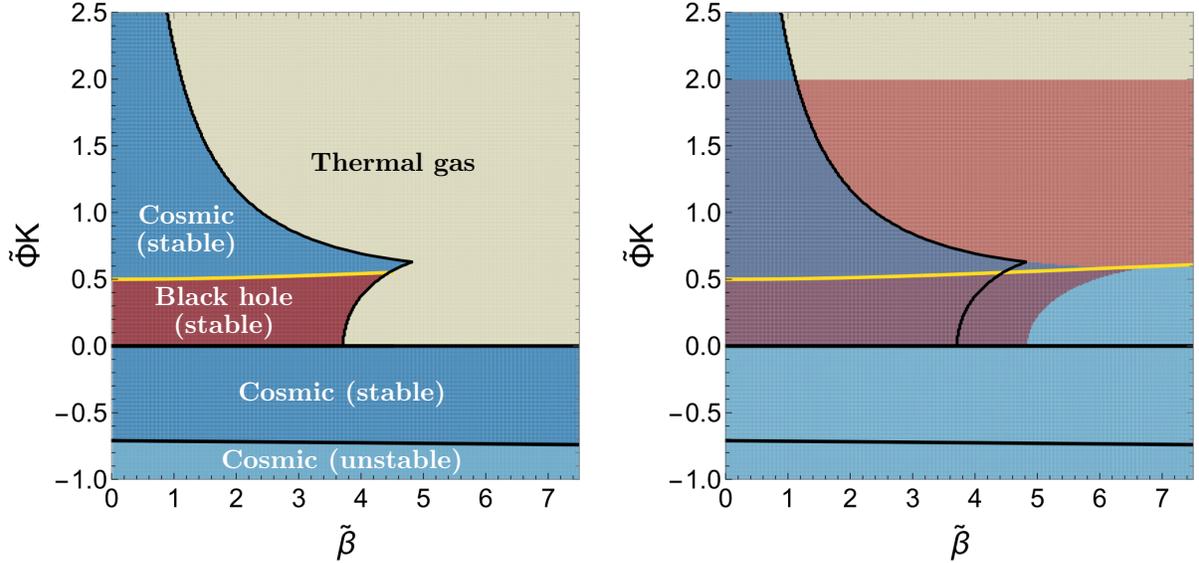
\begin{figure}
\centering
\begin{tikzpicture}
\node at (0,0){\includegraphics[scale=0.6]{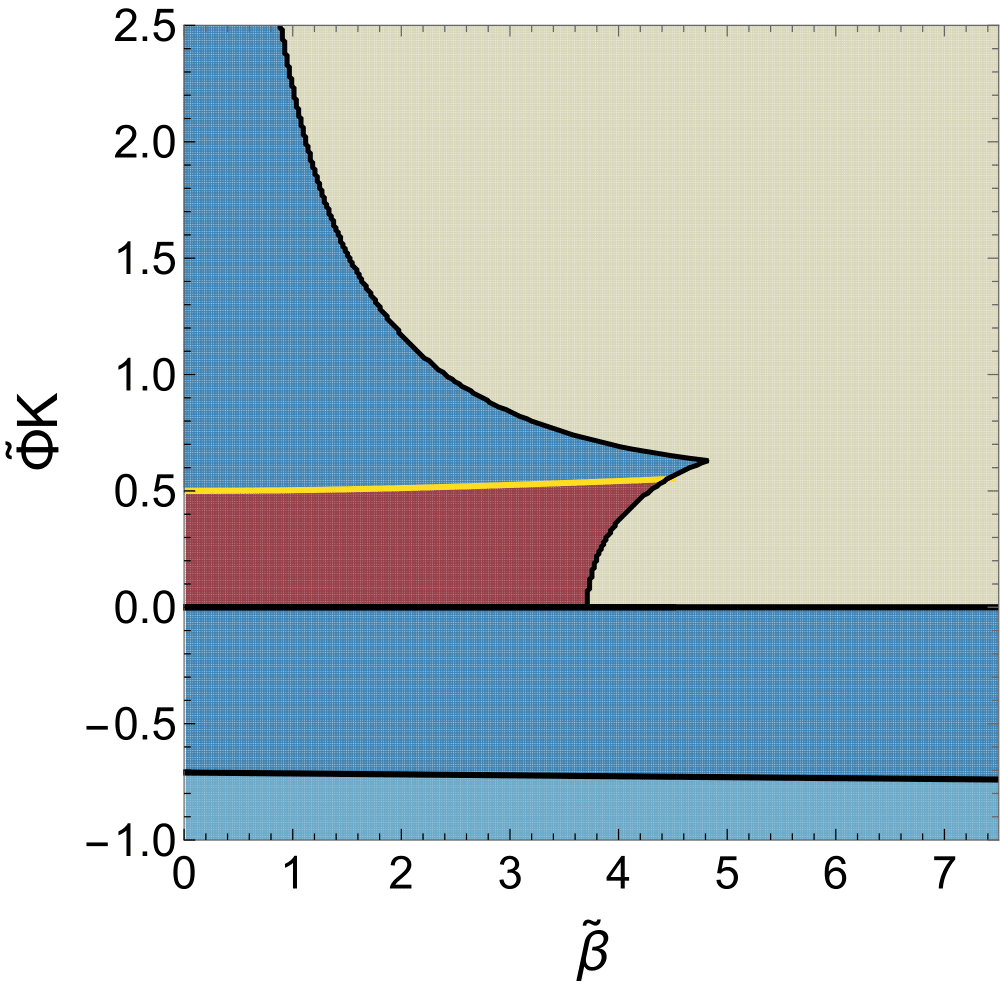}\ \ \ \ 
\includegraphics[scale=0.6]{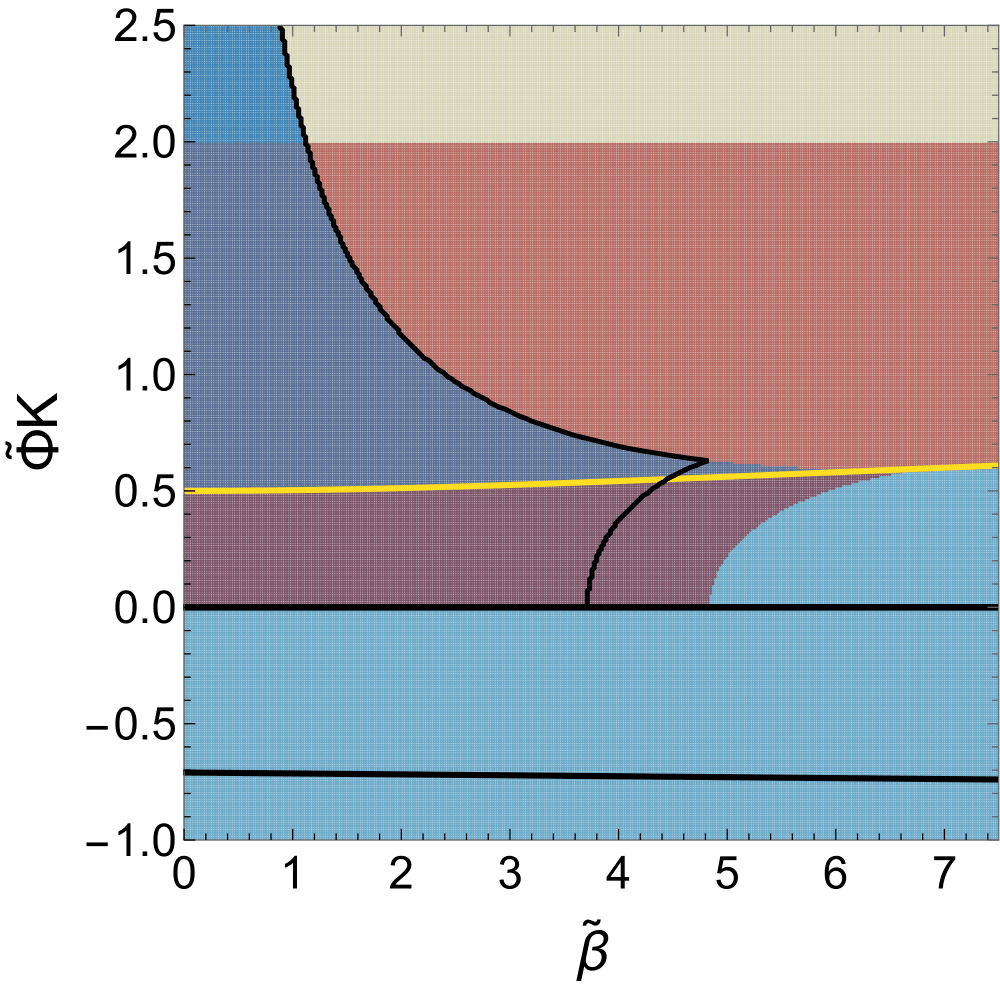}};
\node at (-3.425,-2.4) {\footnotesize{\textcolor{white}{\textbf{Cosmic (unstable)}}}};
\node at (-3.425,-1.5) {\footnotesize{\textcolor{white}{\textbf{Cosmic (stable)}}}};
\node at (-5,-0.425) {\footnotesize\shortstack{\textcolor{white}{\textbf{Black hole}}\\\textcolor{white}{\textbf{(stable)}}}};
\node at (-5.5,0.65) {\footnotesize\shortstack{\textcolor{white}{\textbf{Cosmic}}\\\textcolor{white}{\textbf{(stable)}}}};
\node at (-2.75,1.55) {\footnotesize{\textcolor{black}{\textbf{Thermal gas}}}};
\end{tikzpicture}
\caption{$\tilde{\Phi} \neq 0$, $\L = 0$, $k=1$. On the left, the phase diagram for $d = 3$; on the right, the boundary of the phase diagram superimposed on the solution space (Figure \ref{figs:solutionsFlatSphereCharged}). For small $\tilde{\Phi}K$, we have a Hawking--Page transition between a black hole and the thermal gas. As we increase $\tilde{\Phi}K$, the ensemble at high temperatures becomes dominated by a cosmic geometry.}
\label{figs:phasesFlatSphereCharged}
\end{figure}

%% file: figs/flatCharged/texCode/penroseDiagrams/penroseFlatNonSphereCharged.tex
\begin{figure}
\centering
\begin{tikzpicture}[scale=1.4]

\draw[draw=none,pattern = crosshatch,pattern color=penroseblue] (1,1) .. controls (0.45,0.25) and (0.45,-0.25) .. (1,-1) to (0,0) -- cycle;

\draw[-,penroseblue!80!black,dashed,very thick] (-1,-1) to (1,1);
\draw[-,penroseblue!80!black,dashed,very thick] (-1,1) to (1,-1);

\draw[-] (-1,-1) to (0,-2) to (1,-1);
\draw[-] (-1,1) to (0,2) to (1,1);

\draw[-,decorate,decoration={zigzag,amplitude=0.5mm,segment length=2.5mm}] (1,-1) to (1,1);
\draw[-,decorate,decoration={zigzag,amplitude=0.5mm,segment length=2.5mm}] (-1,1) to (-1,-1);

\node[penroseblue!80!black] at (0.35,0.7) {\large$r_h$};
\node[penroseblue!80!black] at (0.35,-0.7) {\large$r_h$};

\draw[-] (1,1) .. controls (0.45,0.25) and (0.45,-0.25) .. (1,-1);

\node at (0.775,0) {\large$r_c$};
\end{tikzpicture}
\caption{The Penrose diagram for both the hyperbolic and planar horizon geometries for $\L = 0$, $\tilde{\Phi} \neq 0$. Only cosmic solutions, represented by the blue patch, are possible.}
\label{figs:penroseFlatNonSphereCharged}
\end{figure}
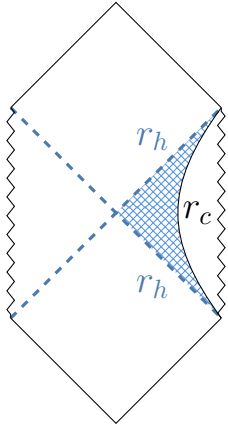

%% file: figs/flatCharged/texCode/solutionSpaces/solutionsFlatHyperCharged.tex
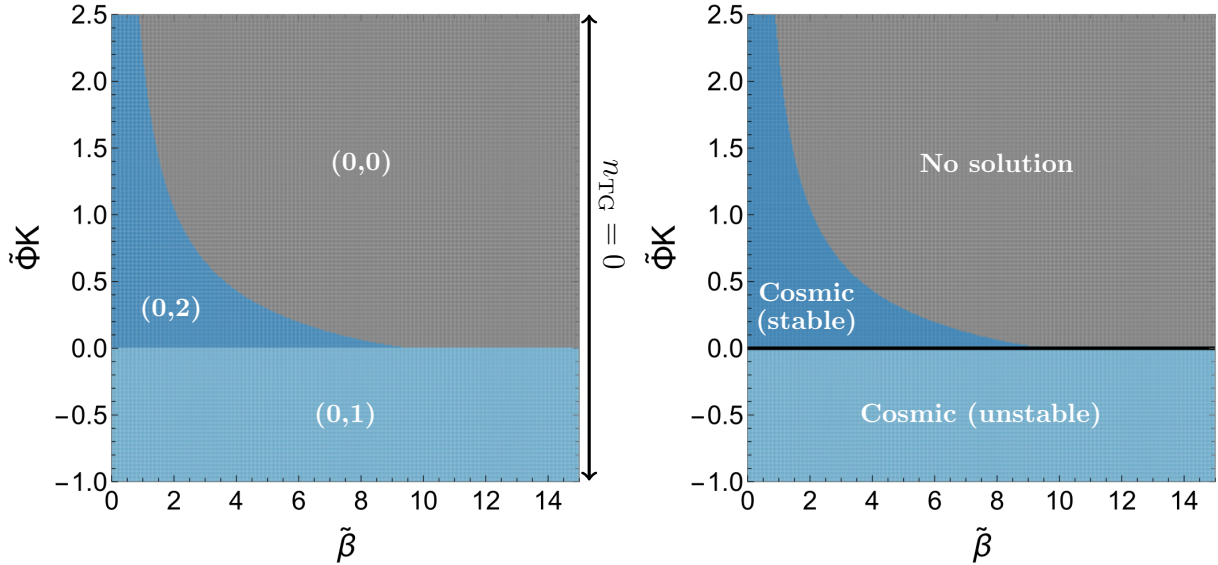
\begin{figure}
\centering
\begin{tikzpicture}
\node at (0,0){\includegraphics[scale=0.6]{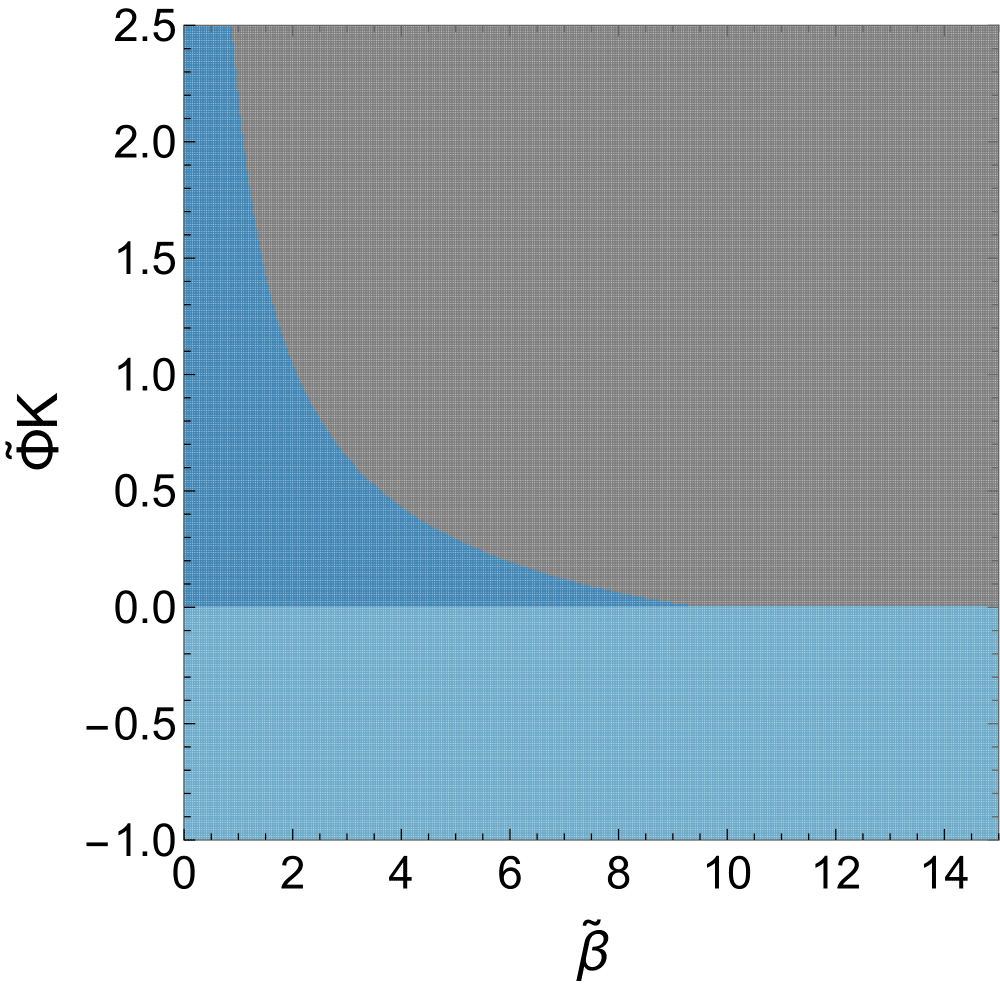}\ \ \ \ \ \ 
\includegraphics[scale=0.6]{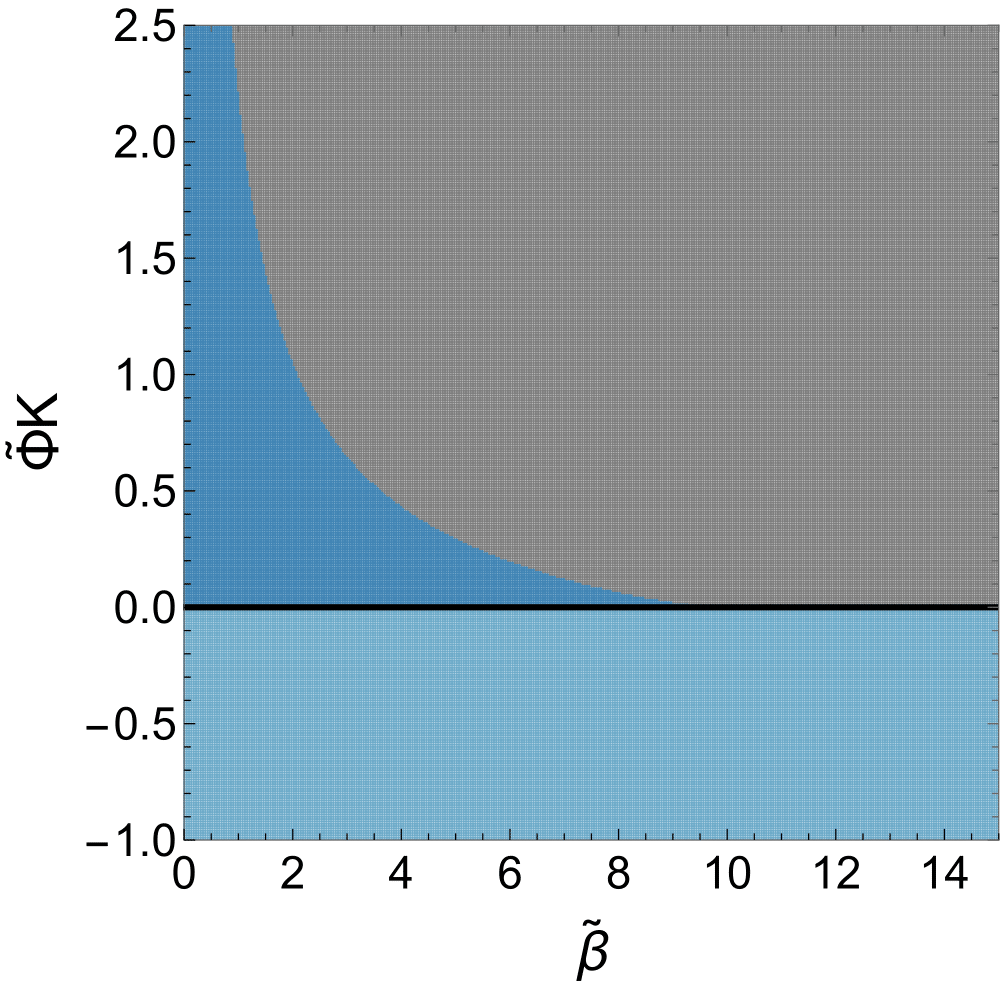}};

\node[white] at (-3.32,1.595) {$\footnotesize\textbf{(0,0)}$};
\node[white] at (-3.53,-2.1375+0.37) {$\footnotesize\textbf{(0,1)}$};
\node[white] at (-5.85,-1+0.65) {$\footnotesize\textbf{(0,2)}$};

\node at (8.45-3.32,1.595) {\footnotesize{\textcolor{white}{\textbf{No solution}}}};
\node[white] at (8.45-3.53,-2.1375+0.37)  {\footnotesize{\textcolor{white}{\textbf{Cosmic (unstable)}}}};
\node[white] at (8.45-5.85,-1+0.65) {\footnotesize\shortstack{\textcolor{white}{\textbf{Cosmic}}\\\textcolor{white}{\textbf{(stable)}}}};

\draw[<->,very thick] (-3.74+3.3+0.15,0.44+3.13) to (-3.74+3.3+0.15,0.44-3.07);

\node[rotate=-90] at (-3.44+3.3+0.15,0.91) {$n_{\text{TG}} = 0$};

%\node at (-3.53,0.475-3.15) {$\circ$};
%\node at (-6.6,1.33) {$\circ$};
%\node at (8.45-3.53,0.475-3.15) {$\circ$};

\end{tikzpicture}
\caption{$\tilde{\Phi} \neq 0$, $\L = 0$, $k=-1$. We have fixed $\tilde{\Phi} \geq 0$. On the left, the number of each phase with hyperbolic horizon across various values of $(\tilde{\beta},\tilde{\Phi}K)$. Each color corresponds to a particular pair $(n_{\text{BH}},n_{\text{CH}})$, where $n_{\text{BH}}$ is the number of black holes and $n_{\text{CH}}$ is the number of cosmic geometries. On the right, the phase diagram showing the dominant phases and their stability for this solution space. Due to the lack of a thermal gas solution, we have no candidate solution for part of the phase diagram.}
\label{figs:phasesFlatHyperCharged}
\end{figure}

%% file: figs/flatCharged/texCode/solutionSpaces/solutionsFlatPlanarCharged.tex
\begin{figure}
\centering
\begin{tikzpicture}[scale=0.933]
\node at (0,0) {\includegraphics[scale=0.75]{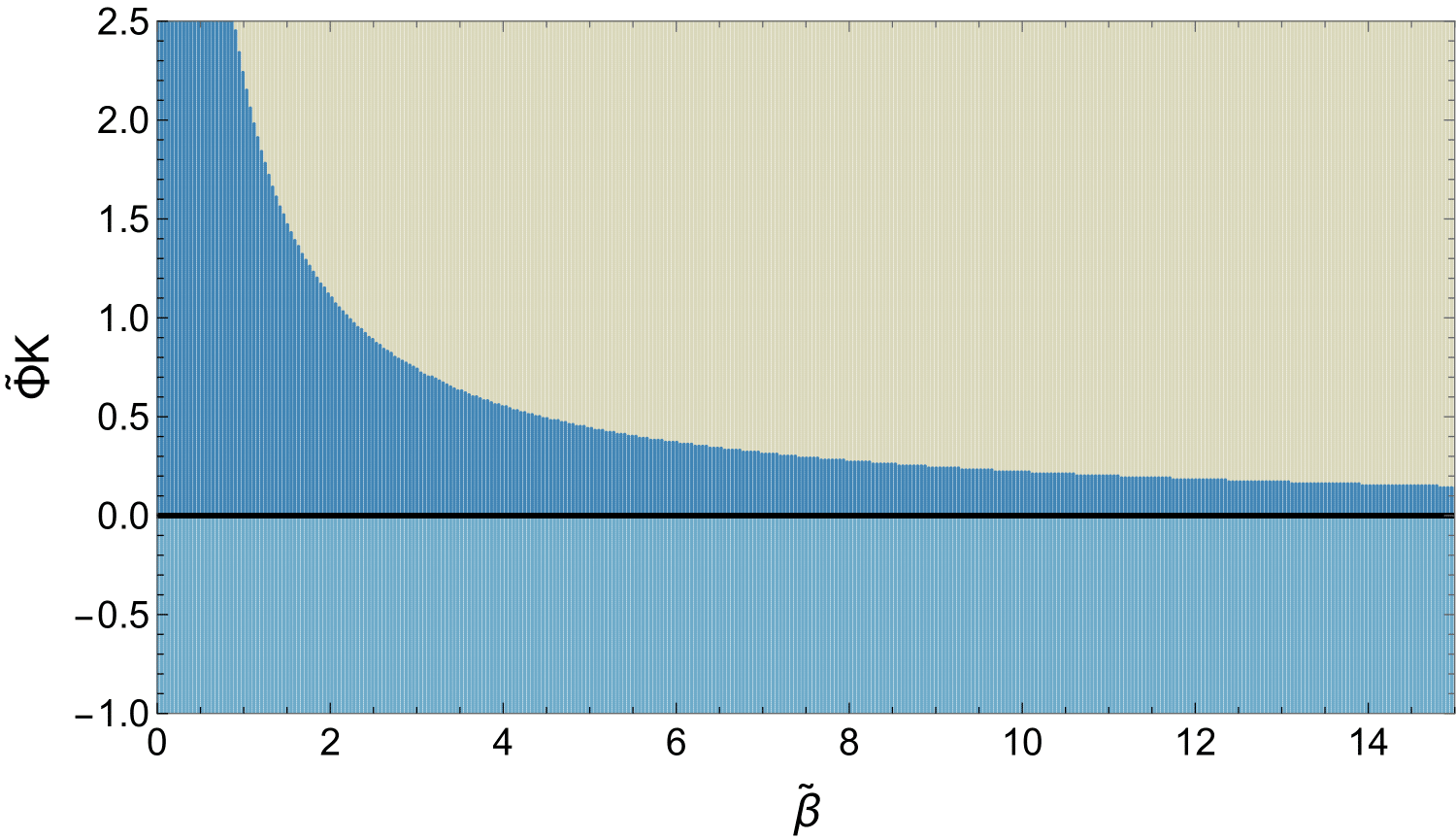}};
\node at (8.75,0) {};

\node[white] at (0.75,-1.9) {\footnotesize\textbf{(0,1)}};
\node[white] at (-2.35-3.25+0.775,0.05) {\footnotesize\textbf{(0,2)}};
\node[black] at (2-0.6,0.45+0.1+1.025) {\footnotesize\textbf{(0,0)}};

\draw[<->,very thick] (7.3+0.5,-1) to (7.3+0.5,-0.95-1.9375-0.2);
\draw[<->,very thick] (7.3+0.5,-1) to (7.3+0.5,-0.95+1.9375*2.5+0.3);

\node[rotate=-90] at (7.6+0.5,-2.04375) {$n_{\text{TG}} = 0$};
\node[rotate=-90] at (7.6+0.5,1.596875) {$n_{\text{TG}} = 2$};
\end{tikzpicture}
\caption{$\tilde{\Phi} \neq 0$, $\L = 0$, $k=0$, toroidal compactification. The number of phases with  a toroidal horizon across various values of $(\tilde{\beta},\tilde{\Phi}K)$ for $d=3$. We have fixed $\tilde{\Phi} \geq 0$. Each color corresponds to a particular pair $(n_{\text{BH}},n_{\text{CH}})$, where $n_{\text{BH}}$ is the number of black holes and $n_{\text{CH}}$ is the number of cosmic geometries. The critical curve separating the $(0,2)$ and $(0,0)$ regions can be found as discussed in the text, and behaves as $\tilde{\Phi} K \propto \tilde{\b}^{-1}$. We also have the thermal gas geometries  \eqref{flatsoliton}  above $\tilde{\Phi}K = 0$. }
\label{figs:solutionsFlatPlanarCharged}
\end{figure}
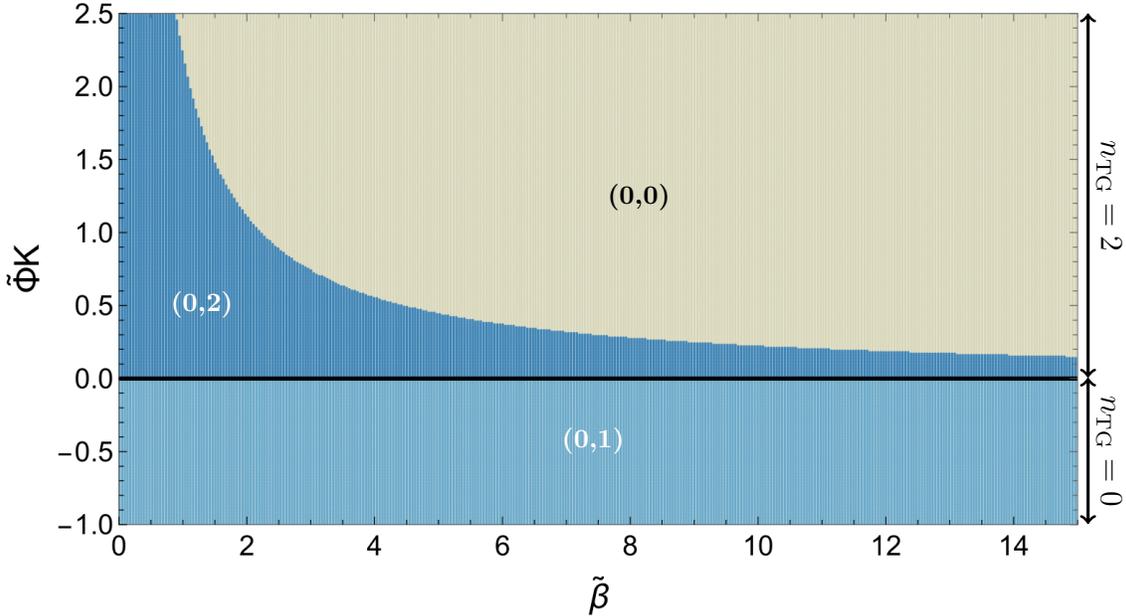

%% file: figs/flatCharged/texCode/phasePlots/phasesFlatPlanarCharged.tex
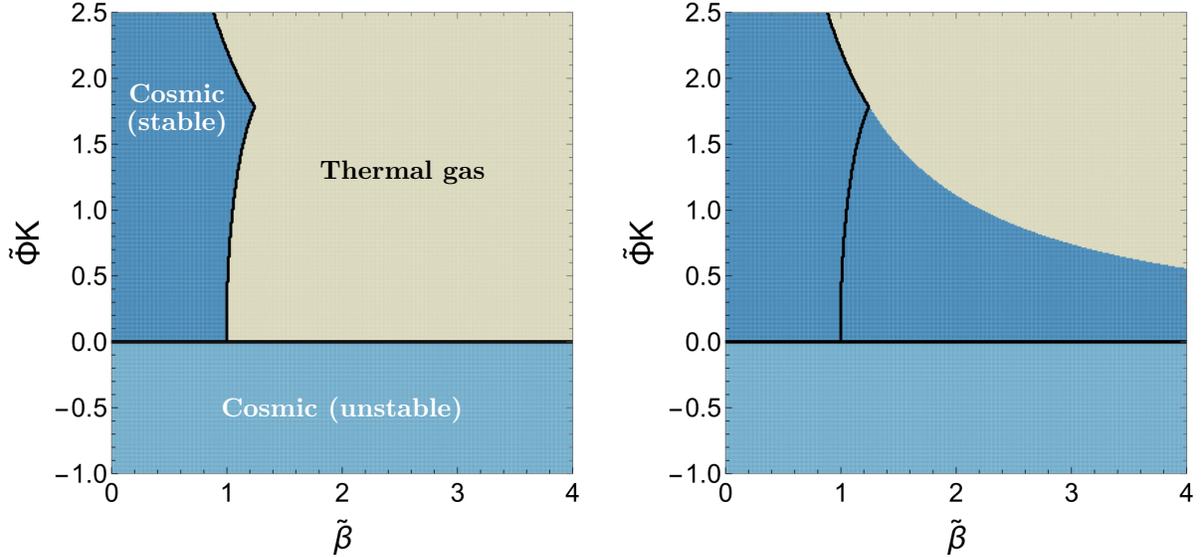
\begin{figure}
\centering
\begin{tikzpicture}
\node[white] at (4.5,0) {$\bullet$};
\node at (0,0){\includegraphics[scale=0.6]{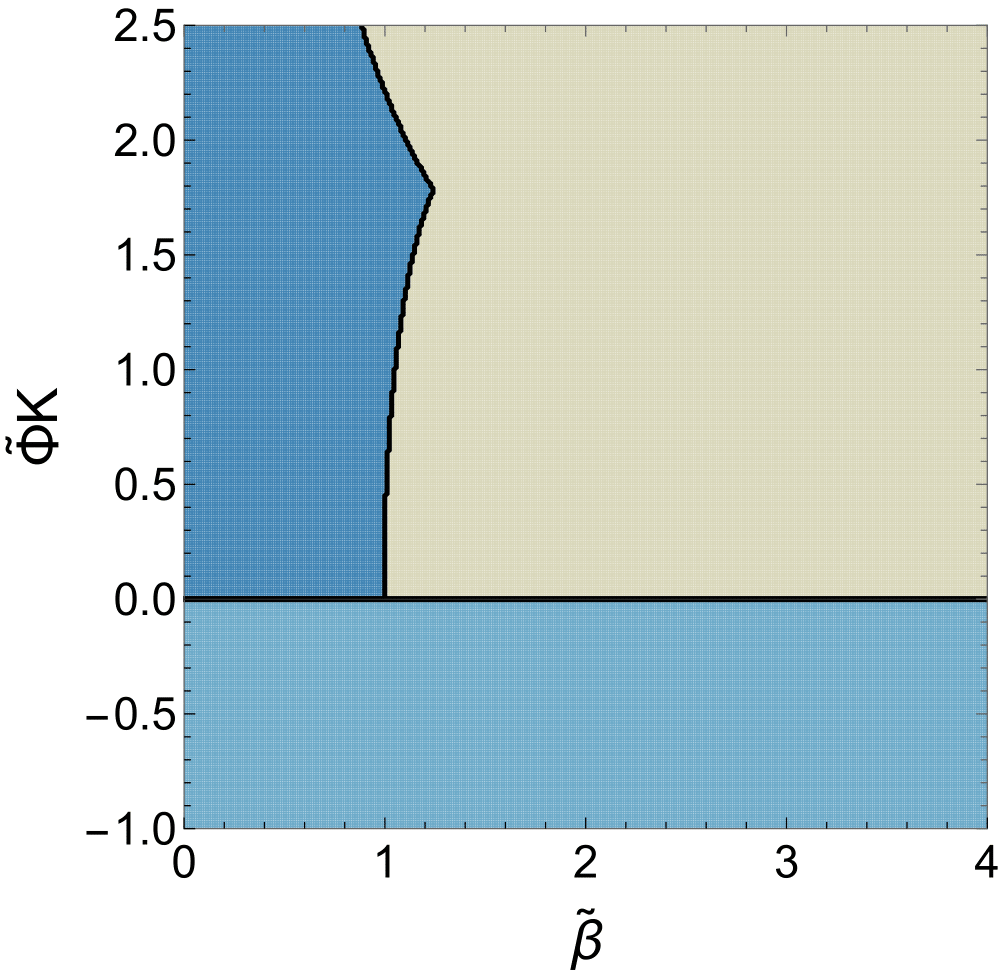}\ \ \ \ 
\includegraphics[scale=0.6]{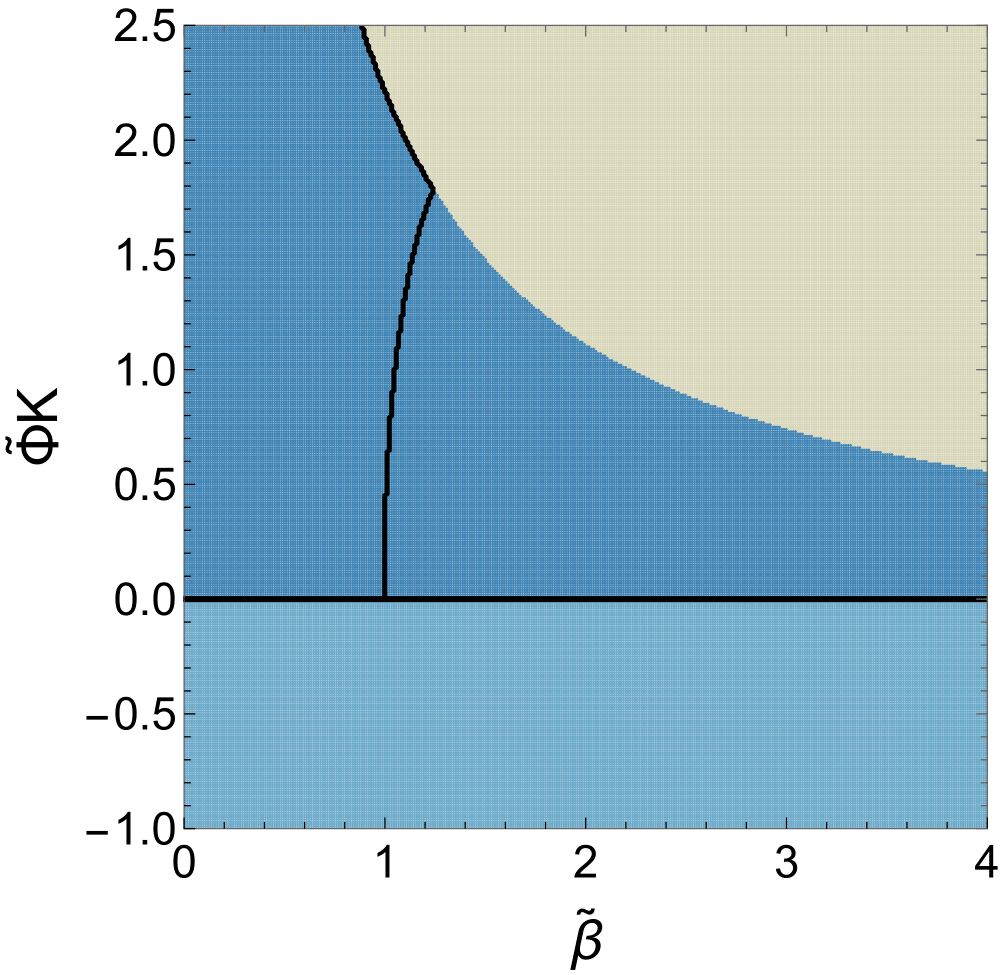}};
\node at (1.5-4.125,1.4) 
{\footnotesize{\textcolor{black}{\textbf{Thermal gas}}}};
\node at (0.7-4.125,-2.1375+0.37) {\footnotesize{\textcolor{white}{\textbf{Cosmic (unstable)}}}};
\node at (-1.5-4.125,2.22) {\footnotesize\shortstack{\textcolor{white}{\textbf{Cosmic}}\\\textcolor{white}{\textbf{(stable)}}}};
\end{tikzpicture}

\caption{$\tilde{\Phi} \neq 0$, $\L = 0$, $k=0$, toroidal compactification. On the left, the phase diagram  in $d = 3$; on the right, the boundary of this phase diagram superimposed on the solution space (Figure \ref{figs:solutionsFlatPlanarCharged}). For small $\tilde{\Phi}K$, we have a Hawking--Page transition between the stable cosmic solution and the thermal gas with minimum cycle length $\tilde{L}_i = \tilde{L}_{\text{min}}$}
\label{figs:phasesFlatPlanarCharged}
\end{figure}

%% file: figs/AdSUncharged/texCode/penroseDiagrams/penroseAdS.tex
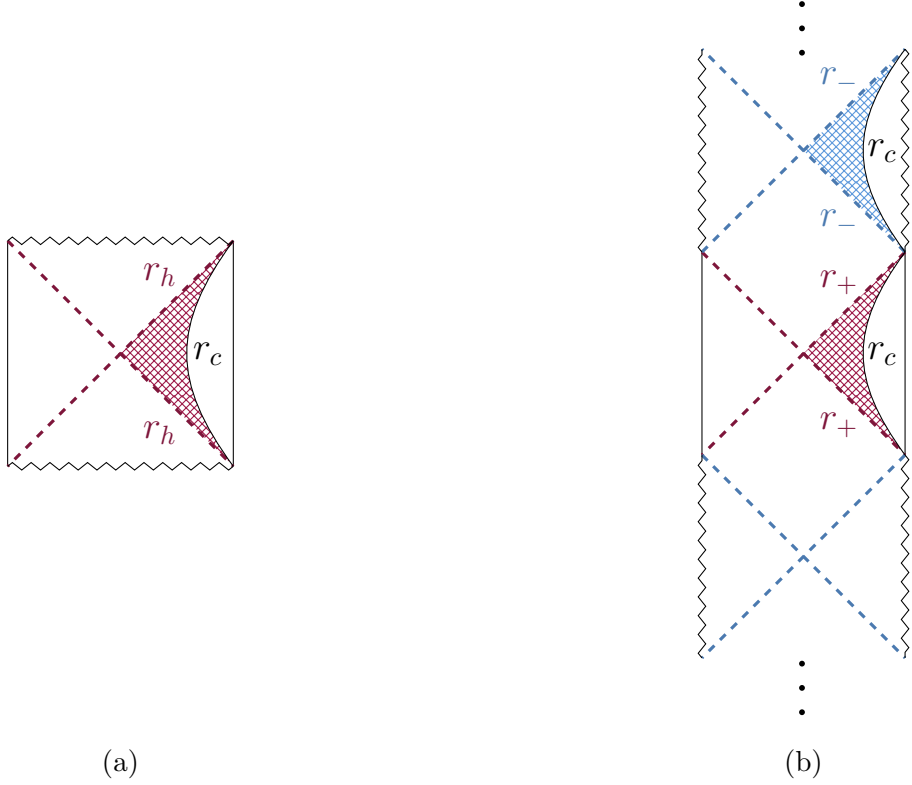
\begin{figure}[h]
  \centering
    % First TikZ figure
\begin{subfigure}{0.45\textwidth}
        \centering 
%\raisebox{20mm}
{\begin{tikzpicture}[scale=1.5]%1.95 before
\node[rotate=90,white] at (0,2.925) {\LARGE$\cdots$};
\node[rotate=90,white] at (0,-2.925) {\LARGE$\cdots$};

\draw[draw=none,pattern = crosshatch,pattern color=penrosered] (1,1) .. controls (0.45,0.25) and (0.45,-0.25) .. (1,-1) to (0,0) -- cycle;

% conf boundaries
\draw[-] (-1,-1) to (-1,1);
\draw[-] (1,-1) to (1,1);
% outer horizons
\draw[-,penrosered!80!black,dashed,very thick] (-1,1) to (0,0) to (-1,-1);
\draw[-,penrosered!80!black,dashed,very thick] (1,1) to (0,0) to (1,-1);
% inner horizons
% singularities
\draw[-,decorate,decoration={zigzag,amplitude=0.5mm,segment length=2.5mm}] (-1,-1) to (1,-1);
\draw[-,decorate,decoration={zigzag,amplitude=0.5mm,segment length=2.5mm}] (-1,1) to (1,1);
% labels
\node[penrosered!80!black] at (0.35,0.7) {\large$r_h$};
\node[penrosered!80!black] at (0.35,-0.7) {\large$r_h$};

\draw[-] (1,1) .. controls (0.45,0.25) and (0.45,-0.25) .. (1,-1);

\node at (0.775,0) {\large$r_c$};

\node at (0,-1.5) {};
\node at (0,1.5) {};
\end{tikzpicture}}
        \caption{}
    \end{subfigure}
    \hfill
    % Second TikZ figure
    \begin{subfigure}{0.45\textwidth}
        \centering
\begin{tikzpicture}[scale=1.35]
%% AdS Hyperbolic BH
\draw[draw=none,pattern = crosshatch,pattern color=penroseblue] (1,3) .. controls (0.45,2.25) and (0.45,1.75) .. (1,1) to (0,2) -- cycle;
\draw[draw=none,pattern = crosshatch,pattern color=penrosered] (1,1) .. controls (0.45,0.25) and (0.45,-0.25) .. (1,-1) to (0,0) -- cycle;

% conf boundaries
\draw[-] (-1,-1) to (-1,1);
\draw[-] (1,-1) to (1,1);
% outer horizons
\draw[-,penrosered!80!black,dashed,very thick] (-1,1) to (0,0) to (-1,-1);
\draw[-,penrosered!80!black,dashed,very thick] (1,1) to (0,0) to (1,-1);
% inner horizons
\draw[-,penroseblue!80!black,dashed,very thick] (-1,1) to (1,3);
\draw[-,penroseblue!80!black,dashed,very thick] (1,1) to (-1,3);
\draw[-,penroseblue!80!black,dashed,very thick] (-1,-1) to (1,-3);
\draw[-,penroseblue!80!black,dashed,very thick] (1,-1) to (-1,-3);
% singularities
\draw[-,decorate,decoration={zigzag,amplitude=0.5mm,segment length=2.5mm}] (-1,1) to (-1,3);
\draw[-,decorate,decoration={zigzag,amplitude=0.5mm,segment length=2.5mm}] (1,3) to (1,1);
\draw[-,decorate,decoration={zigzag,amplitude=0.5mm,segment length=2.5mm}] (-1,-3) to (-1,-1);
\draw[-,decorate,decoration={zigzag,amplitude=0.5mm,segment length=2.5mm}] (1,-1) to (1,-3);
% labels
\node[penrosered!80!black] at (0.35,0.7) {\large$r_+$};
\node[penrosered!80!black] at (0.35,-0.7) {\large$r_+$};

\node[penroseblue!80!black] at (0.35,2+0.7) {\large$r_-$};
\node[penroseblue!80!black] at (0.35,2-0.7) {\large$r_-$};

\node[rotate=90] at (0,3.25) {\LARGE$\cdots$};
\node[rotate=90] at (0,-3.25) {\LARGE$\cdots$};

\draw[-] (1,3) .. controls (0.45,2.25) and (0.45,1.75) .. (1,1);
\draw[-] (1,1) .. controls (0.45,0.25) and (0.45,-0.25) .. (1,-1);

\node at (0.775,2) {\large$r_c$};
\node at (0.775,0) {\large$r_c$};

\node at (-1.75,0) {};
\node at (1.75,0) {};
\end{tikzpicture}
 \caption{}
    \end{subfigure}

\caption{(a) The Penrose diagram for the spherical, planar, and positive mass $\m > 0$ hyperbolic AdS black holes. Only black-hole type solutions, represented by the red patch, are possible. (b) The Penrose diagram for the hyperbolic AdS black hole with negative mass $\m < 0$. We can construct both black-hole patches (in red) and  cosmic patches (in blue).}
\label{figs:penroseAdS}
\end{figure}

%% file: figs/AdSUncharged/texCode/solutionSpaces/solutionsAdSSphere.tex
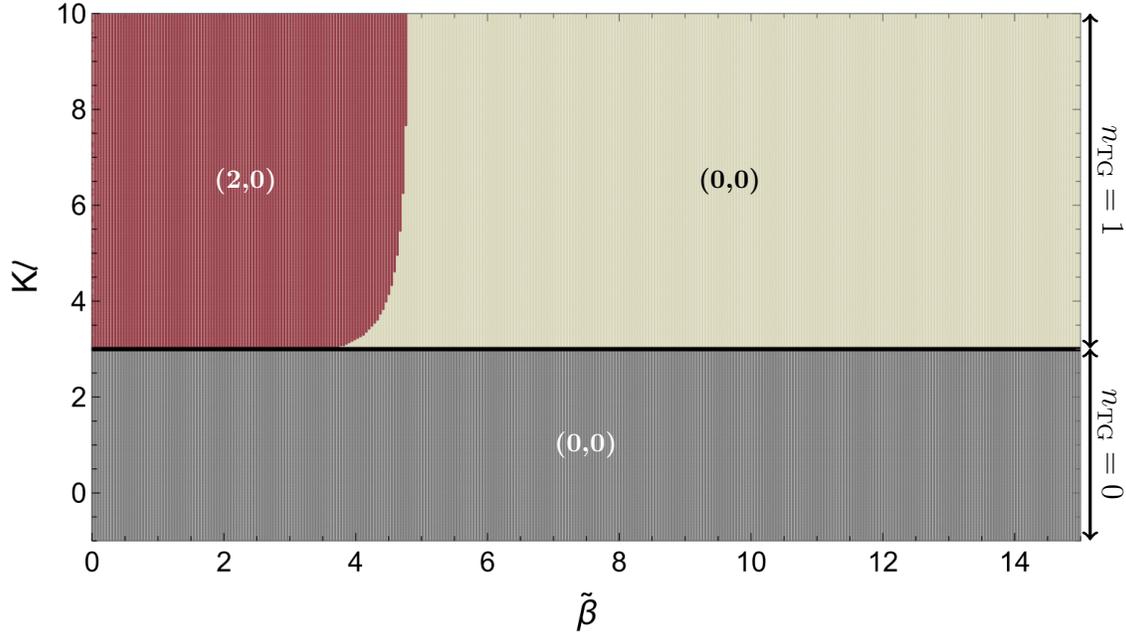
\begin{figure}[h]
\centering
{
\begin{tikzpicture}[scale=0.9]
\node[white] at (0,0) {\includegraphics[scale=0.75]{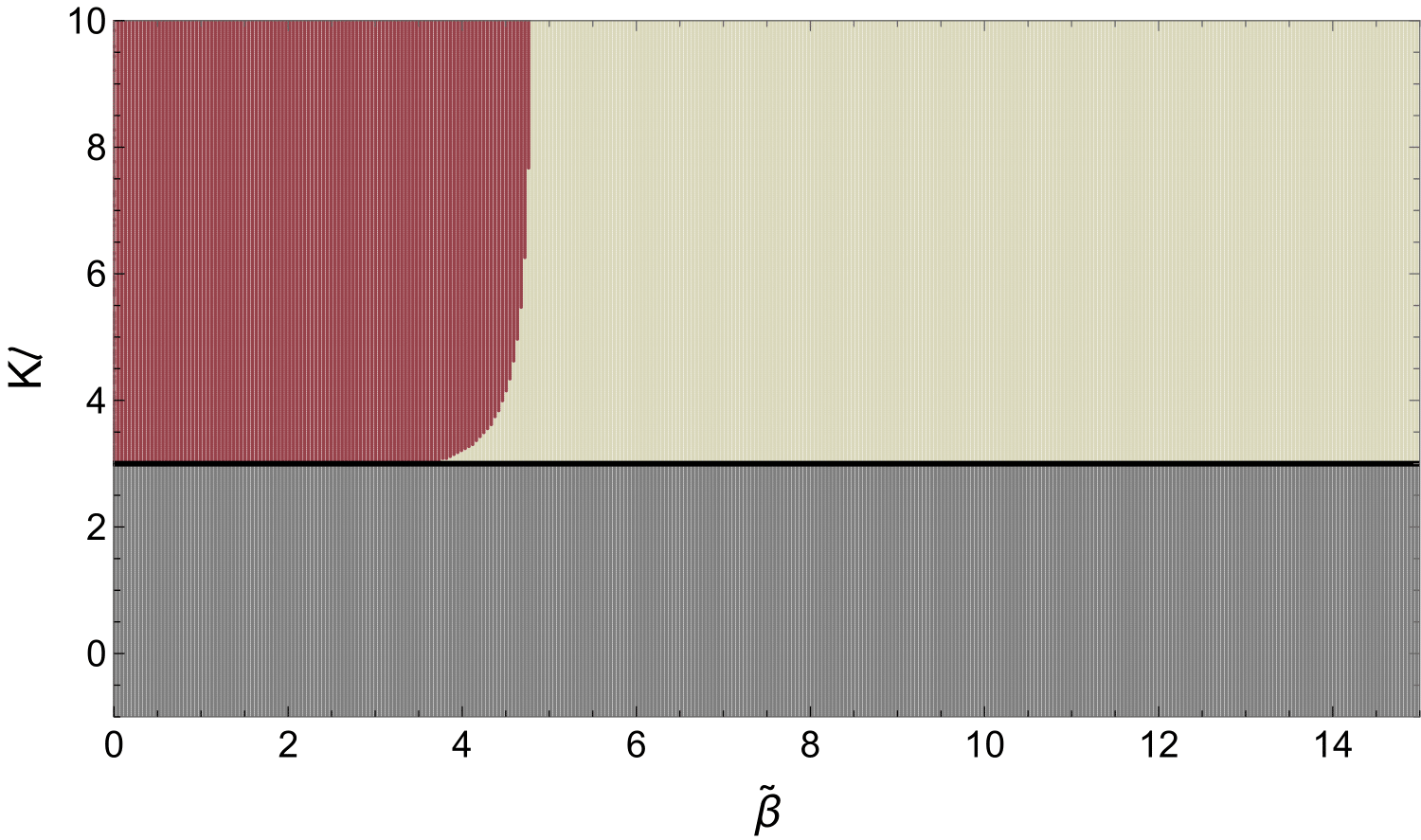}};

\node at (8.75,0) {};

\node[white] at (-4.4,1.991875) {\footnotesize\textbf{(2,0)}};

\node[black] at (2.75,1.991875) {\footnotesize\textbf{(0,0)}};
\node[white] at (0.625,-1.90275) {\footnotesize\textbf{(0,0)}};

\draw[<->,very thick] (7.3+0.775,-0.485) to (7.3+0.775,-0.95-1.9375-0.433);
\draw[<->,very thick] (7.3+0.775,-0.485) to (7.3+0.775,-0.95+1.9375*2.5+0.575);

\node[rotate=-90] at (7.6+0.775,-1.90275) {$n_{\text{TG}} = 0$};
\node[rotate=-90] at (7.6+0.775,1.991875) {$n_{\text{TG}} = 1$};
\end{tikzpicture}}

\caption{$\L < 0$, $k=1$. The number of solutions with spherical horizons across various values of $(\tilde{\beta}, K \ell)$ for $d=3$. Each color corresponds to a particular pair $(n_{\text{BH}},n_{\text{CH}})$, where $n_{\text{BH}}$ is the number of black holes and $n_{\text{CH}}$ is the number of cosmic geometries. The $(2,0)$ and $(0,0)$ regions are separated by a critical curve that approaches $\tilde{\beta} = 2\pi/\sqrt{3}$ as $K\ell \to 3^+$ and $\tilde{\beta} = 8\pi/\sqrt{27}$ as $K\ell \to \infty$. These are the $d = 3$ maximal values of the inverse black hole temperature in AdS with Dirichlet boundary conditions (at infinity) \cite{Hawking:1982dh} and flat space with conformal boundary conditions \cite{Anninos:2023epi}, respectively.}
\label{figs:solutionsAdSSphere}
\end{figure}

%% file: figs/AdSUncharged/texCode/phasePlots/phasesAdSSphere.tex
\begin{figure}
\centering
\begin{tikzpicture}
\node at (0,0){\includegraphics[scale=0.6]{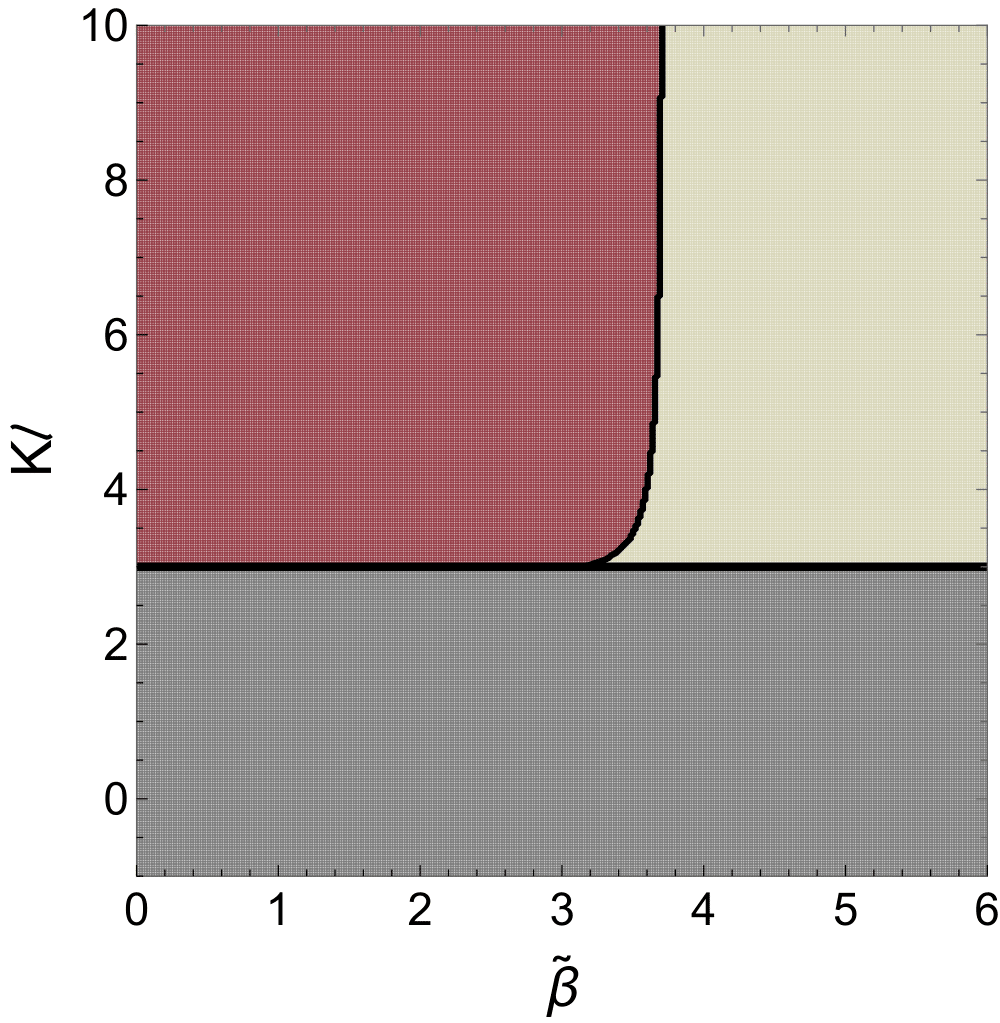}\ \ \ \ 
\includegraphics[scale=0.6]{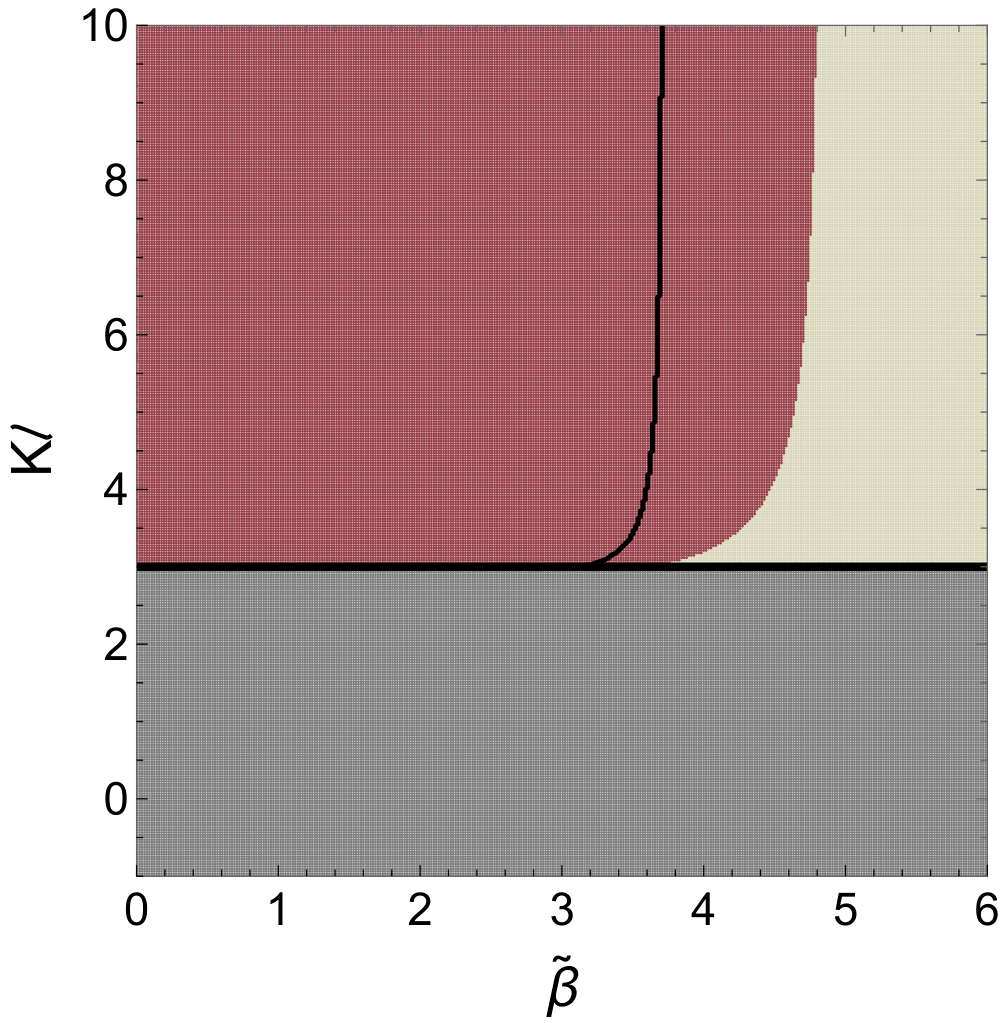}};
\node at (-3.6,-2.125+0.515) {\footnotesize{\textcolor{white}{\textbf{No solutions}}}};
\node at (-4.85,1.2+0.4) {\footnotesize{\shortstack{\textcolor{white}{\textbf{Black hole}}\\\textcolor{white}{\textbf{(stable)}}}}};
\node at (-1.6,1.2+0.4) {\footnotesize{\textcolor{black}{\textbf{Thermal gas}}}};
\end{tikzpicture}
\caption{$\L < 0$, $k=1$. On the left, the phase diagram for spherical horizon topology in $d = 3$; on the right, the boundary of this phase diagram superimposed on the  solution space (Figure \ref{figs:solutionsAdSSphere}).}
\label{figs:phasesAdSSphere}
\end{figure}

%% file: figs/AdSUncharged/texCode/solutionSpaces/solutionsAdSHyper.tex
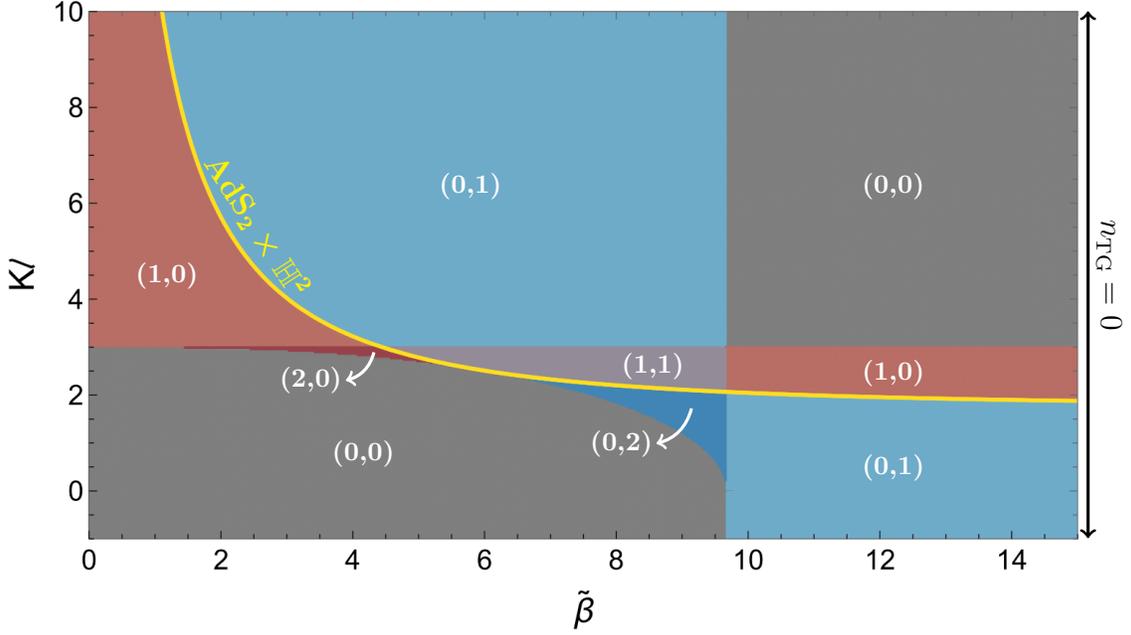
\begin{figure}
\centering
\begin{tikzpicture}[scale=0.933]
\node at (0,0) {\includegraphics[scale=0.75]{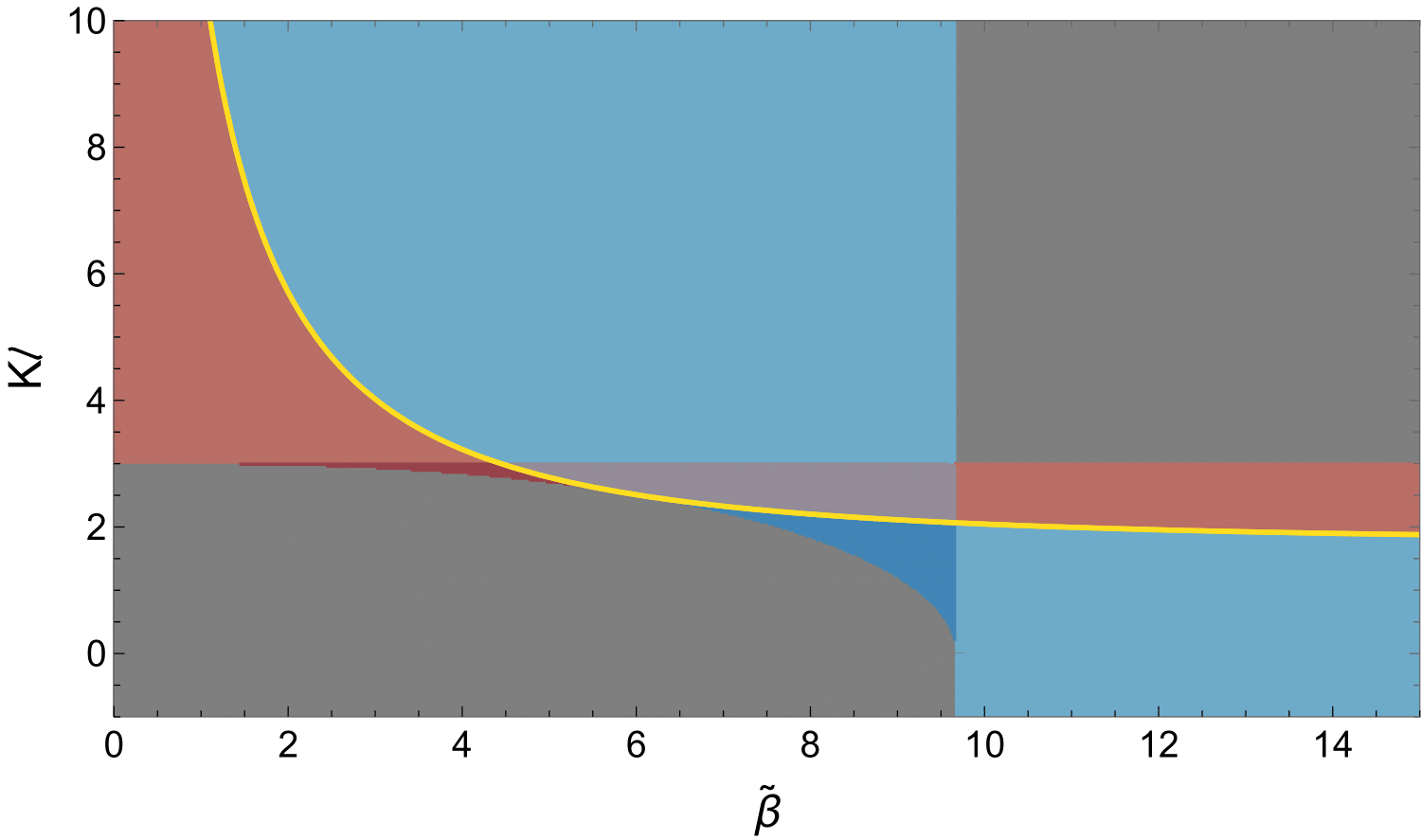}};
\node at (8.75,0) {};

\node[white] at (-1.775-0.75,-2) {\footnotesize\textbf{(0,0)}};
\node[white] at (9.55/2+0.25,1.825) {\footnotesize\textbf{(0,0)}};

\node[white] at (-2.6-3.5+0.775,0.55) {\footnotesize\textbf{(1,0)}};
\node[white] at (9.55/2+0.25,-0.85) {\footnotesize\textbf{(1,0)}};

\node[white] at (9.55/2+0.25,-2.19) {\footnotesize\textbf{(0,1)}};
\node[white] at (-1,1.825) {\footnotesize\textbf{(0,1)}};

\node[white] at (1.6,-0.75) {\footnotesize\textbf{(1,1)}};

\draw[->,very thick,white] (2.15,-1.35) to[bend left] (2.15-0.5,-1.35-0.5);
\node[white] at (2.15-0.5-0.5,-1.35-0.5) {\footnotesize\textbf{(0,2)}};

\draw[->,very thick,white] (-2.375,-0.55) to[bend left] (-2.275-0.5,-0.45-0.5);
\node[white] at (-2.275-0.5-0.5,-0.45-0.5) {\footnotesize\textbf{(2,0)}};

\draw[<->,very thick] (7.3+0.5,-0.95-1.9375-0.3125) to (7.3+0.5,-0.95+1.9375*2.5+0.42);

\node[rotate=-90] at (7.6+0.5,0.556875) {$n_{\text{TG}} = 0$};

\node[yellow,rotate=-57] at (-4.05,1.25) {{$\boldsymbol{\mathrm{AdS}_2 \times \mathbb{H}^2}$}};
\end{tikzpicture}
\caption{$\L < 0$, $k=-1$. The number of each solution with hyperbolic horizon across various values of $(\tilde{\beta}, K\ell)$ for $d=3$. Each color corresponds to a particular pair $(n_{\text{BH}},n_{\text{CH}})$, where $n_{\text{BH}}$ is the number of black hole solutions and $n_{\text{CH}}$ is the number of cosmic solutions. The gold curve has equation given by \eqref{extremalcurveads2h2}  and represents an extremal solution across which a cosmic and black hole solution interchange.}
\label{fig:solutionsAdSHyper}
\end{figure}

%% file: figs/AdSUncharged/texCode/phasePlots/phasesAdSHyper.tex
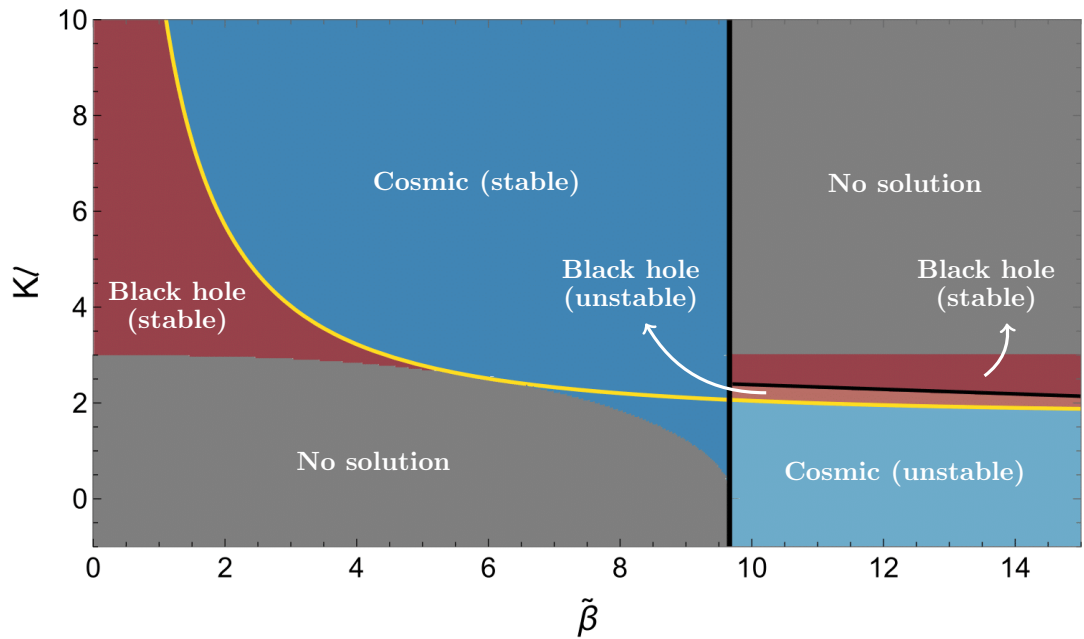
\begin{figure}
\centering
{
\begin{tikzpicture}[scale=0.9]
\node at (0,0) {\includegraphics[scale=0.75]{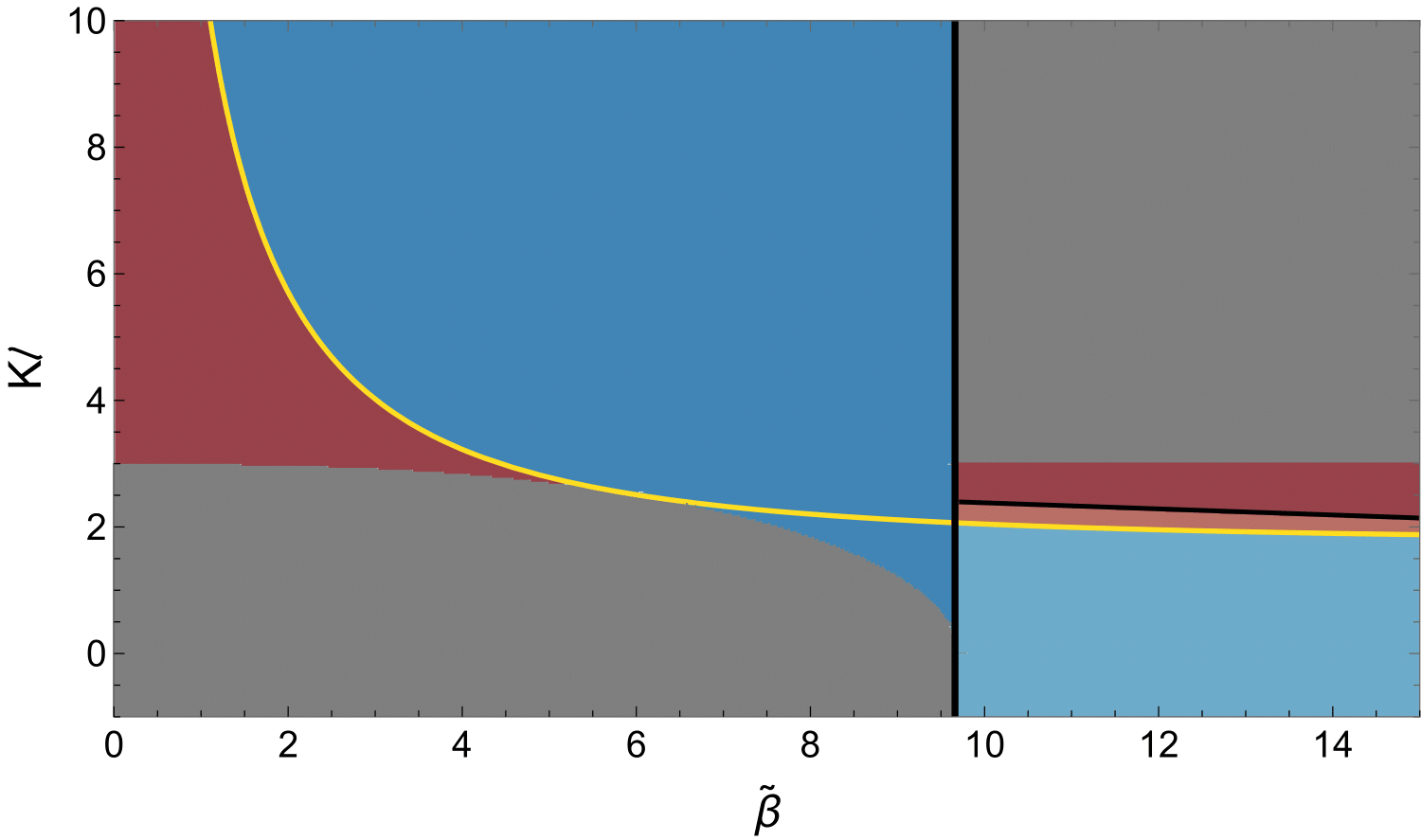}};

\node at (8.75,0) {};

\node[black] at (-1.775-0.75,-2.05) {\footnotesize\textcolor{white}{\textbf{No solution}}};
\node[black] at (9.55/2+0.55,2.05) {\footnotesize\textcolor{white}{\textbf{No solution}}};

\node[white] at (-2.6-3.6+0.775,0.2) {\shortstack{\footnotesize{\textcolor{white}{\textbf{Black hole}}}\\\footnotesize{\textcolor{white}{\textbf{(stable)}}}}};

\node[white] at (9.55/2+0.55,-2.24) {\footnotesize\textbf{Cosmic (unstable)}};
\node[white] at (-1,2.05) {\footnotesize\textbf{Cosmic (stable)}};

\node (A) at (9.55/2+1.75,-0.85+1.375) {\shortstack{\footnotesize{\textcolor{white}{\textbf{Black hole}}}\\\footnotesize{\textcolor{white}{\textbf{(stable)}}}}};
\node (B) at (9.55/2-3.5,-0.85+1.375) {\shortstack{\footnotesize{\textcolor{white}{\textbf{Black hole}}}\\\footnotesize{\textcolor{white}{\textbf{(unstable)}}}}};

\draw[->,very thick,white] (9/2+2,-0.85+0.05) to[bend right] (A);
\draw[->,very thick,white] (9.55/2-1.5,-0.85-0.2) to[bend left] (B);

\end{tikzpicture}}

\caption{$\L < 0$, $k=-1$. the phase diagram showing the dominant phases and their stability corresponding to the solution space for hyperbolic horizons in $d=3$ (Figure \ref{fig:solutionsAdSHyper}).}
\label{figs:phasesAdSHyper}
\end{figure}

%% file: figs/AdSUncharged/texCode/solutionSpaces/solutionsAdSPlanar.tex
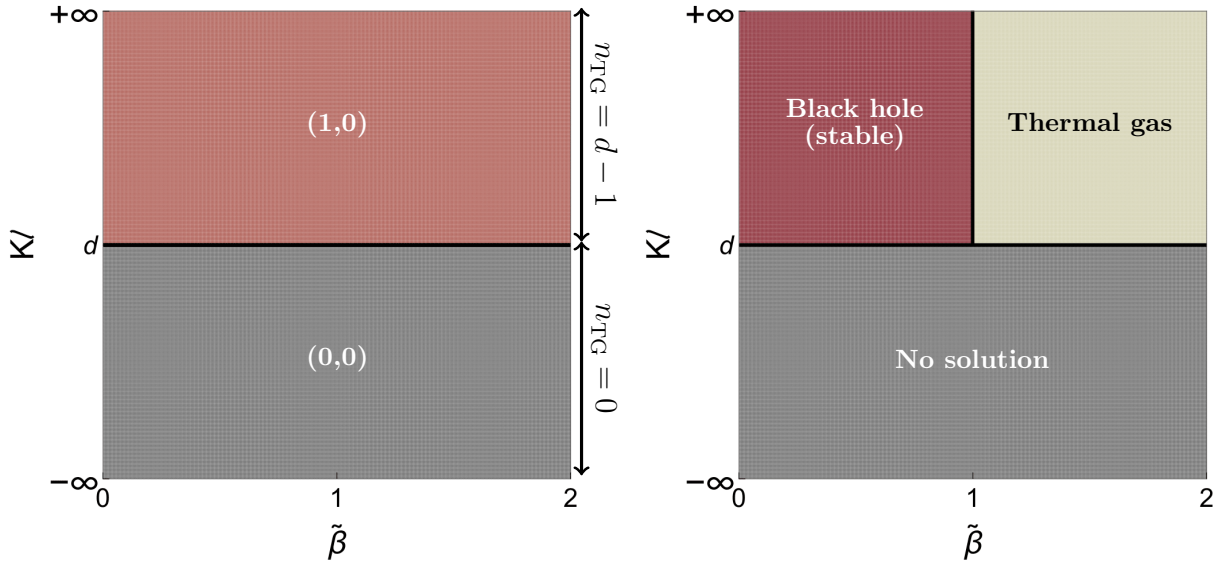
\begin{figure}
\centering
\begin{tikzpicture}
\node at (0,0){\includegraphics[scale=0.6]{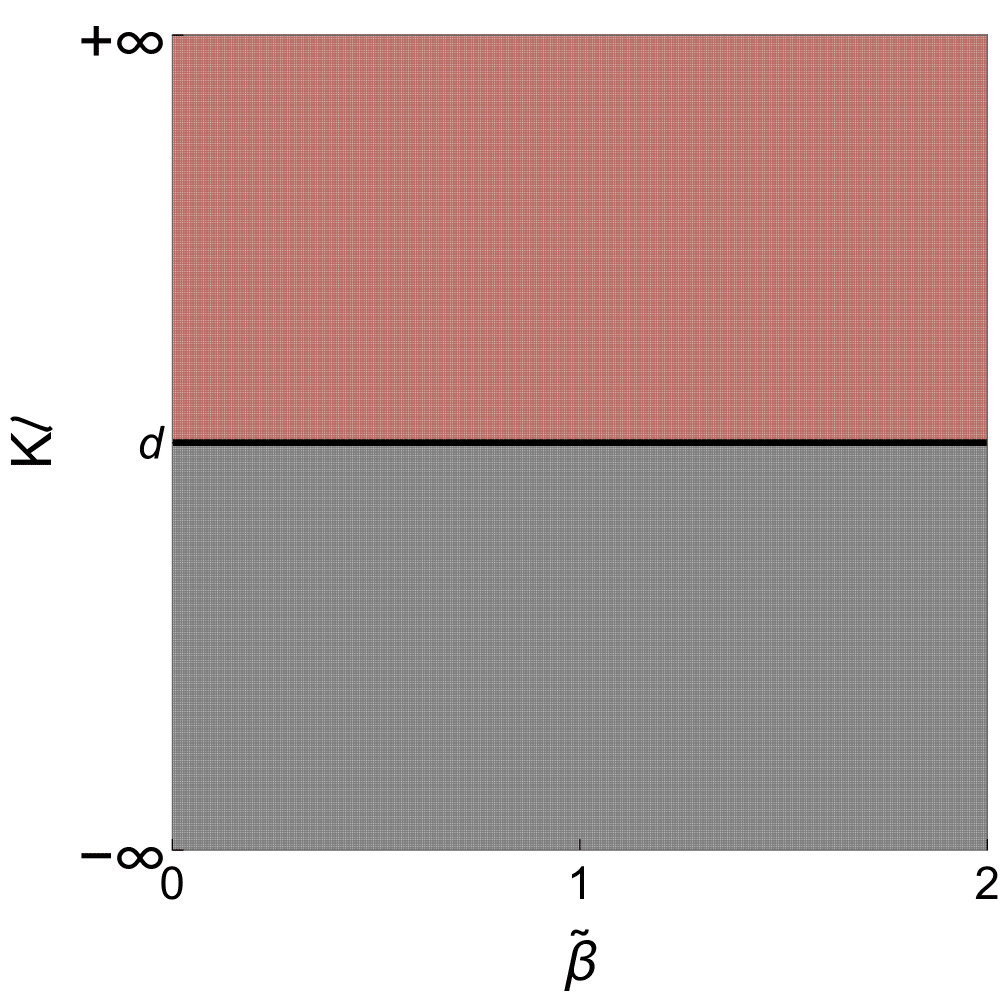}\ \ \ \ \ \ 
\includegraphics[scale=0.6]{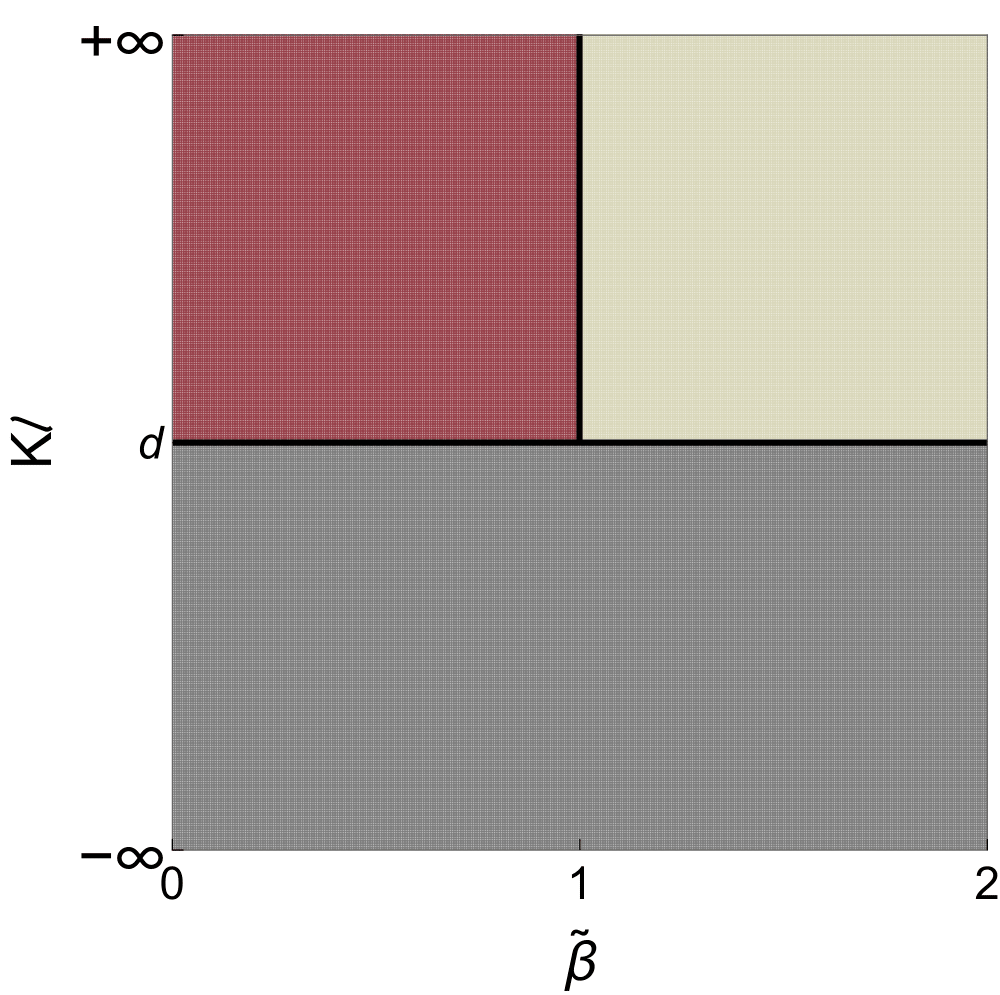}};

\node[white] at (-3.6125,2.015) {$\footnotesize\textbf{(1,0)}$};
\node[white] at (-3.6125,-1.095) {$\footnotesize\textbf{(0,0)}$};

\node at (8.4425-5.175,2.015) {\shortstack{\footnotesize{\textcolor{white}{\textbf{Black hole}}}\\\footnotesize{\textcolor{white}{\textbf{(stable)}}}}};
\node at (8.4425-2.05,2.015) 
{\footnotesize{\textcolor{black}{\textbf{Thermal gas}}}};
\node at (8.4425-3.6125,-1.095) 
{\footnotesize{\textcolor{white}{\textbf{No solution}}}};

\draw[<->,very thick] (-3.74+3.375,3.57) to (-3.74+3.375,0.46);
\draw[<->,very thick] (-3.74+3.375,0.46) to (-3.74+3.375,-2.65);

\node[rotate=-90] at (-3.44+3.375,2.015) {$n_{\text{TG}} = d-1$};
\node[rotate=-90] at (-3.44+3.375,-1.095) {$n_{\text{TG}} = 0$};

%\node at (-5.175,-2.675) {$\circ$};
%\node at (-2.05,-2.675) {$\circ$};
%\node at (8.4425-5.175,-2.675) {$\circ$};
%\node at (8.4425-2.05,-2.675) {$\circ$};
%\node at (8.4425-3.6125,-2.675) {$\circ$};
\end{tikzpicture}
\caption{$\L < 0$, $k=0$, toroidal compactification.  On the left, we have the number of solutions with horizons  across various values of $(\tilde{\beta},K\ell)$. Each color corresponds to a particular pair $(n_{\text{BH}},n_{\text{CH}})$, where $n_{\text{BH}}$ is the number of black holes and $n_{\text{CH}}$ is the number of cosmic geometries.  On the right, we have the phase diagram. There is a first-order phase transition between the black hole and the AdS soliton at $\tilde{\b}/\tilde{L}_{\text{min}} = 1$, consistent with modular invariance, and there are no solutions below $K\ell = d$.}
\label{figs:solutionsAdSPlanar}
\end{figure}

%% file: figs/dSUncharged/texCode/penroseDiagram/penrosedS.tex
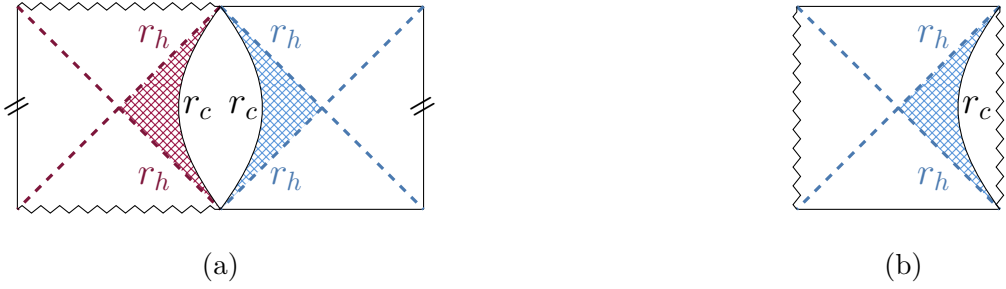
\begin{figure}
\vspace{-5mm}    \centering
    % First TikZ figure
    \begin{subfigure}{0.45\textwidth}
        \centering
       \begin{tikzpicture}[scale=1.35]
\draw[draw=none,pattern = crosshatch,pattern color=penrosered] (1,1) .. controls (0.45,0.25) and (0.45,-0.25) .. (1,-1) to (0,0) -- cycle;
\draw[draw=none,pattern = crosshatch,pattern color=penroseblue] (2-1,1) .. controls (2-0.45,0.25) and (2-0.45,-0.25) .. (2-1,-1) to (2-0,0) -- cycle;

\node[rotate=30] at (-1,0) {$\boldsymbol{=}$};
\node[rotate=30] at (3,0) {$\boldsymbol{=}$};

\draw[-] (1,-1) to (3,-1);
\draw[-] (1,1) to (3,1);
\draw[-] (-1,1) to (-1,-1);
\draw[-] (3,1) to (3,-1);
%\draw[-] (-1,-1) to (-3,-1);
%\draw[-] (-1,1) to (-3,1);
% outer horizons
\draw[-,penrosered!80!black,dashed,very thick] (-1,1) to (0,0) to (-1,-1);
\draw[-,penrosered!80!black,dashed,very thick] (1,1) to (0,0) to (1,-1);

\draw[-,penroseblue!80!black,dashed,very thick] (1,1) to (2,0) to (1,-1);
\draw[-,penroseblue!80!black,dashed,very thick] (3,1) to (2,0) to (3,-1);

%\draw[-,penroseblue!80!black,dashed,very thick] (-1,1) to (-2,0) to (-1,-1);
%\draw[-,penroseblue!80!black,dashed,very thick] (-3,1) to (-2,0) to (-3,-1);

% singularities
\draw[-,decorate,decoration={zigzag,amplitude=0.5mm,segment length=2.5mm}] (-1,-1) to (1,-1);
\draw[-,decorate,decoration={zigzag,amplitude=0.5mm,segment length=2.5mm}] (-1,1) to (1,1);
% labels
\node[penrosered!80!black] at (0.35,0.7) {\large$r_h$};
\node[penrosered!80!black] at (0.35,-0.7) {\large$r_h$};

\node[penroseblue!80!black] at (2-0.35,0.7) {\large$r_h$};
\node[penroseblue!80!black] at (2-0.35,-0.7) {\large$r_h$};

\draw[-] (1,1) .. controls (0.45,0.25) and (0.45,-0.25) .. (1,-1);
\draw[-] (2-1,1) .. controls (2-0.45,0.25) and (2-0.45,-0.25) .. (2-1,-1);

\node at (0.775,0) {\large$r_c$};
\node at (2-0.775,0) {\large$r_c$};

%\node[] at (3.25,0) {\LARGE$\cdots$};
%\node[] at (-3.25,0) {\LARGE$\cdots$};

\node at (0,-1.5) {};
\node at (0,1.5) {};
\end{tikzpicture}\vspace{-5mm}
        \caption{}
    \end{subfigure}
    \hfill
    % Second TikZ figure
    \begin{subfigure}{0.45\textwidth}
        \centering
      \begin{tikzpicture}[scale=1.35]
\draw[draw=none,pattern = crosshatch,pattern color=penroseblue] (1,1) .. controls (0.45,0.25) and (0.45,-0.25) .. (1,-1) to (0,0) -- cycle;

% conf boundaries
\draw[-] (-1,-1) to (1,-1);
\draw[-] (-1,1) to (1,1);
% outer horizons
\draw[-,penroseblue!80!black,dashed,very thick] (-1,1) to (0,0) to (-1,-1);
\draw[-,penroseblue!80!black,dashed,very thick] (1,1) to (0,0) to (1,-1);
% singularities
\draw[-,decorate,decoration={zigzag,amplitude=0.5mm,segment length=2.5mm}] (-1,-1) to (-1,1);
\draw[-,decorate,decoration={zigzag,amplitude=0.5mm,segment length=2.5mm}] (1,-1) to (1,1);
% labels
\node[penroseblue!80!black] at (0.35,0.7) {\large$r_h$};
\node[penroseblue!80!black] at (0.35,-0.7) {\large$r_h$};

\draw[-] (1,1) .. controls (0.45,0.25) and (0.45,-0.25) .. (1,-1);

\node at (0.775,0) {\large$r_c$};

\node at (0,-1.5) {};
\node at (0,1.5) {};
\end{tikzpicture}
     \vspace{-5mm}   \caption{}
    \end{subfigure}

\caption{(a) The Penrose diagram for the $\m > 0$ spherical dS black hole. This geometry accommodates both black holes and cosmic solutions, respectively represented by the red and the blue patches. (b) The Penrose diagram for the $\m < 0$ spherical, planar and hyperbolic dS black holes. The only solutions consistent with conformal boundary conditions are cosmic patches, shown in blue.}
\label{figs:penrosedS}
\end{figure}

%% file: figs/dSUncharged/texCode/solutionSpaces/solutionsdSSphere.tex
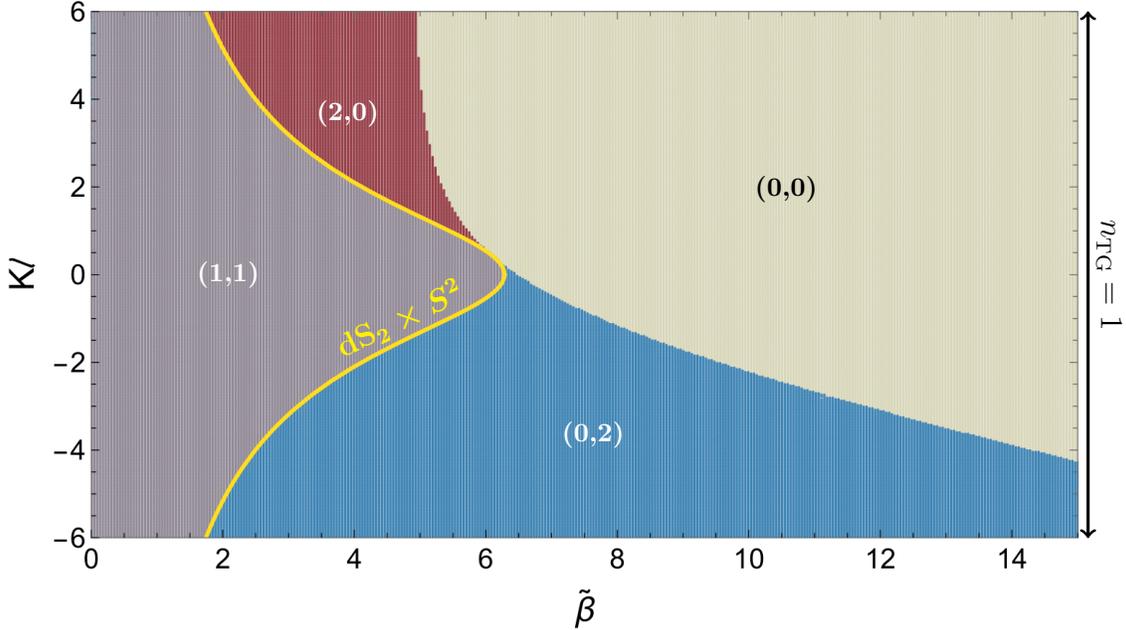
\begin{figure}
\centering
\begin{tikzpicture}[scale=0.933]
\node at (0,0) {\includegraphics[scale=0.75]{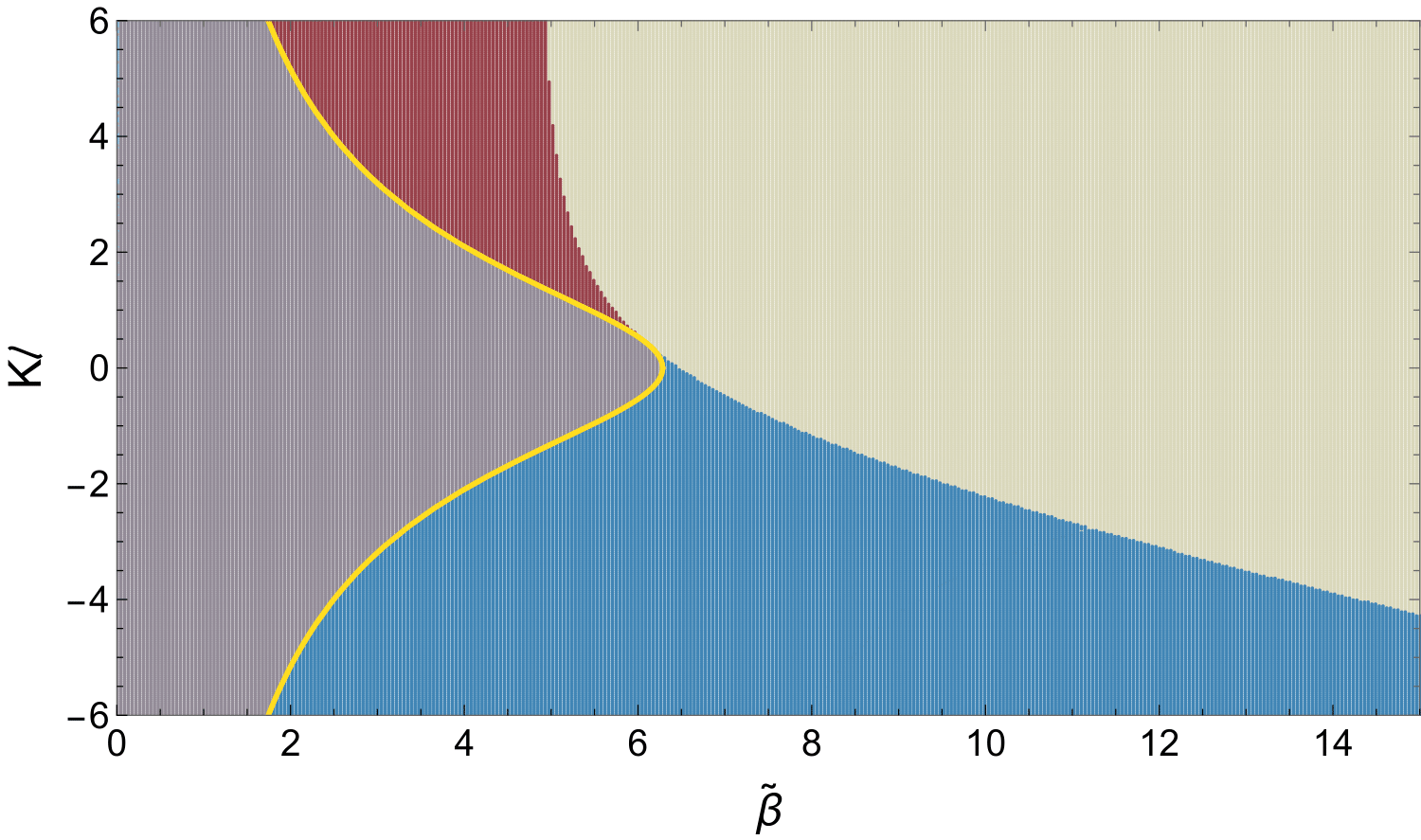}};
\node at (8.75,0) {};

\node[black] at (3.5,1.775) {\footnotesize\textbf{(0,0)}};

\node[white] at (-2.75,2.85) {\footnotesize\textbf{(2,0)}};

\node[white] at (-6.9/2-1,0.56) {\footnotesize\textbf{(1,1)}};

\node[white] at (0.75,-1.725) {\footnotesize\textbf{(0,2)}};

\draw[<->,very thick] (7.3+0.5,-0.95-1.9375-0.3125) to (7.3+0.5,-0.95+1.9375*2.5+0.42);

\node[rotate=-90] at (7.6+0.5,0.556875) {$n_{\text{TG}} = 1$};
\node[yellow,rotate=25.5] at (-2,-0.075) {{$\boldsymbol{\mathrm{dS}_2 \times S^2}$}};
\end{tikzpicture}
\caption{ $\L > 0$, $k=1$. The number of each solution with spherical horizon across various values of $(\tilde{\beta},K\ell)$ for $d=3$. Each color corresponds to a particular pair $(n_{\text{BH}},n_{\text{CH}})$, where $n_{\text{BH}}$ is the number of black hole solutions and $n_{\text{CH}}$ is the number of cosmic solutions.  As usual, the extremal solution dS$_2 \times S^2$ sits at the transition between cosmic and black-hole solutions, with functional form given by \eqref{extremalcurveds2s2}. We also have a thermal gas solution for all $K\ell \in \mathbb{R}$ corresponding to a patch of empty dS space. This plot was found previously in \cite{Anninos:2024wpy}.}
\label{figs:solutionsdSSphere}
\end{figure}

%% file: figs/dSUncharged/texCode/phasePlots/phasesdSSphere.tex
\begin{figure}
\centering
\begin{tikzpicture}
\node at (0,0){\includegraphics[scale=0.6]{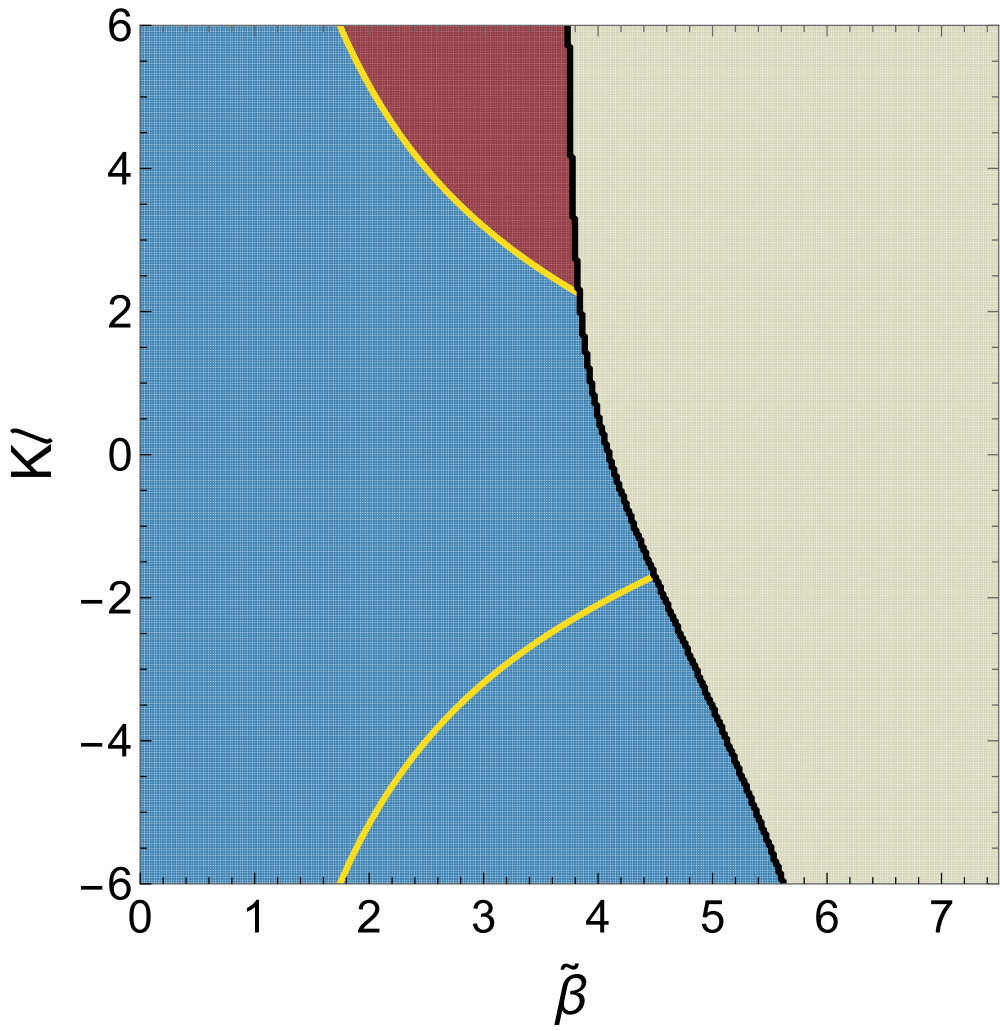}\ \ \ \ 
\includegraphics[scale=0.6]{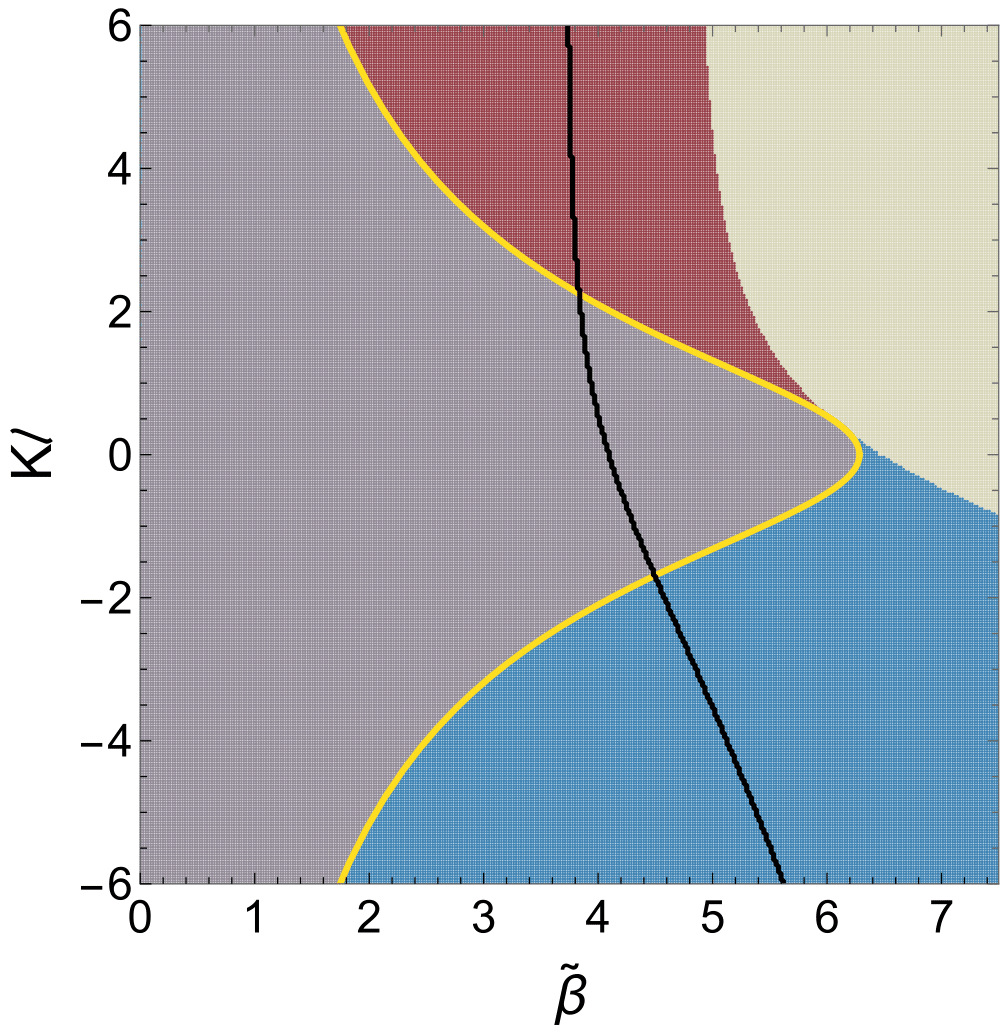}};
\node at (-5.65,-1.2){\shortstack{\footnotesize{\textcolor{white}{\textbf{Cosmic}}}\\\footnotesize{\textcolor{white}{\textbf{(stable)}}}}};

\node (A) at (-5.65,2.05) {\footnotesize{\shortstack{\textcolor{white}{\textbf{Black hole}}\\\textcolor{white}{\textbf{(stable)}}}}};
\draw[->,very thick,white] (-4.25,2.75) to[bend right] (A);

\node at (-1.8,0.4375) {\footnotesize{\textcolor{black}{\textbf{Thermal gas}}}};
\end{tikzpicture}
\caption{$\L > 0$, $k=1$. On the left, the phase diagram for $d = 3$; on the right, the boundary of this phase diagram superimposed on the spherical solution space (Figure \ref{figs:solutionsdSSphere}).}
\label{figs:phasesdSSphere}
\end{figure}

%% file: figs/dSUncharged/texCode/solutionSpaces/solutionsdSHyper.tex
\begin{figure}[h]
\centering
\begin{tikzpicture}
\node at (0,0){\includegraphics[scale=0.6]{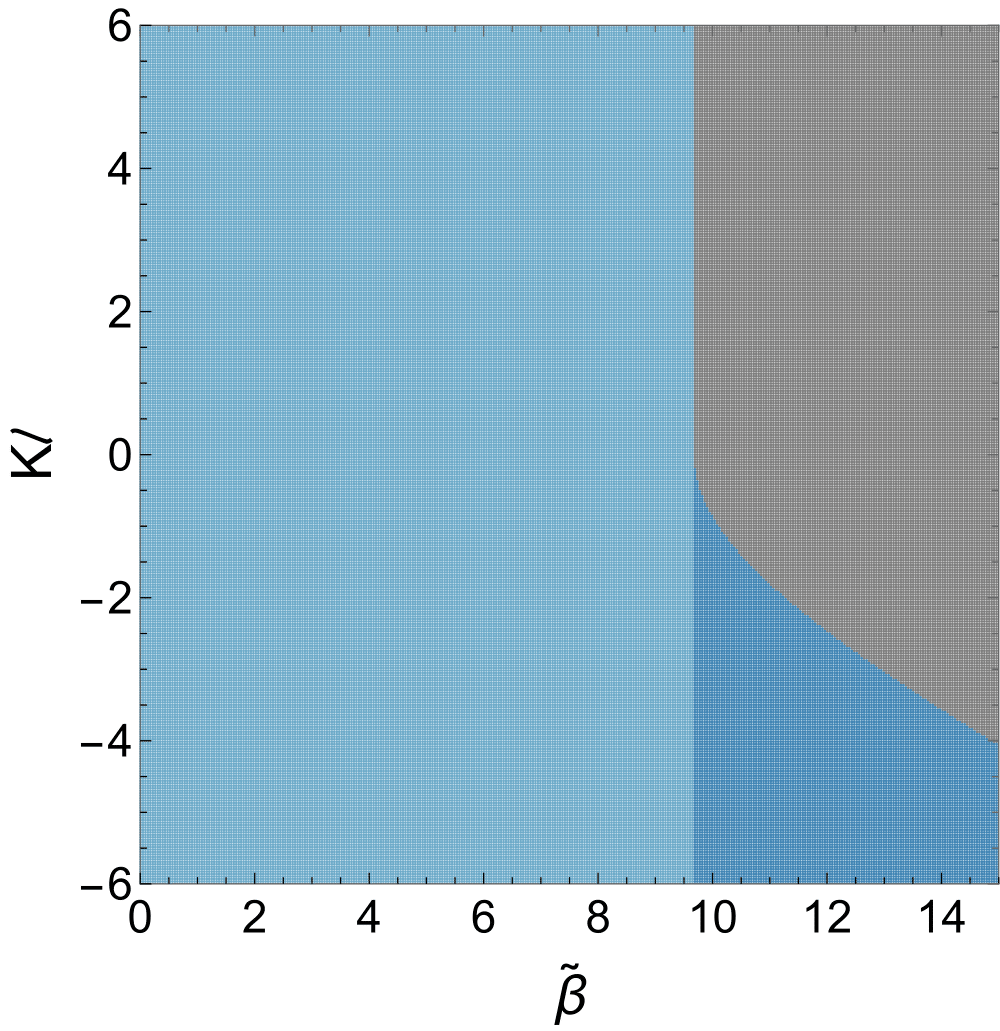}\ \ \ \ \ \ 
\includegraphics[scale=0.6]{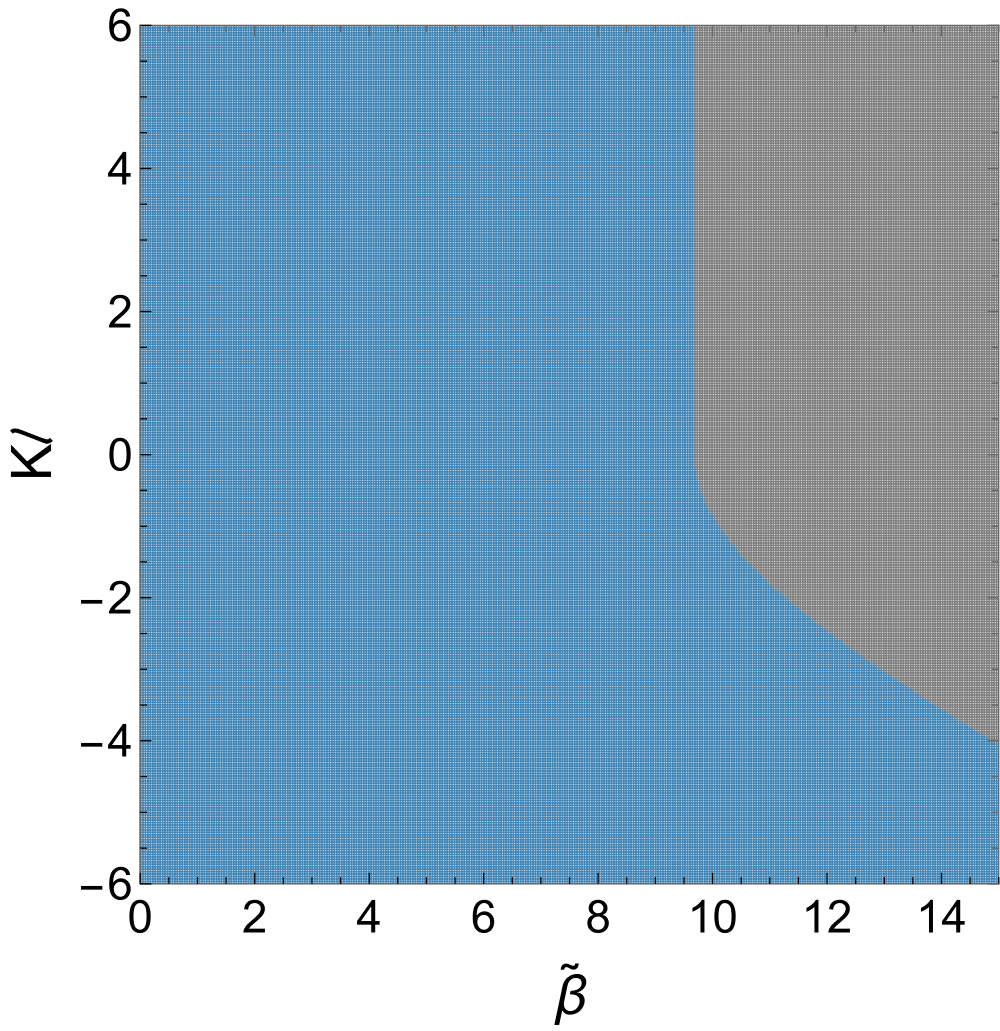}};

\node[white] at (-1.57,1.55) {$\footnotesize\textbf{(0,0)}$};
\node[white] at (-4.85,0.45) {$\footnotesize\textbf{(0,1)}$};
\node[white] at (-1.62,-2.3+0.55/2) {$\footnotesize\textbf{(0,2)}$};

\node at (8.45-1.57,1.55) {\footnotesize{\textcolor{white}{\textbf{No solution}}}};
\node[white] at (8.45-4.85,0.45) {\footnotesize\shortstack{\textcolor{white}{\textbf{Cosmic}}\\\textcolor{white}{\textbf{(stable)}}}};

\draw[<->,very thick] (-3.74+3.3+0.15,3.74) to (-3.74+3.3+0.15,-2.81);

\node[rotate=-90] at (-3.44+3.3+0.15,0.465) {$n_{\text{TG}} = 0$};

%\node at (-6.95,1.55) {$\circ$};
%\node at (-6.95,0.45) {$\circ$};
%\node at (-6.95,-1.75) {$\circ$};
%\node at (-4.85,-2.8) {$\circ$};
%\node at (-1.57,-2.8) {$\circ$};
%\node at (8.45-6.95,1.55) {$\circ$};

\end{tikzpicture}
\caption{$\L > 0$, $k=-1$. On the left is the solution space for $d=3$. Each color corresponds to a particular pair $(n_{\text{BH}},n_{\text{CH}})$, where $n_{\text{BH}}$ is the number of black hole solutions and $n_{\text{CH}}$ is the number of cosmic solutions. We only have cosmic solutions, and for any $K\ell$ we have no solutions at sufficiently low temperature. On the right, the phase diagram showing the dominant phases and their stability for this solution space. Due to the lack of a thermal gas solution, we have no candidate solution for part of the phase diagram.}
\label{fig:solutionsdSHyper}
\end{figure}
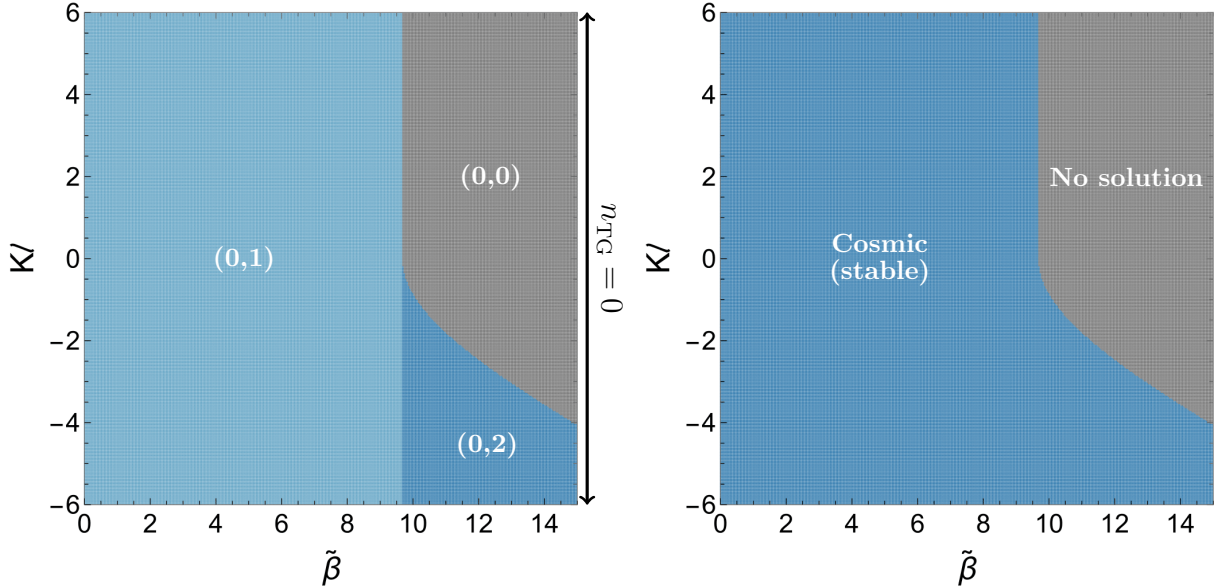

%% file: figs/dSUncharged/texCode/solutionSpaces/solutionsdSPlanar.tex
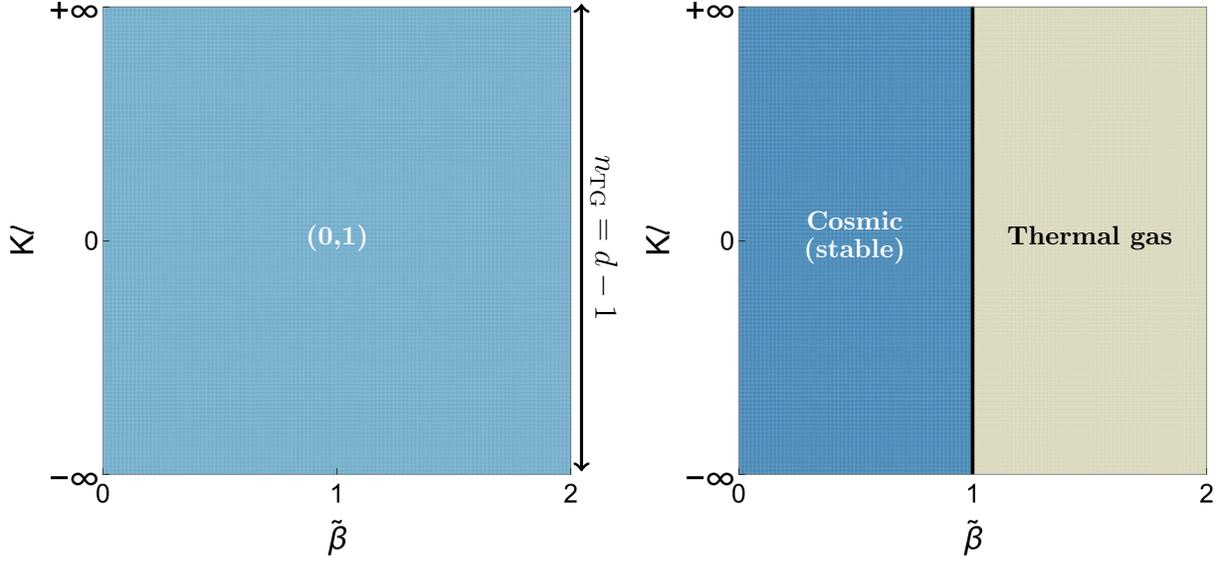
\begin{figure}
\centering
\begin{tikzpicture}
\node at (0,0){\includegraphics[scale=0.6]{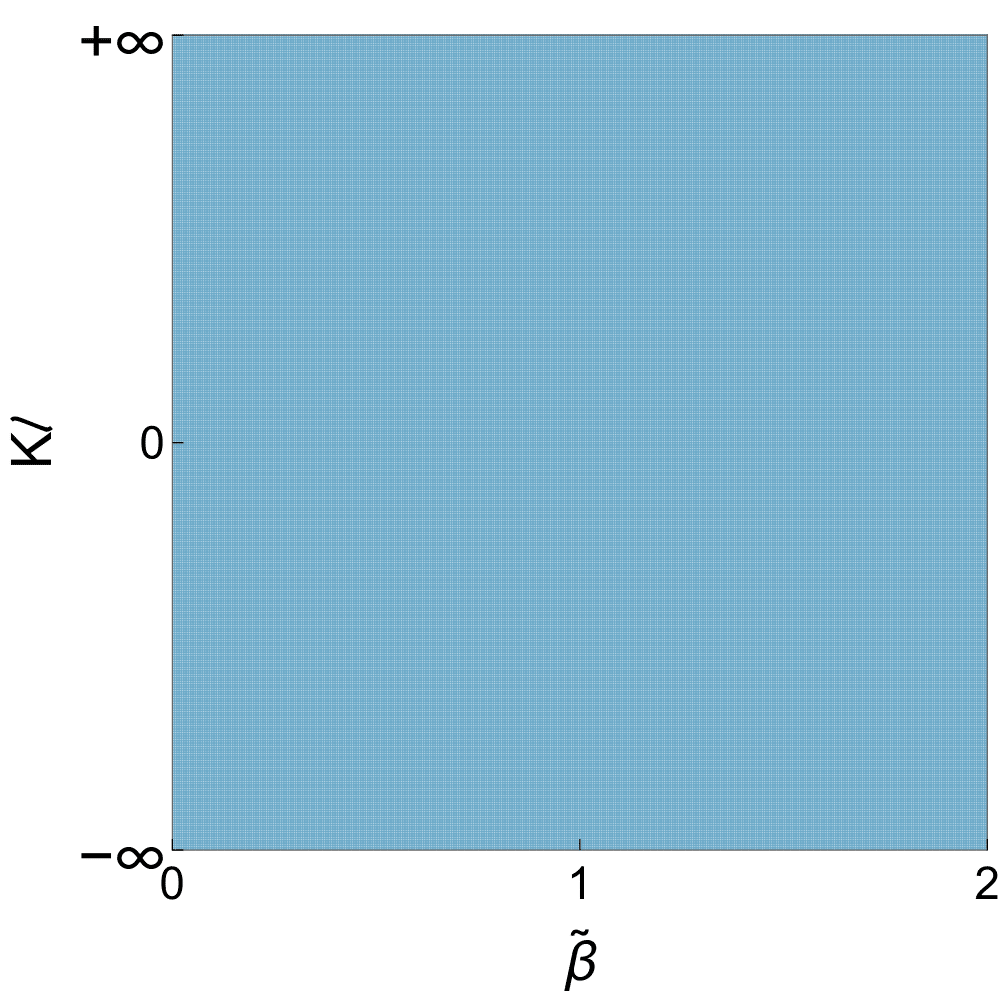}\ \ \ \ \ \ 
\includegraphics[scale=0.6]{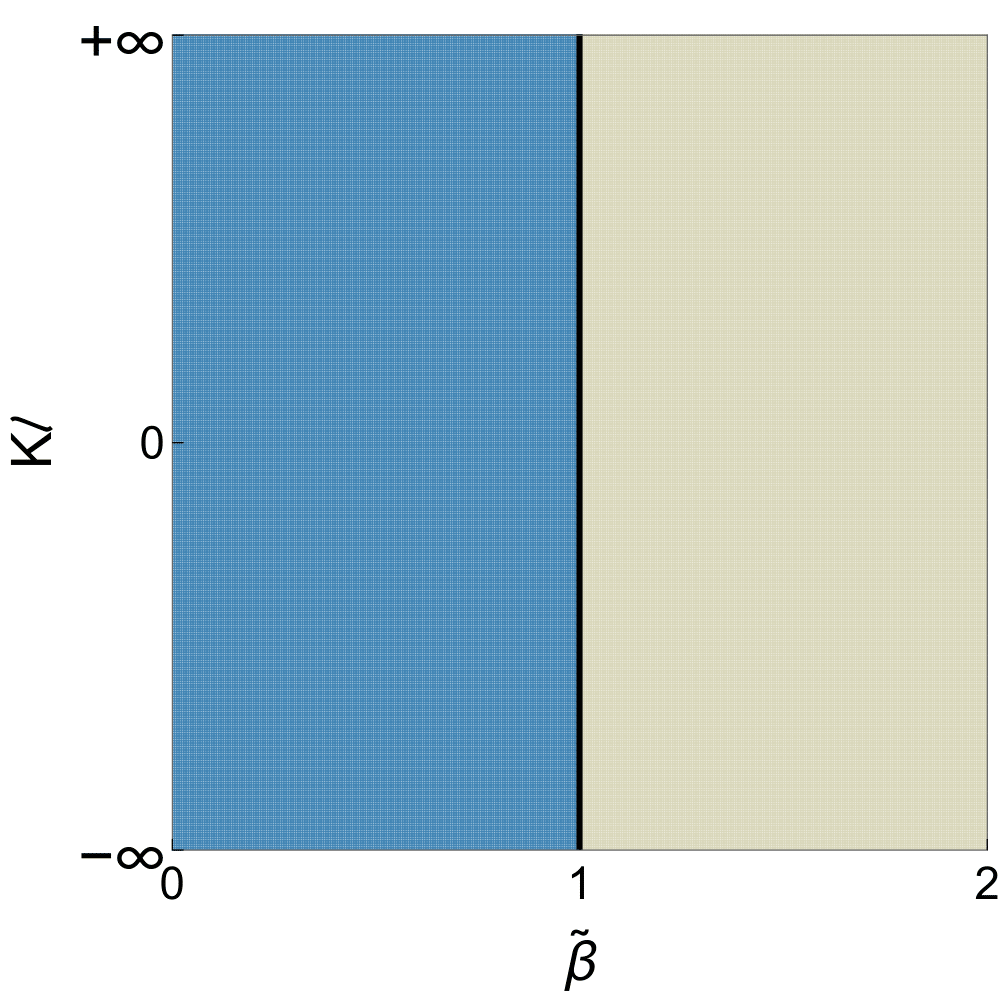}};

\node[white] at (-3.6125,0.46) {$\footnotesize\textbf{(0,1)}$};
\node at (8.4425-5.175,0.46) {\shortstack{\footnotesize{\textcolor{white}{\textbf{Cosmic}}}\\\footnotesize{\textcolor{white}{\textbf{(stable)}}}}};
%\node at (4.567+1.575,0.46) {\shortstack{\footnotesize{\textcolor{black}{\textbf{Ground-state}}}\\\footnotesize{\textcolor{black}{\textbf{thermal gas}}}}};
\node at (8.4425-2.05,0.46) 
{\footnotesize{\textcolor{black}{\textbf{Thermal gas}}}};

\draw[<->,very thick] (-3.74+3.375,3.57) to (-3.74+3.375,-2.65);

\node[rotate=-90] at (-3.44+3.375,0.46) {$n_{\text{TG}} = d-1$};

%\node at (-5.175,-2.675) {$\circ$};
%\node at (-2.05,-2.675) {$\circ$};
%\node at (8.4425-5.175,-2.675) {$\circ$};
%\node at (8.4425-2.05,-2.675) {$\circ$};
%\node at (8.4425-3.6125,-2.675) {$\circ$};

\end{tikzpicture}
\caption{$\L > 0$, $k=0$, toroidal compactification.  On the left, we have the number of solutions with horizons  across various values of $(\tilde{\beta},K\ell)$. We have one cosmic horizon solution and $d-1$ thermal gas solutions for all $K\ell \in d$. On the right, we have the phase diagram. There is a first-order phase transition between the cosmic-horizon solution and the thermal gas at $\tilde{\b}/\tilde{L}_{\text{min}} = 1$, consistent with modular invariance.}
\label{fig:phasesdSPlanar}
\end{figure}

%% file: figs/penroseAnalyticCont/penroseAnalyticCont.tex
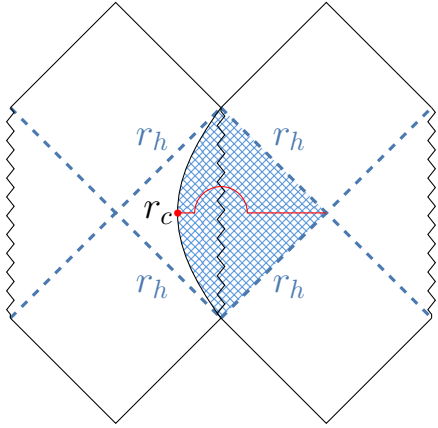
\begin{figure}
\centering
\begin{tikzpicture}[scale=1.4]

\draw[draw=none,pattern = crosshatch,pattern color=penroseblue] (1,1) .. controls (0.45,0.25) and (0.45,-0.25) .. (1,-1) to (2,0) -- cycle;

\draw[-,penroseblue!80!black,dashed,very thick] (-1,-1) to (1,1);
\draw[-,penroseblue!80!black,dashed,very thick] (-1,1) to (1,-1);

\draw[-,penroseblue!80!black,dashed,very thick] (2-1,-1) to (2+1,1);
\draw[-,penroseblue!80!black,dashed,very thick] (2-1,1) to (2+1,-1);

\draw[-] (-1,-1) to (0,-2) to (1,-1);
\draw[-] (-1,1) to (0,2) to (1,1);

\draw[-] (2-1,-1) to (2+0,-2) to (2+1,-1);
\draw[-] (2-1,1) to (2+0,2) to (2+1,1);

\node[penroseblue!80!black] at (0.35,0.7) {\large$r_h$};
\node[penroseblue!80!black] at (0.35,-0.7) {\large$r_h$};

\node[penroseblue!80!black] at (2-0.35,0.7) {\large$r_h$};
\node[penroseblue!80!black] at (2-0.35,-0.7) {\large$r_h$};

\draw[-,decorate,decoration={zigzag,amplitude=0.5mm,segment length=2.5mm}] (1,-1) to (1,1);
\draw[-,decorate,decoration={zigzag,amplitude=0.5mm,segment length=2.5mm}] (-1,1) to (-1,-1);
\draw[-,decorate,decoration={zigzag,amplitude=0.5mm,segment length=2.5mm}] (4-1,1) to (4-1,-1);

\draw[-] (1,1) .. controls (0.45,0.25) and (0.45,-0.25) .. (1,-1);

\node[red] at (0.59,0) {\tiny$\bullet$};
\draw[-,red] (0.59,0) to (0.75,0) arc (180:0:0.25) to (2,0);

\node at (0.58-0.175,0) {\large$r_c$};

\end{tikzpicture}
\caption{We interpret cosmic solutions with a sigularity at $r=0$ and boundary at $r_c < 0$ as living on a ``doubled" Penrose diagram glued across the singularity. The above figure is an example of the situation for vanishing cosmological constant and nonzero electric potential. We deform the contour from $r_h$ to $r_c$ into the complex $r$-plane to avoid the $r=0$ singularity.}
\label{figs:penroseDouble}
\end{figure}

%% file: beyondFlatCBC-draft.bbl
\providecommand{\href}[2]{#2}\begingroup\raggedright\begin{thebibliography}{10}

\bibitem{Anninos:2023epi}
D.~Anninos, D.~A. Galante and C.~Maneerat, \emph{{Gravitational
  observatories}}, \href{https://doi.org/10.1007/JHEP12(2023)024}{\emph{JHEP}
  {\bfseries 12} (2023) 024}
  [\href{https://arxiv.org/abs/2310.08648}{{\ttfamily 2310.08648}}].

\bibitem{Banihashemi:2024yye}
B.~Banihashemi, E.~Shaghoulian and S.~Shashi, \emph{{Flat space gravity at
  finite cutoff}}, \href{https://doi.org/10.1088/1361-6382/ada2d7}{\emph{Class.
  Quant. Grav.} {\bfseries 42} (2025) 035010}
  [\href{https://arxiv.org/abs/2409.07643}{{\ttfamily 2409.07643}}].

\bibitem{York:1972sj}
J.~W. York, Jr., \emph{{Role of conformal three geometry in the dynamics of
  gravitation}}, \href{https://doi.org/10.1103/PhysRevLett.28.1082}{\emph{Phys.
  Rev. Lett.} {\bfseries 28} (1972) 1082}.

\bibitem{York:1986it}
J.~W. York, Jr., \emph{{Black hole thermodynamics and the Euclidean Einstein
  action}}, \href{https://doi.org/10.1103/PhysRevD.33.2092}{\emph{Phys. Rev. D}
  {\bfseries 33} (1986) 2092}.

\bibitem{York:1986lje}
J.~York, \emph{{Boundary terms in the action principles of general
  relativity}}, \href{https://doi.org/10.1007/BF01889475}{\emph{Found. Phys.}
  {\bfseries 16} (1986) 249}.

\bibitem{Anderson:2006lqb}
M.~T. Anderson, \emph{{On boundary value problems for Einstein metrics}},
  \href{https://doi.org/10.2140/gt.2008.12.2009}{\emph{Geom. Topol.} {\bfseries
  12} (2008) 2009} [\href{https://arxiv.org/abs/math/0612647}{{\ttfamily
  math/0612647}}].

\bibitem{An:2021fcq}
Z.~An and M.~T. Anderson, \emph{{The initial boundary value problem and
  quasi-local Hamiltonians in General Relativity}},
  \href{https://arxiv.org/abs/2103.15673}{{\ttfamily 2103.15673}}.

\bibitem{An:2025rlw}
Z.~An and M.~T. Anderson, \emph{{Well-posed geometric boundary data in General
  Relativity, I: Conformal-mean curvature boundary data}},
  \href{https://arxiv.org/abs/2503.12599}{{\ttfamily 2503.12599}}.

\bibitem{Anninos:2024wpy}
D.~Anninos, D.~A. Galante and C.~Maneerat, \emph{{Cosmological observatories}},
  \href{https://doi.org/10.1088/1361-6382/ad5824}{\emph{Class. Quant. Grav.}
  {\bfseries 41} (2024) 165009}
  [\href{https://arxiv.org/abs/2402.04305}{{\ttfamily 2402.04305}}].

\bibitem{Gorbenko:2018oov}
V.~Gorbenko, E.~Silverstein and G.~Torroba, \emph{{dS/dS and $ T\overline{T}
  $}}, \href{https://doi.org/10.1007/JHEP03(2019)085}{\emph{JHEP} {\bfseries
  03} (2019) 085} [\href{https://arxiv.org/abs/1811.07965}{{\ttfamily
  1811.07965}}].

\bibitem{Lewkowycz:2019xse}
A.~Lewkowycz, J.~Liu, E.~Silverstein and G.~Torroba, \emph{{$ T\overline{T} $
  and EE, with implications for (A)dS subregion encodings}},
  \href{https://doi.org/10.1007/JHEP04(2020)152}{\emph{JHEP} {\bfseries 04}
  (2020) 152} [\href{https://arxiv.org/abs/1909.13808}{{\ttfamily
  1909.13808}}].

\bibitem{Coleman:2021nor}
E.~Coleman, E.~A. Mazenc, V.~Shyam, E.~Silverstein, R.~M. Soni, G.~Torroba and
  S.~Yang, \emph{{De Sitter microstates from T$ \overline{T} $ +
  \ensuremath{\Lambda}$_{2}$ and the Hawking-Page transition}},
  \href{https://doi.org/10.1007/JHEP07(2022)140}{\emph{JHEP} {\bfseries 07}
  (2022) 140} [\href{https://arxiv.org/abs/2110.14670}{{\ttfamily
  2110.14670}}].

\bibitem{Banihashemi:2022jys}
B.~Banihashemi and T.~Jacobson, \emph{{Thermodynamic ensembles with
  cosmological horizons}},
  \href{https://doi.org/10.1007/JHEP07(2022)042}{\emph{JHEP} {\bfseries 07}
  (2022) 042} [\href{https://arxiv.org/abs/2204.05324}{{\ttfamily
  2204.05324}}].

\bibitem{Banihashemi:2022htw}
B.~Banihashemi, T.~Jacobson, A.~Svesko and M.~Visser, \emph{{The minus sign in
  the first law of de Sitter horizons}},
  \href{https://doi.org/10.1007/JHEP01(2023)054}{\emph{JHEP} {\bfseries 01}
  (2023) 054} [\href{https://arxiv.org/abs/2208.11706}{{\ttfamily
  2208.11706}}].

\bibitem{Batra:2024kjl}
G.~Batra, G.~B. De~Luca, E.~Silverstein, G.~Torroba and S.~Yang,
  \emph{{Bulk-local dS$_3$ holography: the Matter with $T\bar T+\Lambda_2$}},
  \href{https://arxiv.org/abs/2403.01040}{{\ttfamily 2403.01040}}.

\bibitem{Silverstein:2024xnr}
E.~Silverstein and G.~Torroba, \emph{{Timelike-bounded $dS_4$ holography from a
  solvable sector of the $T^2$ deformation}},
  \href{https://arxiv.org/abs/2409.08709}{{\ttfamily 2409.08709}}.

\bibitem{Liu:2024ymn}
X.~Liu, J.~E. Santos and T.~Wiseman, \emph{{New Well-Posed Boundary Conditions
  for Semi-Classical Euclidean Gravity}},
  \href{https://arxiv.org/abs/2402.04308}{{\ttfamily 2402.04308}}.

\bibitem{Anninos:2024xhc}
D.~Anninos, R.~Arias, D.~A. Galante and C.~Maneerat, \emph{{Gravitational
  Observatories in AdS$_4$}},
  \href{https://arxiv.org/abs/2412.16305}{{\ttfamily 2412.16305}}.

\bibitem{McGough:2016lol}
L.~McGough, M.~Mezei and H.~Verlinde, \emph{{Moving the CFT into the bulk with
  $ T\overline{T} $}},
  \href{https://doi.org/10.1007/JHEP04(2018)010}{\emph{JHEP} {\bfseries 04}
  (2018) 010} [\href{https://arxiv.org/abs/1611.03470}{{\ttfamily
  1611.03470}}].

\bibitem{Giveon:2017nie}
A.~Giveon, N.~Itzhaki and D.~Kutasov, \emph{{$ \mathrm{T}\overline{\mathrm{T}}
  $ and LST}}, \href{https://doi.org/10.1007/JHEP07(2017)122}{\emph{JHEP}
  {\bfseries 07} (2017) 122}
  [\href{https://arxiv.org/abs/1701.05576}{{\ttfamily 1701.05576}}].

\bibitem{Taylor:2018xcy}
M.~Taylor, \emph{{$T \bar{T}$ deformations in general dimensions}},
  \href{https://doi.org/10.4310/ATMP.2023.v27.n1.a2}{\emph{Adv. Theor. Math.
  Phys.} {\bfseries 27} (2023) 37}
  [\href{https://arxiv.org/abs/1805.10287}{{\ttfamily 1805.10287}}].

\bibitem{Hartman:2018tkw}
T.~Hartman, J.~Kruthoff, E.~Shaghoulian and A.~Tajdini, \emph{{Holography at
  finite cutoff with a $T^2$ deformation}},
  \href{https://doi.org/10.1007/JHEP03(2019)004}{\emph{JHEP} {\bfseries 03}
  (2019) 004} [\href{https://arxiv.org/abs/1807.11401}{{\ttfamily
  1807.11401}}].

\bibitem{Gross:2019ach}
D.~J. Gross, J.~Kruthoff, A.~Rolph and E.~Shaghoulian, \emph{{$T\overline{T}$
  in AdS$_2$ and Quantum Mechanics}},
  \href{https://doi.org/10.1103/PhysRevD.101.026011}{\emph{Phys. Rev. D}
  {\bfseries 101} (2020) 026011}
  [\href{https://arxiv.org/abs/1907.04873}{{\ttfamily 1907.04873}}].

\bibitem{Gross:2019uxi}
D.~J. Gross, J.~Kruthoff, A.~Rolph and E.~Shaghoulian, \emph{{Hamiltonian
  deformations in quantum mechanics, $T\bar T$, and the SYK model}},
  \href{https://doi.org/10.1103/PhysRevD.102.046019}{\emph{Phys. Rev. D}
  {\bfseries 102} (2020) 046019}
  [\href{https://arxiv.org/abs/1912.06132}{{\ttfamily 1912.06132}}].

\bibitem{Guica:2019nzm}
M.~Guica and R.~Monten, \emph{{$T\bar T$ and the mirage of a bulk cutoff}},
  \href{https://doi.org/10.21468/SciPostPhys.10.2.024}{\emph{SciPost Phys.}
  {\bfseries 10} (2021) 024}
  [\href{https://arxiv.org/abs/1906.11251}{{\ttfamily 1906.11251}}].

\bibitem{Araujo-Regado:2022gvw}
G.~Araujo-Regado, R.~Khan and A.~C. Wall, \emph{{Cauchy slice holography: a new
  AdS/CFT dictionary}},
  \href{https://doi.org/10.1007/JHEP03(2023)026}{\emph{JHEP} {\bfseries 03}
  (2023) 026} [\href{https://arxiv.org/abs/2204.00591}{{\ttfamily
  2204.00591}}].

\bibitem{Anninos:2022ujl}
D.~Anninos, D.~A. Galante and B.~M\"uhlmann, \emph{{Finite features of quantum
  de Sitter space}},
  \href{https://doi.org/10.1088/1361-6382/acaba5}{\emph{Class. Quant. Grav.}
  {\bfseries 40} (2023) 025009}
  [\href{https://arxiv.org/abs/2206.14146}{{\ttfamily 2206.14146}}].

\bibitem{Anninos:2011af}
D.~Anninos, S.~A. Hartnoll and D.~M. Hofman, \emph{{Static Patch Solipsism:
  Conformal Symmetry of the de Sitter Worldline}},
  \href{https://doi.org/10.1088/0264-9381/29/7/075002}{\emph{Class. Quant.
  Grav.} {\bfseries 29} (2012) 075002}
  [\href{https://arxiv.org/abs/1109.4942}{{\ttfamily 1109.4942}}].

\bibitem{Anninos:2011zn}
D.~Anninos, T.~Anous, I.~Bredberg and G.~S. Ng, \emph{{Incompressible Fluids of
  the de Sitter Horizon and Beyond}},
  \href{https://doi.org/10.1007/JHEP05(2012)107}{\emph{JHEP} {\bfseries 05}
  (2012) 107} [\href{https://arxiv.org/abs/1110.3792}{{\ttfamily 1110.3792}}].

\bibitem{Carrozza:2021gju}
S.~Carrozza and P.~A. Hoehn, \emph{{Edge modes as reference frames and boundary
  actions from post-selection}},
  \href{https://doi.org/10.1007/JHEP02(2022)172}{\emph{JHEP} {\bfseries 02}
  (2022) 172} [\href{https://arxiv.org/abs/2109.06184}{{\ttfamily
  2109.06184}}].

\bibitem{Carrozza:2022xut}
S.~Carrozza, S.~Eccles and P.~A. Hoehn, \emph{{Edge modes as dynamical frames:
  charges from post-selection in generally covariant theories}},
  \href{https://doi.org/10.21468/SciPostPhys.17.2.048}{\emph{SciPost Phys.}
  {\bfseries 17} (2024) 048}
  [\href{https://arxiv.org/abs/2205.00913}{{\ttfamily 2205.00913}}].

\bibitem{Goeller:2022rsx}
C.~Goeller, P.~A. Hoehn and J.~Kirklin, \emph{{Diffeomorphism-invariant
  observables and dynamical frames in gravity: reconciling bulk locality with
  general covariance}},  \href{https://arxiv.org/abs/2206.01193}{{\ttfamily
  2206.01193}}.

\bibitem{Chandrasekaran:2022cip}
V.~Chandrasekaran, R.~Longo, G.~Penington and E.~Witten, \emph{{An algebra of
  observables for de Sitter space}},
  \href{https://doi.org/10.1007/JHEP02(2023)082}{\emph{JHEP} {\bfseries 02}
  (2023) 082} [\href{https://arxiv.org/abs/2206.10780}{{\ttfamily
  2206.10780}}].

\bibitem{Loganayagam:2023pfb}
R.~Loganayagam and O.~Shetye, \emph{{Influence phase of a dS observer. Part I.
  Scalar exchange}}, \href{https://doi.org/10.1007/JHEP01(2024)138}{\emph{JHEP}
  {\bfseries 01} (2024) 138}
  [\href{https://arxiv.org/abs/2309.07290}{{\ttfamily 2309.07290}}].

\bibitem{Shaghoulian:2023odo}
E.~Shaghoulian, \emph{{Quantum gravity and the measurement problem in quantum
  mechanics}},  \href{https://arxiv.org/abs/2305.10635}{{\ttfamily
  2305.10635}}.

\bibitem{Blacker:2023oan}
M.~J. Blacker and S.~A. Hartnoll, \emph{{Cosmological quantum states of de
  Sitter-Schwarzschild are static patch partition functions}},
  \href{https://doi.org/10.1007/JHEP12(2023)025}{\emph{JHEP} {\bfseries 12}
  (2023) 025} [\href{https://arxiv.org/abs/2304.06865}{{\ttfamily
  2304.06865}}].

\bibitem{Witten:2023qsv}
E.~Witten, \emph{{Algebras, regions, and observers.}},
  \href{https://doi.org/10.1090/pspum/107/01954}{\emph{Proc. Symp. Pure Math.}
  {\bfseries 107} (2024) 247}
  [\href{https://arxiv.org/abs/2303.02837}{{\ttfamily 2303.02837}}].

\bibitem{Kolchmeyer:2024fly}
D.~K. Kolchmeyer and H.~Liu, \emph{{Chaos and the Emergence of the Cosmological
  Horizon}},  \href{https://arxiv.org/abs/2411.08090}{{\ttfamily 2411.08090}}.

\bibitem{Abdalla:2025gzn}
A.~I. Abdalla, S.~Antonini, L.~V. Iliesiu and A.~Levine, \emph{{The
  gravitational path integral from an observer's point of view}},
  \href{https://arxiv.org/abs/2501.02632}{{\ttfamily 2501.02632}}.

\bibitem{Harlow:2025pvj}
D.~Harlow, M.~Usatyuk and Y.~Zhao, \emph{{Quantum mechanics and observers for
  gravity in a closed universe}},
  \href{https://arxiv.org/abs/2501.02359}{{\ttfamily 2501.02359}}.

\bibitem{Akers:2025ahe}
C.~Akers, G.~Bueller, O.~DeWolfe, K.~Higginbotham, J.~Reinking and
  R.~Rodriguez, \emph{{On observers in holographic maps}},
  \href{https://arxiv.org/abs/2503.09681}{{\ttfamily 2503.09681}}.

\bibitem{Shaghoulian:2021cef}
E.~Shaghoulian, \emph{{The central dogma and cosmological horizons}},
  \href{https://doi.org/10.1007/JHEP01(2022)132}{\emph{JHEP} {\bfseries 01}
  (2022) 132} [\href{https://arxiv.org/abs/2110.13210}{{\ttfamily
  2110.13210}}].

\bibitem{Levine:2022wos}
A.~Levine and E.~Shaghoulian, \emph{{Encoding beyond cosmological horizons in
  de Sitter JT gravity}},
  \href{https://doi.org/10.1007/JHEP02(2023)179}{\emph{JHEP} {\bfseries 02}
  (2023) 179} [\href{https://arxiv.org/abs/2204.08503}{{\ttfamily
  2204.08503}}].

\bibitem{Svesko:2022txo}
A.~Svesko, E.~Verheijden, E.~P. Verlinde and M.~R. Visser, \emph{{Quasi-local
  energy and microcanonical entropy in two-dimensional nearly de Sitter
  gravity}}, \href{https://doi.org/10.1007/JHEP08(2022)075}{\emph{JHEP}
  {\bfseries 08} (2022) 075}
  [\href{https://arxiv.org/abs/2203.00700}{{\ttfamily 2203.00700}}].

\bibitem{Maldacena:2024spf}
J.~Maldacena, \emph{{Real observers solving imaginary problems}},
  \href{https://arxiv.org/abs/2412.14014}{{\ttfamily 2412.14014}}.

\bibitem{Wall:2012uf}
A.~C. Wall, \emph{{Maximin Surfaces, and the Strong Subadditivity of the
  Covariant Holographic Entanglement Entropy}},
  \href{https://doi.org/10.1088/0264-9381/31/22/225007}{\emph{Class. Quant.
  Grav.} {\bfseries 31} (2014) 225007}
  [\href{https://arxiv.org/abs/1211.3494}{{\ttfamily 1211.3494}}].

\bibitem{allameh}
K.~Allameh and E.~Shaghoulian, \emph{{Modular invariance and thermal effective
  field theory in CFT}},  \href{https://arxiv.org/abs/2402.13337}{{\ttfamily
  2402.13337}}.

\bibitem{Odak:2021axr}
G.~Odak and S.~Speziale, \emph{{Brown-York charges with mixed boundary
  conditions}}, \href{https://doi.org/10.1007/JHEP11(2021)224}{\emph{JHEP}
  {\bfseries 11} (2021) 224}
  [\href{https://arxiv.org/abs/2109.02883}{{\ttfamily 2109.02883}}].

\bibitem{Jensen:2012jh}
K.~Jensen, M.~Kaminski, P.~Kovtun, R.~Meyer, A.~Ritz and A.~Yarom,
  \emph{{Towards hydrodynamics without an entropy current}},
  \href{https://doi.org/10.1103/PhysRevLett.109.101601}{\emph{Phys. Rev. Lett.}
  {\bfseries 109} (2012) 101601}
  [\href{https://arxiv.org/abs/1203.3556}{{\ttfamily 1203.3556}}].

\bibitem{Banerjee:2012iz}
N.~Banerjee, J.~Bhattacharya, S.~Bhattacharyya, S.~Jain, S.~Minwalla and
  T.~Sharma, \emph{{Constraints on Fluid Dynamics from Equilibrium Partition
  Functions}}, \href{https://doi.org/10.1007/JHEP09(2012)046}{\emph{JHEP}
  {\bfseries 09} (2012) 046} [\href{https://arxiv.org/abs/1203.3544}{{\ttfamily
  1203.3544}}].

\bibitem{horowitz}
G.~T. Horowitz and E.~Shaghoulian, \emph{{Detachable circles and
  temperature-inversion dualities for CFT$_{d}$}},
  \href{https://doi.org/10.1007/JHEP01(2018)135}{\emph{JHEP} {\bfseries 01}
  (2018) 135} [\href{https://arxiv.org/abs/1709.06084}{{\ttfamily
  1709.06084}}].

\bibitem{Benjamin:2023qsc}
N.~Benjamin, J.~Lee, H.~Ooguri and D.~Simmons-Duffin, \emph{{Universal
  Asymptotics for High Energy CFT Data}},
  \href{https://arxiv.org/abs/2306.08031}{{\ttfamily 2306.08031}}.

\bibitem{Chamblin:1999tk}
A.~Chamblin, R.~Emparan, C.~V. Johnson and R.~C. Myers, \emph{{Charged AdS
  black holes and catastrophic holography}},
  \href{https://doi.org/10.1103/PhysRevD.60.064018}{\emph{Phys. Rev. D}
  {\bfseries 60} (1999) 064018}
  [\href{https://arxiv.org/abs/hep-th/9902170}{{\ttfamily hep-th/9902170}}].

\bibitem{Hawking:1982dh}
S.~W. Hawking and D.~N. Page, \emph{{Thermodynamics of Black Holes in anti-De
  Sitter Space}}, \href{https://doi.org/10.1007/BF01208266}{\emph{Commun. Math.
  Phys.} {\bfseries 87} (1983) 577}.

\bibitem{Horowitz:1998ha}
G.~T. Horowitz and R.~C. Myers, \emph{{The AdS / CFT correspondence and a new
  positive energy conjecture for general relativity}},
  \href{https://doi.org/10.1103/PhysRevD.59.026005}{\emph{Phys. Rev. D}
  {\bfseries 59} (1998) 026005}
  [\href{https://arxiv.org/abs/hep-th/9808079}{{\ttfamily hep-th/9808079}}].

\bibitem{Shaghoulian:2015lcn}
E.~Shaghoulian, \emph{{Black hole microstates in AdS}},
  \href{https://doi.org/10.1103/PhysRevD.94.104044}{\emph{Phys. Rev. D}
  {\bfseries 94} (2016) 104044}
  [\href{https://arxiv.org/abs/1512.06855}{{\ttfamily 1512.06855}}].

\bibitem{Adam:2011dn}
A.~Adam, S.~Kitchen and T.~Wiseman, \emph{{A numerical approach to finding
  general stationary vacuum black holes}},
  \href{https://doi.org/10.1088/0264-9381/29/16/165002}{\emph{Class. Quant.
  Grav.} {\bfseries 29} (2012) 165002}
  [\href{https://arxiv.org/abs/1105.6347}{{\ttfamily 1105.6347}}].

\bibitem{Hubeny:2011hd}
V.~E. Hubeny, S.~Minwalla and M.~Rangamani, \emph{{The fluid/gravity
  correspondence}},  in \emph{{Theoretical Advanced Study Institute in
  Elementary Particle Physics}: {String theory and its Applications: From meV
  to the Planck Scale}}, pp.~348--383, 2012,
  \href{https://arxiv.org/abs/1107.5780}{{\ttfamily 1107.5780}}.

\bibitem{Banados:1998ys}
M.~Banados and F.~Mendez, \emph{{A Note on covariant action integrals in
  three-dimensions}},
  \href{https://doi.org/10.1103/PhysRevD.58.104014}{\emph{Phys. Rev. D}
  {\bfseries 58} (1998) 104014}
  [\href{https://arxiv.org/abs/hep-th/9806065}{{\ttfamily hep-th/9806065}}].

\bibitem{Parvizi:2025shq}
A.~Parvizi, M.~M. Sheikh-Jabbari and V.~Taghiloo, \emph{{Freelance Holography,
  Part I: Setting Boundary Conditions Free in Gauge/Gravity Correspondence}},
  \href{https://arxiv.org/abs/2503.09371}{{\ttfamily 2503.09371}}.

\bibitem{Parvizi:2025wsg}
A.~Parvizi, M.~M. Sheikh-Jabbari and V.~Taghiloo, \emph{{Freelance Holography,
  Part II: Moving Boundary in Gauge/Gravity Correspondence}},
  \href{https://arxiv.org/abs/2503.09372}{{\ttfamily 2503.09372}}.

\bibitem{Mertens:2022irh}
T.~G. Mertens and G.~J. Turiaci, \emph{{Solvable models of quantum black holes:
  a review on Jackiw\textendash{}Teitelboim gravity}},
  \href{https://doi.org/10.1007/s41114-023-00046-1}{\emph{Living Rev. Rel.}
  {\bfseries 26} (2023) 4} [\href{https://arxiv.org/abs/2210.10846}{{\ttfamily
  2210.10846}}].

\bibitem{Nayak:2018qej}
P.~Nayak, A.~Shukla, R.~M. Soni, S.~P. Trivedi and V.~Vishal, \emph{{On the
  Dynamics of Near-Extremal Black Holes}},
  \href{https://doi.org/10.1007/JHEP09(2018)048}{\emph{JHEP} {\bfseries 09}
  (2018) 048} [\href{https://arxiv.org/abs/1802.09547}{{\ttfamily
  1802.09547}}].

\bibitem{Coleman:2020jte}
E.~Coleman and V.~Shyam, \emph{{Conformal boundary conditions from cutoff
  AdS$_{3}$}}, \href{https://doi.org/10.1007/JHEP09(2021)079}{\emph{JHEP}
  {\bfseries 09} (2021) 079}
  [\href{https://arxiv.org/abs/2010.08504}{{\ttfamily 2010.08504}}].

\bibitem{Miskovic:2006tm}
O.~Miskovic and R.~Olea, \emph{{On boundary conditions in three-dimensional AdS
  gravity}}, \href{https://doi.org/10.1016/j.physletb.2006.07.045}{\emph{Phys.
  Lett. B} {\bfseries 640} (2006) 101}
  [\href{https://arxiv.org/abs/hep-th/0603092}{{\ttfamily hep-th/0603092}}].

\bibitem{Detournay:2014fva}
S.~Detournay, D.~Grumiller, F.~Sch\"oller and J.~Sim\'on, \emph{{Variational
  principle and one-point functions in three-dimensional flat space Einstein
  gravity}}, \href{https://doi.org/10.1103/PhysRevD.89.084061}{\emph{Phys. Rev.
  D} {\bfseries 89} (2014) 084061}
  [\href{https://arxiv.org/abs/1402.3687}{{\ttfamily 1402.3687}}].

\bibitem{Krishnan:2016mcj}
C.~Krishnan and A.~Raju, \emph{{A Neumann Boundary Term for Gravity}},
  \href{https://doi.org/10.1142/S0217732317500778}{\emph{Mod. Phys. Lett. A}
  {\bfseries 32} (2017) 1750077}
  [\href{https://arxiv.org/abs/1605.01603}{{\ttfamily 1605.01603}}].

\bibitem{Maldacena:2016upp}
J.~Maldacena, D.~Stanford and Z.~Yang, \emph{{Conformal symmetry and its
  breaking in two dimensional Nearly Anti-de-Sitter space}},
  \href{https://doi.org/10.1093/ptep/ptw124}{\emph{PTEP} {\bfseries 2016}
  (2016) 12C104} [\href{https://arxiv.org/abs/1606.01857}{{\ttfamily
  1606.01857}}].

\bibitem{Achucarro:1986uwr}
A.~Achucarro and P.~K. Townsend, \emph{{A Chern-Simons Action for
  Three-Dimensional anti-De Sitter Supergravity Theories}},
  \href{https://doi.org/10.1016/0370-2693(86)90140-1}{\emph{Phys. Lett. B}
  {\bfseries 180} (1986) 89}.

\bibitem{Witten:1988hc}
E.~Witten, \emph{{(2+1)-Dimensional Gravity as an Exactly Soluble System}},
  \href{https://doi.org/10.1016/0550-3213(88)90143-5}{\emph{Nucl. Phys. B}
  {\bfseries 311} (1988) 46}.

\bibitem{Saad:2019lba}
P.~Saad, S.~H. Shenker and D.~Stanford, \emph{{JT gravity as a matrix
  integral}},  \href{https://arxiv.org/abs/1903.11115}{{\ttfamily 1903.11115}}.

\bibitem{Coussaert:1995zp}
O.~Coussaert, M.~Henneaux and P.~van Driel, \emph{{The Asymptotic dynamics of
  three-dimensional Einstein gravity with a negative cosmological constant}},
  \href{https://doi.org/10.1088/0264-9381/12/12/012}{\emph{Class. Quant. Grav.}
  {\bfseries 12} (1995) 2961}
  [\href{https://arxiv.org/abs/gr-qc/9506019}{{\ttfamily gr-qc/9506019}}].

\bibitem{allameh2}
K.~Allameh and E.~Shaghoulian, \emph{{work in progress}}, .

\bibitem{Donos:2011bh}
A.~Donos and J.~P. Gauntlett, \emph{{Holographic striped phases}},
  \href{https://doi.org/10.1007/JHEP08(2011)140}{\emph{JHEP} {\bfseries 08}
  (2011) 140} [\href{https://arxiv.org/abs/1106.2004}{{\ttfamily 1106.2004}}].

\end{thebibliography}\endgroup
